\documentclass[final,3p,times,authoryear]{elsarticle}
\usepackage{geometry}
\usepackage{amssymb}
\usepackage{mathrsfs}
\usepackage{rotating}
\UseRawInputEncoding
\usepackage{epsfig}
\usepackage{graphicx}
\usepackage[table,dvipsnames]{xcolor}
\usepackage{dcolumn}
\usepackage{tcolorbox}
\usepackage{bm}
\usepackage{times}
\usepackage{xcolor}
\usepackage{makecell}%
\usepackage[percent]{overpic}
\usepackage{cancel}
\usepackage{amsmath}
\usepackage{multirow}
\usepackage{booktabs} 
\usepackage{hyperref}
\usepackage[hyphenbreaks]{breakurl}
\usepackage{mathrsfs}
\usepackage{multirow}
\usepackage{longtable}
\usepackage{tabu}
\usepackage{lscape} 
\usepackage{setspace}
\usepackage{dsfont}
\usepackage{extarrows}
\usepackage{bbm}
\usepackage{fancybox}
\usepackage{subfigure}
\usepackage{natbib}
\setcitestyle{numbers,square}
\usepackage{pdflscape}
\usepackage[utf8]{inputenc}     
\usepackage{tabularx}
\usepackage{xltabular}

\begin{document}
\begin{sloppypar}

\begin{frontmatter}

\title{Large language models for spreading dynamics in complex systems}

\author{Shuyu Jiang$^{1,2,3}$, Hao Ren$^{1,2,3}$, Yichang Gao$^{4}$, Yi-Cheng Zhang$^{5}$, Li Qi$^{6}$, Dayong Xiao$^{6}$, Jie Fan$^{*7}$, Rui Tang$^{*1,2,3}$, Wei Wang$^{*8}$} 
\cortext[cor1]{Corresponding Authors: 241031@cqmpc.edu.cn (Jie Fan), tangrscu@scu.edu.cn (Rui Tang), wwzqbx@hotmail.com (Wei Wang)}

\setcounter{footnote}{1}
\footnotetext{This paper has been accepted for publication in Physics Reports.}

\address{1. School of Cyber Science and Engineering, Sichuan University, Chengdu, 610065, China}
\address{2.Key Laboratory of Data Protection and Intelligent Management (Sichuan University), Ministry of Education, Chengdu, 610065, China}
\address{3. Cyber Science Research Institute, Sichuan University, Chengdu, 610065, China}
\address{4. Royal Melbourne Institute of Technology University, GPO Box 2476, Melbourne VIC 3001, Australia}
\address{5. Physics Department, University of Fribourg, Chemin du Musée 3, 1700 Fribourg, Switzerland}
\address{6. Chongqing Center for Disease Control and Prevention (Chongqing Academy of Preventive Medicine), Chongqing 400070, China}
\address{7. School of Public Health and Emergency Management, Chongqing Medical and Pharmaceutical College, Chongqing 401331, China}

\address{8. School of Public Health, Chongqing Medical University, Chongqing, 400016, China}

\begin{abstract}
Spreading dynamics is a central topic in the physics of complex systems and network science, providing a unified framework for understanding how information, behaviors, and diseases propagate through interactions among system units. In many propagation contexts, spreading processes are influenced by multiple interacting factors, such as information expression patterns, cultural contexts, living environments, cognitive preferences, and public policies, which are difficult to incorporate directly into classical modeling frameworks. Recently, large language models (LLMs) have exhibited strong capabilities in natural language understanding, reasoning, and generation, enabling explicit perception of semantic content and contextual cues in spreading processes, thereby supporting the analysis of the different influencing factors. Beyond serving as external analytical tools, LLMs can also act as interactive agents embedded in propagation systems, potentially influencing spreading pathways and feedback structures. Consequently, the roles and impacts of LLMs on spreading dynamics have become an active and rapidly growing research area across multiple research disciplines. This review provides a comprehensive overview of recent advances in applying LLMs to the study of spreading dynamics across two representative domains: digital epidemics, such as misinformation and rumors, and biological epidemics, including infectious disease outbreaks. We first examine the foundations of epidemic modeling from a complex-systems perspective and discuss how LLM-based approaches relate to traditional frameworks. We then systematically review recent studies from three key perspectives, which are epidemic modeling, epidemic detection and surveillance, and epidemic prediction and  management, to clarify how LLMs enhance these areas. Finally, open challenges and potential research directions are discussed.
\end{abstract}


\end{frontmatter}

\newpage
\tableofcontents
\newpage
\section{Introduction} 
\label{intro}
In our world, a wide variety of physical, social, biological, and technological systems exhibit complex structures and
interactions~\cite{strogatz2001exploring}. Due to their ubiquity and complexity, these systems have attracted sustained
attention from physicists, biologists, ecologists, economists, and social scientists~\cite{ottino2004engineering}.
Researchers are interested not only in individual systems, but also in classes of systems that, while differing in
nature, share similar structural characteristics~\cite{bar2019dynamics}. This interest stems from the enduring
challenge of achieving an accurate and comprehensive description of different complex systems, which has long been
considered one of humanity's greatest scientific problems~\cite{wilson1999consilience}. To meet this challenge,
scientists have sought to reassemble these systems, at least through mathematical models that encapsulate the essential
characteristics of the entire ensemble~\cite{strogatz2001exploring,wilson1999consilience}. In fact, although different
complex systems exhibit diverse behaviors and phenomena, their underlying mathematical representations often rely on
similar modeling paradigms~\cite{vespignani2012modelling}. Among these paradigms, the complex network stands out as a
particularly effective and representative approach. In its simplest form, the fundamental units
of a system are represented as nodes, while interactions among them are described by edges,
providing a concise abstraction for analyzing structural and dynamical
properties~\cite{boccaletti2006complex,de2016physics}. 

\begin{figure*}[htbp]
        \centering
    \includegraphics[width=1.\linewidth]{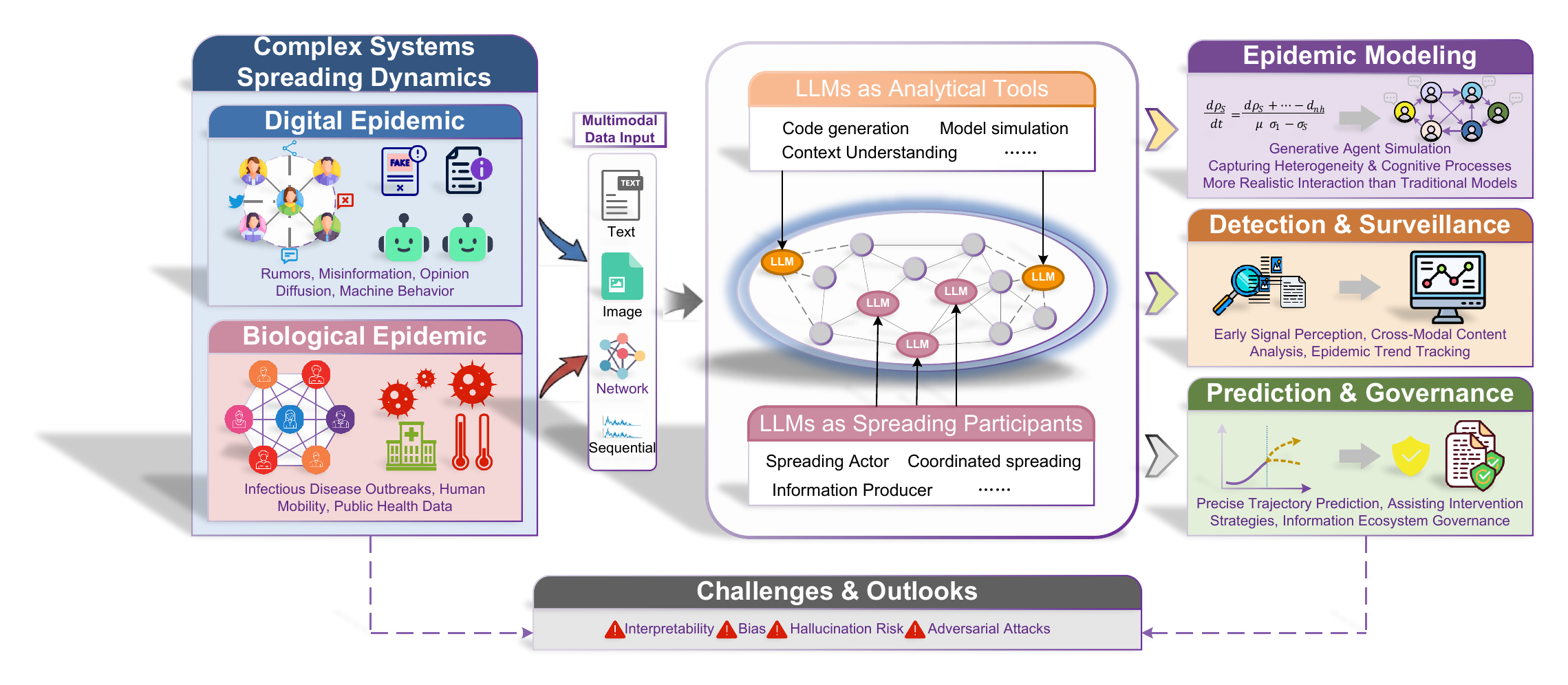}
    \caption{\textcolor{black}{Overview of the paper structure, illustrating  the Spreading dynamics spreading analysis that integrates LLMs.
    On the left, biological epidemics and digital epidemics are shown as two representative classes of spreading processes. In the center, heterogeneous data sources, including textual, network, and multimodal signals, provide observations of spreading processes. LLMs are depicted as a core component that interacts with these data, acting both as analytical tools for processing and reasoning over spreading information and as active entities that may participate in information generation and dissemination. On the right, the figure organizes LLM-enhanced applications across key stages of spreading workflows, including modeling, detection, surveillance, prediction, and governance. The bottom part summarizes representative challenges and risks associated with LLM-enhanced spreading systems.}}
    \label{fig1-1}
\end{figure*}

Since complex networks serve as powerful representations of complex systems, the propagation phenomena intrinsic to these systems can be theoretically characterized through spreading dynamics that unfold over their corresponding network structures, with propagating items, such as infectious diseases { or information (especially low-risk information)} being treated as ``epidemics'' that flow or diffuse between nodes~\cite{notarmuzi2022universality, centola2007complex, guilbeault2018complex,mariani2024collective,keeling2008modeling,centola2018behavior,hethcote2000mathematics,ferraz2024contagion}. More specifically, interactions among different individual nodes enable the spreading process to exhibit a structurally organized temporal pattern and group characteristics. Characterizing spreading dynamics is not only a description of individual behavior, but also a crucial approach to understanding how group systems evolve with time. Through analyzing the spreading dynamics, scientists can observe how groups form consensus or diverge, how networks form breakouts or decay, how risks move between regions and communities, and how local perturbations trigger systemic responses. Based on this, systematically uncovering the underlying patterns and key mechanisms of spreading processes lays the foundations for predicting their future trajectories, { mitigating} undesirable propagation, formulating effective management strategies, and so on~\cite{wang2019coevolution}. 

Classical spreading dynamics research typically draws on theories from statistical physics, network science, applied mathematics, and epidemiology to simplify and abstract complex spreading processes, yielding models ranging from simple compartmental models like susceptible--infected--susceptible (SIS) and susceptible--infected--recovered (SIR) models to large-scale agent-based models (ABMs) that simulate individuals' movements within cities or across countries~\cite {kiss2017mathematics, wang2024epidemic, de2016physics}.
These models enable a theoretical explanation of how primary mechanisms influence important characteristics, playing a crucial role in revealing overall system evolution patterns and key behaviors.
As research has progressed, it has become increasingly recognized that spreading processes in real-world complex systems are influenced by a variety of intertwined factors, such as information expression patterns, cultural backgrounds, living environments, cognitive preferences, and public policies~\cite{gao2024large, macal2005tutorial}. 
Many of these factors are difficult to incorporate explicitly using fixed parameters or predefined rules, and are therefore often sacrificed to preserve analytical tractability and computational efficiency~\cite{gao2024large,davis2020phase}. { In response, researchers have continuously explored alternative modeling and analysis approaches.}

{ With the rapid development of artificial intelligence (AI), particularly the widespread deployment and growing visibility of LLMs in recent years, exemplified by ChatGPT and followed by products such as Google Gemini and DeepSeek~\cite{GPT-4, DeepSeek-R1, oh2024llm, romera2024mathematical, kumar2024large}, new opportunities have emerged for perceiving and simulating the influence of human cognition, evolving policy measures, daily behavioral habits, and even collective responses such as rumor-induced panic on spreading processes, offering new perspectives for understanding the mechanisms and dynamical patterns underlying spreading phenomena.} 

{ These developments can be attributed to two aspects. First, LLMs are trained in a self-supervised manner~\cite{gui2024survey} through large-scale contextual prediction on diverse textual corpora, including a wide spectrum of textual sources~\cite{chowdhery2023palm}, ranging from canonical works and authoritative writings that reflect accumulated human knowledge, to formal records of professional and institutional activities across diverse domains, as well as large volumes of informal, multilingual online texts capturing everyday communication, personal expression, and social interaction. Through this process, high-dimensional neural representations capture rich statistical regularities and implicitly integrate broad, cross-domain knowledge related to language use, real-world contexts, and patterns of human behavior, thereby supporting advanced capabilities in natural language understanding and generation. As a result, LLMs have been adopted by many fields and even begun to influence labor demand~\cite{eloundou2024gpts,fenoaltea2024follow}.} 
For example, in natural sciences, LLMs have been employed for literature synthesis, experimental process assistance, and cross-disciplinary concept association~\cite{swanson2025virtual}; in biomedicine, LLMs can be used for disease knowledge interpretation, drug target description, and treatment recommendation generation~\cite{de2024assessing}; in economics and finance, LLMs can integrate semantic information from market narratives, policy documents, and enterprise data to enable richer contextual understanding and strategic comparison~\cite{kong2024large}; in social environments, they can participate in content generation, thematic extension, and context-aware dialogue~\cite{ yang2025socialmind, kumar2024large}. Within the study of spreading dynamics, these capabilities position LLMs as versatile and effective tools for analyzing propagation-related information and behaviors.

{ Second, beyond their role as analytical tools, LLMs are increasingly shaping research across multiple disciplines, including behavioral science~\cite{brinkmann2023machine}, sociology~\cite{brinkmann2023machine,anthisposition,kozlowski2024simulating}, political science~\cite{schroeder2025malicious,li2024political}, and cognitive science~\cite{lu2024llms}, and are beginning to affect economic activities, social practices, and cultural production in contemporary society. From human-like accounts controlled by LLMs on social platforms to conversational agents embedded in autonomous vehicles and digital services, LLM-driven agents are rapidly expanding their presence across communication channels, social interactions, economic activities, and everyday life. As a consequence, complex social systems can no longer be viewed as composed solely of human agents, but increasingly involve heterogeneous populations of humans and AI-driven agents that interact and co-evolve~\cite{tsvetkova2024new}. Moreover, interactions between humans and LLMs can give rise to sustained feedback loops, in which human-generated data influence model training, while model-generated content, in turn, shapes subsequent human preferences and behaviors~\cite{pedreschi2025human}. Such human–AI co-evolutionary processes introduce new forms of coupling and nonlinearity into spreading dynamics, with potential implications for both information diffusion and disease-related behaviors.} Consequently, LLMs not only serve as tools for external observation and analysis, but can also become active participants in spreading processes, directly coupling with and influencing the dynamics of propagation itself~\cite{bai2025llm, radivojevic2024llms}.

Against this background, research on spreading dynamics has exhibited new shifts. On the one hand, the content being propagated is no longer viewed as abstract labels or simple payloads. Instead, it can be modeled through the semantics, narrative patterns, and expression strategies captured by LLMs, making the content itself a modelable component of the spreading process~\cite{xi2025rise, lu2024llms}. 
For example, in agent-based modeling based on LLMs, agents can express their intentions through language and generate context-sensitive behaviors based on their perception of a multimodal environment. This allows behavioral updates during the spreading process to no longer rely on fixed transition rules but instead to be determined by cognition, experience, and social context, reflecting the heterogeneity and dynamic uncertainty that more closely resemble real-world interaction situations~\cite{lu2024llms, gao2024large,park2023generative}. 

On the other hand, the interaction between LLMs and humans or multi-agent systems can affect how individuals acquire information, form understandings, and respond behaviorally, thus affecting the pathways, speed, and feedback structure of information spreading to some extent~\cite{shen2025understanding,bai2025llm, radivojevic2024llms,piao2025social}. For instance, in digital information spreading on social media, { populations of agents driven by LLMs can exhibit rich and context-dependent heterogeneity through language-based reasoning and interaction, without relying on prespecified behavioral rules or extensive empirical calibration. As such agents increasingly participate directly in spreading processes, they introduce new forms of dynamically generated heterogeneity compared with the traditional simple contagion models~\cite{smilkov2014beyond,seymour2022bayesian,aral2018social,gomez2012inferring} and complex contagion models~\cite{valente1996social,tanase2024integrating}, which raise new questions and challenges for the study of spreading dynamics.} In the context of biological epidemic spreading, individuals may obtain explanations, suggestions, or treatment comparisons through interactions with LLMs when facing uncertainty, and this feedback may influence their behavioral choices. For instance, in health-related scenarios, users may describe symptoms in natural language and receive suggestions regarding risk and protective strategies. This can influence their decisions regarding travel, social interaction, medical consultation, or isolation, thereby indirectly reshaping the population's contact network.

Therefore, it is necessary to systematically review and summarize the relationship between LLMs and spreading dynamics to clarify their roles and potential impact within these processes. Specifically, there are three reasons for us to write this review. First, LLMs can not only facilitate the revelation and analysis of propagation mechanisms and phenomena but may also act as new participants in spreading processes within complex systems, thereby introducing novel patterns of interaction and diffusion. However, it remains unclear how LLMs can effectively enhance our understanding of spreading dynamics and what new changes emerge when they themselves become active agents in these processes, an issue that has not yet been systematically examined. Second, in recent years, a growing number of researchers have devoted substantial effort to this emerging area, uncovering intriguing phenomena and producing a variety of novel findings. Despite this progress, there is still a lack of systematic synthesis of these studies, making it difficult for new researchers or interested scholars to quickly grasp the current state and latest developments.
In addition, it remains an open question how LLMs can transform traditional propagation-related tasks into generative paradigms in which these models excel, thereby providing more effective analytical or decision-support services. At the same time, the use of LLMs in spreading-related research introduces several limitations, risks, and challenges that have not yet been comprehensively analyzed.

Motivated by these considerations, this review focuses on two representative categories of complex systems: cyberspace complex systems, which encompass digital epidemics such as rumors, misinformation, and ideological diffusion, and { physical} complex systems, which involve biological epidemics such as influenza~\cite{peacock2025global}, pneumonia~\cite{budia2024international}, dengue~\cite{ni2024epidemiological,aguiar2022mathematical}, and Ebola~\cite{nasrin2025dynamical,keita2021resurgence}, among others. By examining these two paradigmatic domains, this review aims to advance research on how LLMs can enhance the study of spreading dynamics and to deepen the understanding of the intrinsic characteristics of spreading processes when LLMs themselves become active participants. Recent related studies are systematically surveyed and organized into three major aspects: epidemic modeling, epidemic detection and surveillance, and epidemic prediction and { management. Figure \ref{fig1-1} illustrates the architecture of this paper, and Table \ref{tab:all_method} summarizes representative studies that incorporate LLMs to enhance different stages of spreading processes. For each study, the table highlights the primary stage addressed and the functional role played by LLMs within the corresponding framework.}
In Section~\ref{sec2}, the background of traditional spreading models and LLMs is introduced. Section~\ref{sec3} reviews LLM-based epidemic modeling methods and their influence on spreading processes. Section~\ref{sec4} investigates the application of LLMs in epidemic detection and surveillance, illustrating how they perceive spreading objects and processes. Section~\ref{sec5} discusses the use of LLMs in the prediction of epidemics and the support of management. Finally, Section~\ref{sec6} identifies open issues and outlines potential directions for future research.

\begin{xltabular}{\textwidth}{p{1.5cm}lp{3.2cm}X}
\caption{\textcolor{black}{Overview of LLM-related studies about spreading processes.}}
\label{tab:all_method} \\
\toprule
\rowcolor{blue!10}
\textbf{Type} & \textbf{Research} & \textbf{LLM} & \textbf{Role of LLM} \\ 
\midrule
\endfirsthead

\caption[]{\textcolor{black}{Overview of LLM-related studies about spreading processes.}} \\
\toprule
\rowcolor{blue!10}
\textbf{Type} & \textbf{Research} & \textbf{LLM} & \textbf{Role of LLM} \\ 
\midrule
\endhead

\midrule
\endfoot

\bottomrule
\endlastfoot
Modeling & Williams et al.~\cite{williams2023epidemic} & gpt-3.5-turbo-0301 & Simulate human behavior in epidemic spread\\
& Li et al.~\cite{li2024large} & gpt-3.5-turbo-1106 & Model dynamics of misinformation spread\\
& FPS~\cite{liu2024skepticism} & gpt-3.5-turbo-1106 & Simulate the attitude dynamics toward fake news\\
& GAG~\cite{ji2025llm} & Llama-3, gpt-3.5-turbo & Model graph evolution\\
& SSF~\cite{wang2025decoding} & gpt-4o-mini & Simulate polarization and echo chamber effects\\
& Chuang et al.~\cite{chuang2024simulating} & gpt-3.5-turbo-16k & Simulate opinion dynamics\\
& FUSE~\cite{liu2025stepwise} & gpt-4o-mini & Simulate how true news progressively distorts into fake news\\
& SMIST~\cite{ma2024simulated} & gpt-4 & Simulate misinformation susceptibility\\
& Mou et al.~\cite{mou2024unveiling}  & gpt-3.5-turbo & Simulate core users in social movements\\
& FDE-LLM~\cite{yao2025social} & ChatGLM-4 & Simulate the attitudes and behaviors of opinion leaders\\
& RumorSphere~\cite{liu2025rumorsphere} & gpt-4o-mini & Model core agents in rumor propagation\\
& Cheng et al.~\cite{cheng2025interactive} & Qwen2.5-3B-Instruct, gpt-3.5 & Simulate public event dynamics\\
& EIN~\cite{jiang2025epidemiology} & Gemma-2-9B & Generate stance labels for user responses\\
& BotSim~\cite{qiao2025botsim} & gpt-4o-mini & Construct user nodes that generate and propagate harmful information\\
& Radivojevic et al.~\cite{radivojevic2024llms} & gpt-4, Llama-2, Claude-2 & Simulate interactions with real online users\\
& De et al.~\cite{de2023emergence} & gpt-3.5-turbo & Simulate social network growth and topology formation\\
& Chang et al.~\cite{chang2025llms} & gpt-3.5-turbo, gpt-4o & Generate social networks\\
& Papachristou et al.~\cite{papachristou2025network}  & gpt-3.5, gpt-4 & Generate friendship, telecommunication, and employment networks\\
& Lu et al.~\cite{lu2025understanding} & gpt-4o & Generate adversarial social influence content\\
& Bandara~\cite{bandara2024hallucination} & - & Amplify conspiracy theories and fake news\\
& Villaplana et al.~\cite{villaplana2024application} & gpt-3.5 & Simulate human behavior in the context of an ongoing pandemic \\
& EpiLLM~\cite{gong2025epillm} & gpt-2, DeepSeek-R1, Gemma-3 & Model the temporal evolution of epidemic transmission\\
& MIFlu~\cite{moon2025miflu} & gpt-2 & Fuse data from different modalities\\
& PandemicLLM~\cite{du2025advancing} & Llama-2 & Fuse data from different modalities\\
& LLM-DG~\cite{kang2025llm} & gpt & Enrich representations of patients, diseases, and discharge summaries\\
& Kwok et al.~\cite{kwok2024utilizing} & gpt & Support the development and validation of disease transmission model\\
& Zaslavsky et al.\cite{zaslavsky2024enhancing} & gpt-3.5, Claude-2 & Enhance spatially-disaggregated simulation\\
& Wang et al.~\cite{wang2024large} & gpt-3.5 & Generate individual activity trajectory data\\
& Wu et al.~\cite{wu2025llm} & - & Explain social patterns, construct theories, and provide interpretable foundations for hypothesis generation\\
& Nudo et al.~\cite{nudo2026generative} & Gemini, Mistral, DeepSeek & Simulate political discourse on social media\\
& AgentSociety~\cite{piao2025agentsociety} & DeepSeek-V3 & Simulate large-scale interactions among agents and between agents and the environment\\
& Light Society~\cite{guan2025modeling} & gemini-2.0-flash-001, gpt-4.1-nano & Model social trust and information diffusion\\
& TinyTroupe~\cite{salem2025tinytroupe} & gpt-4o-mini & Simulate real human behavior \\
\midrule
Detection & ARG~\cite{hu2024bad} & gpt-3.5-turbo & Provide multi-perspective instructive rationales \\
& GenFEND~\cite{nan2024let} & GLM-4, gpt-3.5-turbo & Simulate social media users and generate diverse comments\\
& LoCal~\cite{ma2025local} & Flan-T5, gpt-3.5-turbo & Fact check\\
& SynthX~\cite{zong2025empowering} & gpt-4 & Synthesize both misinformation detection and explanation outputs\\
& MiLk-FD~\cite{xie2024multiknowledge} & Llama-2 & Obtain excellent initial feature vectors from news content\\
& SR$^3$~\cite{zhang2025llms} & Llama-3 & Generate both positive and negative reasoning for news\\
& F3~\cite{lucas2023fighting} & gpt-3.5-turbo, Llama-2, Dolly-2, Palm-2 & Generate and detect disinformation\\
& LESS4FD~\cite{ma2024fake} & gpt-3.5, Llama-2 & Extract and embed news topics and real entities\\
& MIL~\cite{yang2025llm} & gpt & Generate stance explanations for a post\\
& Krykoniuk et al.~\cite{krykoniuk2024detecting} & T0, all-Mini-LM-L6-v2 & Generate expected replies\\
& MAGE-fend~\cite{hu2025mage} & Qwen-VL, GLM-4 & Provide inferred rationale texts for classification\\
& DEEP~\cite{chen2025you} & LLaVA-13B & Generate supporting and refuting evidence\\
& FCRV~\cite{bai2024large} & - & Identify and extract claim and entity\\
& Zeng et al.~\cite{zeng2024combining} & BERT-large, DeBERTa-large, RoBERTa-large & Detect misinformation\\
& TED~\cite{liu2025truth} & gpt-4o-mini, BERT & Simulate the debate process\\
& FactAgent~\cite{li2024largefake} & gpt-3.5-turbo &  Emulate human expert behavior in verifying news claims\\
& OKI~\cite{xie2024integrating} & gpt, LLaVA-13B & Integrate open-domain knowledge\\
& Costabile et al.~\cite{costabile2025assessing} & Llama-3.1-8B, Gemma-2-9B, Mistral-7B & Retrieve evidence, assess claims, and determine veracity\\
& DELL~\cite{wan2024dell} & Mistral-7B, Llama2-70B, gpt & Generate news reactions , generate explanations for proxy tasks, merge task-specific experts\\
& NIMGA~\cite{li2025semantic} & gpt-3.5-turbo & Refine diverse perspectives and semantic evolution\\
& SePro~\cite{zeng2025exploring} & gpt-3.5-turbo-0125 & Detect rumors and generate explanations\\
& GLPN-LLM~\cite{hu2025synergizing} & gpt-4o & Generate pseudo labels\\
\midrule
Surveillance & Deiner et al.~\cite{deiner2025use} & gpt-3.5-turbo-0613, gpt-4-0314, gpt-4o-2024-05-13, Claude-3, Mixtral-8x22b, Llama-3-70B & Classify epidemiological characteristics in synthetic and real-world social media posts\\
& Deiner et al.~\cite{deiner2024use} & gpt-3.5, gpt-4 & Assess the likelihood of epidemics from the content of tweets\\
& Kaur et al.~\cite{kaur2025ai} & - & Process multi-source information across diverse linguistic environments \\
& Xie et al.~\cite{xie2025leveraging} & BERT, RoBERTa, GPT-2, BLOOM, Llama-2 & Identify and analyze posts related to COVID-19 infection\\
& Consoli et al.~\cite{consoli2024epidemic} & gpt & Extract valuable epidemic information\\
& Sufi et al.~\cite{sufi2024innovative} & gpt-4 & Analyze COVID topics\\
\midrule
Prediction & Wang et al.~\cite{wang2024news} & gpt-4-turbo & Understand and filter news content, integrate textual information into time series prediction\\
& CARE~\cite{zhong2024information} & - & Infer the next activation of users in social networks \\
& Diffusion-LLM~\cite{shang2025leveraging} & - & Predict diffusion in social networks \\
& Chen et al.~\cite{chen2024predicting} & Qwen &  Predict social media popularity \\
& AutoCas~\cite{zheng2025autocas} & gpt-2, Llama-3.2, Qwen2.5 & Predict cascade popularity\\
& ComPaSC$^3$~\cite{gao2025social} & Llama-3.1-8B & Transform videos and community characteristics into semantically enriched representations\\ 
& ALL-RP~\cite{sun2025learning} & gpt-Neo & Learning temporal user features for repost prediction\\
& S$^3$~\cite{gao2023s3} & gpt-3.5, GLM-6B & Simulate real human behavior in social networks\\
& Chuang et al.~\cite{chuang2024simulating} & gpt-3.5-turbo-16k & Simulate opinion dynamics\\
& Li et al.~\cite{li2024large} & gpt-3.5-turbo-1106 & Model dynamics of misinformation spread\\
& SMIST~\cite{ma2024simulated} & gpt-4 & Simulate misinformation susceptibility\\
& FDE-LLM~\cite{yao2025social} & ChatGLM-4 & Simulate the attitudes and behaviors of opinion leaders\\
& Y Social~\cite{rossetti2024social} & Llama-3-7B & Simulate user interactions, content dissemination, and network dynamics \\
& LLM-AIDSim~\cite{zhang2025llm} & Llama 3 & Simulate influence diffusion in social networks \\
& MOSAIC~\cite{liu2025mosaic} & gpt-4o & Predict user behaviors such as liking, sharing, and reporting content\\
& MIFlu~\cite{moon2025miflu} & gpt-2 & Fuse data from different modalities\\
& PandemicLLM~\cite{du2025advancing} & Llama-2 & Fuse data from different modalities\\
& LLM4Cast~\cite{saeed2024llm4cast} & TinyLlama & Forecast viral disease trends \\
& LLM-DG~\cite{kang2025llm} & gpt & Enrich representations of patients, diseases, and discharge summaries\\
& Priya et al.\cite{priya2024smart} & Medisearch AI & Perform functions like topic modeling, sentiment analysis, and relationship extraction\\
\midrule
Management & Almaliki et al.\cite{almaliki2025combining} & gemini-1.5-pro-latest & Process user comments, identify instances of cyberhate, and generate appropriate responses\\
& Jo et al.\cite{jo2024understanding} & HyperCLOVA & Provide natural language understanding and natural language generation capabilities\\
\end{xltabular}

\section{Background} 
\label{sec2}
This section introduces the foundational context by first reviewing classical models of spreading dynamics, including epidemiological models and social contagion models. It then provides a systematic overview of the core technologies and theoretical basis of LLMs, analyzes the key characteristics of mainstream model architectures, and highlights the unique capabilities of LLMs in representing textual and contextual features relevant to spreading processes. 
Table~\ref{tab:symbols} illustrates the symbols and notations employed throughout this article, and Table \ref{tab:llm} presents an overview of the key fundamental concepts about LLMs.

\begin{table*}[htbp]
\caption{Symbols and notations.}
\label{tab:symbols}
\centering
\begin{tabular}{ll}
\hline
\hline
\textbf{Symbol} & \textbf{Description} \\
\hline
\hline
$S(i)$, $I(i)$, $R(i)$ & Susceptible, infected and recovered states of individual $i$ \\
$\rho_{S}, \rho_{I}, \rho_{R}$ & Fractions of nodes in the susceptible, infected, and recovered states at time $t$ \\
$\beta$ & Transmission (infection) rate \\
$\gamma$ & Recovery rate \\
$N$ & Total population size \\
$w_i$ & The $i$-th token in a sequence \\
$x = \{w_1, \dots, w_{|x|}\}$ & Input token sequence (prompt or observed text) \\
$y = \{w_{|x|+1}, \dots, w_{|x|+T}\}$ & Generated token sequence (response) \\
$|x|$ & Length of the input sequence $x$ \\
$T$ & Length of the generated response sequence $y$ \\
$\mathcal{V}$ & Vocabulary set of all possible tokens \\
$d$ & Embedding or hidden dimension \\
$\mathcal{D}$ & Training dataset or corpus \\
$\theta$ & Model parameters \\
$P_\theta(x)$ & Probability distribution over sequence $x$ under model parameters $\theta$ \\
$P(w_i \mid w_{<i}; \theta)$ & Conditional probability of token $w_i$ given previous tokens $w_{<i}$ \\
$\mathcal{M}$ & LLM mapping prompts to responses \\
$\mathbf{h}_t \in \mathbb{R}^d$ & Hidden state vector at position or layer $t$ \\
$\mathcal{L}$ & Loss function (e.g., cross-entropy) used in LLM training \\
\hline
\hline
\end{tabular}
\end{table*}

\subsection{Primer on spreading dynamics}

\subsubsection{Basic concepts}
To facilitate the understanding of spreading dynamics, we briefly introduce several fundamental concepts from network science and spreading dynamics. Networks provide a structural representation of interacting entities, specifying who can influence whom. Based on this structural substrate, spreading dynamics refers to the process by which states, information, or behaviors propagate through interactions over time. While networks define the pathways of interaction, spreading dynamics describes the temporal evolution of states along these pathways, giving rise to system-level behaviors such as large-scale diffusion or extinction.

\begin{itemize}
    \item Network: A network provides an abstract mathematical representation for systems composed of interacting entities. Formally, a network is defined as a graph \( G = (V, E) \), where \( V \) denotes a finite set of nodes (or vertices) and \( E \subseteq V \times V \) denotes a set of edges encoding pairwise relationships between nodes. The structural connectivity of the network can be equivalently characterized by an adjacency matrix \( \mathbf{A} = [a_{ij}] \), in which each entry \( a_{ij} \) indicates the presence or absence of an edge between nodes \( i \) and \( j \), with \( a_{ij} = 1 \) if the two nodes are connected and \( a_{ij} = 0 \) otherwise~\cite{aleta2025multilayer, wang2024epidemic, tang2025network}.
   \item Key attributes of networks:
    \begin{itemize}
          \item Degree: The degree of a node denotes the number of connections it has.
        \item Average degree:  It is the average value of the degrees of all nodes in a network, and it is used to roughly measure the overall connectivity level of the network.
        \item Degree distribution: It describes the frequency of occurrence of different degree values in a network, and it is an important statistical feature for measuring the heterogeneity of network connections.
        \item Network density: It is the ratio of the actual connections in a network to the maximum possible connections, measuring the closeness of the relationships among nodes in the network, reflecting the density of the overall network connection.
        \item Average path length: It is the mean of shortest path lengths between all pairs of nodes, reflecting the efficiency of global connectivity.
        \item Hubs: In networks, hubs are nodes with exceptionally high connectivity that can disproportionately influence spreading processes.
        \item Community/Cluster: It refers to the tightly connected subset of nodes in a network, where the spreading process typically spreads rapidly within the community and then spreads outward through limited cross-community connections.
    \end{itemize}
    
    \item Network topology: Network topology characterizes the global organization of connections among nodes and provides the structural substrate on which spreading processes take place. In complex systems, spreading typically unfolds on a network \( G=(V,E) \), whose topology determines which contacts are feasible and how the influence is transmitted within the system. Variations in connectivity, heterogeneity and temporal characteristics significantly alter macroscopic spreading behaviors, including epidemic thresholds, spreading speed, and so on. Classic network topologies include random networks, scale-free networks, small-world networks,  regular networks, etc. In the following, we introduce these representative network structures in turn.
    \begin{itemize}
        \item Random network: It refers to the network in which, given a fixed set of nodes, edges are independently generated between arbitrary pairs of nodes with a prescribed probability. Owing to the randomness of the edge formation mechanism, node degrees typically concentrate around the mean value, resulting in relatively weak structural heterogeneity. This property is commonly associated with a finite and non-zero spreading threshold.~\cite{posfai2016network}.
        \item Scale-free network: A scale-free network is a network whose degree distribution follows a power-law distribution. In scale-free networks, the connectivity (degree) of nodes is highly heterogeneous: a small fraction of nodes possess extremely high degrees, while the majority of nodes have only a few connections~\cite{barabasi1999emergence}. Scale-free networks exhibit highly heterogeneous degree distributions, often resulting in a vanishing threshold and enhanced spreading potential~\cite{posfai2016network}.
        \item Small-world network: It is a kind of network which possesses both high local clustering and short average path lengths, enabling information or diseases to spread rapidly locally while also being able to efficiently propagate to the remote regions of the network through a small number of long-range connections.
        
        \item Regular network: It consists of a large number of identical or similar nodes, which interact with each other through regular or fixed connections. In a regular network, each node follows the same computational rules and interacts with its neighboring nodes~\cite{varga2017comparison}.

        \item Multilayer network:  It is composed of a group of shared nodes, but the nodes are connected to each other through different types of connection relationships, thereby forming different network.~\cite{boccaletti2014structure}.
        \item Temporal network: It is a class of network structures that explicitly consider the changes in connections over time, where edges are only valid within a specific time period~\cite{holme2012temporal}. By incorporating temporal information, temporal networks can more realistically reflect the impact of contact order and time constraints on propagation paths and propagation efficiency.
        \item Higher-order network: It is an alternative representation of complex networks that explicitly model interactions beyond pairwise edges and are increasingly used in the study of complex systems~\cite{battiston2020networks,lambiotte2019networks,battiston2021physics,qian2025robustness,peng2025sentinel,gao2026coevolution}.
    \end{itemize}
 
       \item Node states: Node states represent the discrete conditions of entities involved in spreading processes, such as susceptible, infected, recovered, or adopted.
    \item State transitions: It describes how node states change through interactions with neighbors or internal dynamics, such as infection, adoption, recovery, or forgetting.
    \item Spreading rate: The spreading rate $\lambda$ characterizes the effective strength of a contagion process and is commonly quantified by the ratio between the transmission rate \( \beta \) and the recovery rate \( \gamma \)~\cite{pastor2015epidemic}.
    \item Epidemic threshold: The epidemic threshold represents the critical condition under which a disease or information can spread widely in a network. Generally, the pathogen can spread only if its spreading rate exceeds an epidemic threshold $\lambda_c$. The random network has a finite epidemic threshold $\lambda_c$, implying that a pathogen with a small spreading rate ($\lambda < \lambda_c$) must die out. If, however, the spreading rate of the pathogen exceeds $\lambda_c$, the pathogen becomes endemic and a finite fraction of the population is infected at any time. For a scale-free network we have $\lambda_c = 0$, hence even viruses with a very small spreading rate $\lambda$ can persist in the population.
    
    \item Spreading speed: It denotes the speed at which a disease, information, or behavior spreads across a network.
    \item Cascades: A cascade is a sequence of activations generated by a contagion process, in which nodes cause connected nodes to be activated with some probability. In analogy with the spread of an infectious disease on a network, an infected (activated) node exposes his fans to the infection. Disease cascades through the network as exposed fans become infected, thereby exposing their own fans to the disease, and so on~\cite{ghosh2011framework}. 
\end{itemize}

\subsubsection{Classical models in spreading dynamics}
Based on the nature of the spreading entity, spreading dynamics models in complex systems can be divided into two major categories: \textbf{epidemiological models} for \emph{biological epidemics}, which describe the transmission of pathogens among biological individuals, and \textbf{social contagion models} for \emph{digital epidemics}, which focus on the diffusion of digital information across social networks, such as opinions, rumors, and computer worms. The former originated from public health and infectious disease studies, emphasizing physiological infection--recovery processes, and are typically built upon mass-action principles or Markovian assumptions over contact networks. The latter stem from sociology and computer science, highlighting socio-cognitive mechanisms of ``perception-decision--interaction'', and are often based on threshold rules or probabilistic cascade processes. These two categories differ fundamentally in their modeling assumptions, dynamical properties, and transmission mechanisms. Each is introduced in the following.

\paragraph{\textbf{Epidemiological models}}

For biological epidemics, transmission occurs primarily via physical contact, with infection probability often correlated with contact frequency. Such models are commonly based on the homogeneous mixing hypothesis, which implies that contacts between individuals are uniform or can be modeled via contact networks \cite{pastor2015epidemic,wang2017unification}.
Individuals are typically represented by a finite set of states, such as susceptible ($S$), infected ($I$), recovered/removed ($R$), and exposed ($E$) states. Transitions between these states follow specific disease mechanisms. Their dynamics are commonly modeled using ordinary differential equations or continuous-time Markov processes. Usually, different types of biological epidemics require different models, and the most representative models are the SIS and SIR models:

- \textbf{SIS Model}: It describes diseases without permanent immunity (e.g., the common cold). Individuals are either susceptible or infected. Its dynamical reaction process is:
\begin{equation}
S(i) + I(j) \xrightarrow{\beta} I(i) + I(j), \quad
I(i) \xrightarrow{\gamma} S(i),
\end{equation}
where $i$ and $j$ represent different individuals, $\beta$ is the effective contact rate and $\gamma$ is the recovery rate. A susceptible individual $S(i)$ becomes infected upon contact with an infected individual $I(j)$, while an infected individual recovers at rate $\gamma$ and returns to the susceptible state, forming the $S \rightarrow I \rightarrow S$ cycle. 
Under the mass-action assumption, the corresponding dynamical evolution equations are as follows:
\begin{equation}
\begin{aligned}
\frac{d \rho_{S}}{dt} &= -\beta \frac{\rho_{S} \rho_{I}}{N} + \gamma \rho_{I}, \\
\frac{d\rho_{I}}{dt} &= \beta \frac{\rho_{S} \rho_{I}}{N} - \gamma \rho_{I}.
\end{aligned}
\end{equation}
Here,  $N=\rho_{S}+\rho_{I}$ denotes the total population. $\rho_{I}$ and $\rho_{S}$ denote the proportions of susceptible and infected individuals at time $t$, capturing the population-level infection dynamics. The term $\beta \rho_{S} \rho_{I}/N$ quantifies the expected rate of new infections per unit time under uniform mixing.

- \textbf{SIR Model}: It applies to diseases with permanent immunity (e.g., smallpox). Individuals in SIR model occupy three $S$, $I$, and $R$ states, and the dynamical reaction process is:
\begin{equation}
S(i) + I(j) \xrightarrow{\beta} I(i) + I(j), \quad
I(i) \xrightarrow{\gamma} R(i),
\end{equation}
which means susceptible individuals become infected at a rate of $\beta$, while infected individuals recover at a rate of $\gamma$ and enter the recovered state $R$, forming the unidirectional $S \rightarrow I \rightarrow R$ pathway. With $N = \rho_{S} + \rho_{I} + \rho_{R}$, the dynamical evolution is { guided} by:
\begin{equation}
\begin{aligned}
\frac{d\rho_{S}}{dt} &= -\beta \frac{\rho_{S} \rho_{I}}{N}, \\
\frac{d\rho_{I}}{dt} &= \beta \frac{\rho_{S} \rho_{I}}{N} - \gamma \rho_{I}, \\
\frac{d\rho_{R}}{dt} &= \gamma \rho_{I},
\end{aligned}
\end{equation}
where $\rho_{R}$ represents the number of recovered individuals at time $t$.

\paragraph{\textbf{Social contagion models}}
Although social contagion and epidemic spreading are both diffusion processes unfolding over networks, they rely on different behavioral assumptions at the individual level. Classical epidemiological models typically operationalize transmission as a function of exposure along contacts, where a single infectious encounter can, in principle, be sufficient to trigger infection, and additional encounters mainly increase the probability of transmission. In contrast, many socially mediated adoptions are not triggered by a single exposure. Simple contagions, such as simple information and familiar ideas, can spread through a single contact.
Whereas complex contagions, such as changes in workplace culture, the use of online social platforms, and the adoption of innovative technologies, typically require often require social reinforcement, i.e., repeated exposure and/or corroboration from multiple peers, before changing opinions or adopting behaviors, especially when the behavior is costly, risky, or unfamiliar~\cite{guilbeault2018complex, mariani2024collective, centola2021influencers}.

Such spreading process emphasizes decision-making mechanisms and often employ probabilistic or game-theoretic rules, which can be divided into threshold models and independent cascade models~\cite{castellano2009statistical,zhang2016dynamics}.
Threshold-based models are commonly associated with complex contagion, as adoption is modeled as a collective-response phenomenon: an individual becomes active only when the number or fraction of adopting neighbors exceeds a threshold, explicitly encoding reinforcement from multiple contacts. By contrast, independent cascade (IC) models are typically viewed as simple contagion processes, where each exposure acts independently and can trigger adoption with some probability, making the dynamics analogous to biological-virus-like transmission.

- \textbf{Threshold Model}: An individual adopts or spreads information if the proportion of activated neighbors exceeds a personal threshold. Formally, let $a_v \in [0,1]$ be the activation threshold of node $v$, and $f_v$ be the fraction of its neighbors that are active. Then, $v$ becomes active if $f_v \geq \theta_v$. This process is typically deterministic.

- \textbf{Independent Cascade Model}: A probabilistic model where each active node $u$ attempts to activate each inactive neighbor $v$ with activation probability $f(u,v)$, independent of previous attempts and other neighbors. In each round, active nodes attempt activation regardless of outcome, and they do not attempt again in subsequent rounds. The process continues until no new activations occur.

\subsection{Primer on large language models}
\begin{table*}[htbp]
     \caption{Important concepts about LLMs}
     \centering
     \label{tab:llm}
\begin{tabular}{p{15cm}}
\hline
\textbf{Transformer} is a neural network architecture that uses self-attention to model global dependencies in sequential data. It processes all elements of a sequence in parallel, allowing each token to dynamically attend to others. This design enables efficient learning of long-range contextual relationships and has become the foundation for most LLMs.\\
\textbf{Tokens}  are the smallest units of meaning processed by the model, which can be individual words, subwords, or characters, depending on the granularity of the model's tokenization process.\\
\textbf{Embedding} refers to a vector representation that maps tokens into continuous space, allowing semantic and syntactic information to be captured numerically.\\
\textbf{Prompts} refer to the input or instruction provided to an LLM to guide its output generation and determine its behavior in specific tasks. \\
\textbf{Prompt engineering} refers to the design and optimization of prompts to elicit desired responses from LLMs, often without modifying model parameters. \\
\textbf{Zero-shot / Few-shot learning} refers to the ability of an LLM to generalize to new tasks without (zero-shot) or with only a few (few-shot) task-specific examples.\\
\textbf{Fine-tuning} refers to the re-training of a pre-trained or trained model on new datasets or tasks.
\\ 
\textbf{Multimodal model}  refers to a model capable of understanding and integrating information from multiple modalities such as textual, visual, or auditory data.\\
\hline
\end{tabular}
\end{table*}

LLMs are a class of advanced artificial intelligence systems trained on massive natural language corpora to capture latent statistical patterns, syntactic structures, and semantic relations, thereby generating contextually relevant and coherent natural language outputs \cite{vaswani2017attention, GPT3, Llama}. Typically built upon the Transformer architecture, LLMs leverage large-scale training data to achieve strong cross-task transfer and generalization capabilities  \cite{DeepSeek-R1,vaswani2017attention}. These capabilities not only enable LLMs to perform conventional tasks such as text generation, conversational systems, and knowledge retrieval, but also, directly or indirectly, support tasks related to epidemic spreading dynamics in complex systems. For example, LLMs can directly replace the rule-based agents in ABMs, enabling more human-like simulations of individual interactions and transmission pathways. Meanwhile, they can also indirectly support epidemic modeling and prediction by enhancing the perception and representation of multimodal epidemic-related data, such as images, textual reports, and temporal sequences, thereby improving both accuracy and robustness~\cite{gao2024large,lu2024llms,tavana2025generative,du2025advancing,bizel2025extracting,kraemer2025artificial,li2023text,hao2024quantifying}. Generally, the core learning process of an LLM is to estimate the probability distribution of a given input sequence. Formally, its training optimization target can be expressed as maximizing the log-likelihood of the sequence with respect to the model parameters, as follows:
\begin{equation}
\max_\theta \sum_{i=1}^{|x|} \log P(w_i \mid w_{<i}; \theta),
\end{equation}
where \( w_i \) denotes the \( i \)-th token in the input sequence  \( x = \{w_1, ..., w_{|x|}\} \) and \( \theta \) refers to the model parameters. By adjusting its parameters, the LLM aims to maximize the probability of generating the correct sequence.

\subsubsection{Process of functionalizing  LLMs}
\begin{figure*}[htbp]
    \centering
    \includegraphics[width=1.\linewidth]{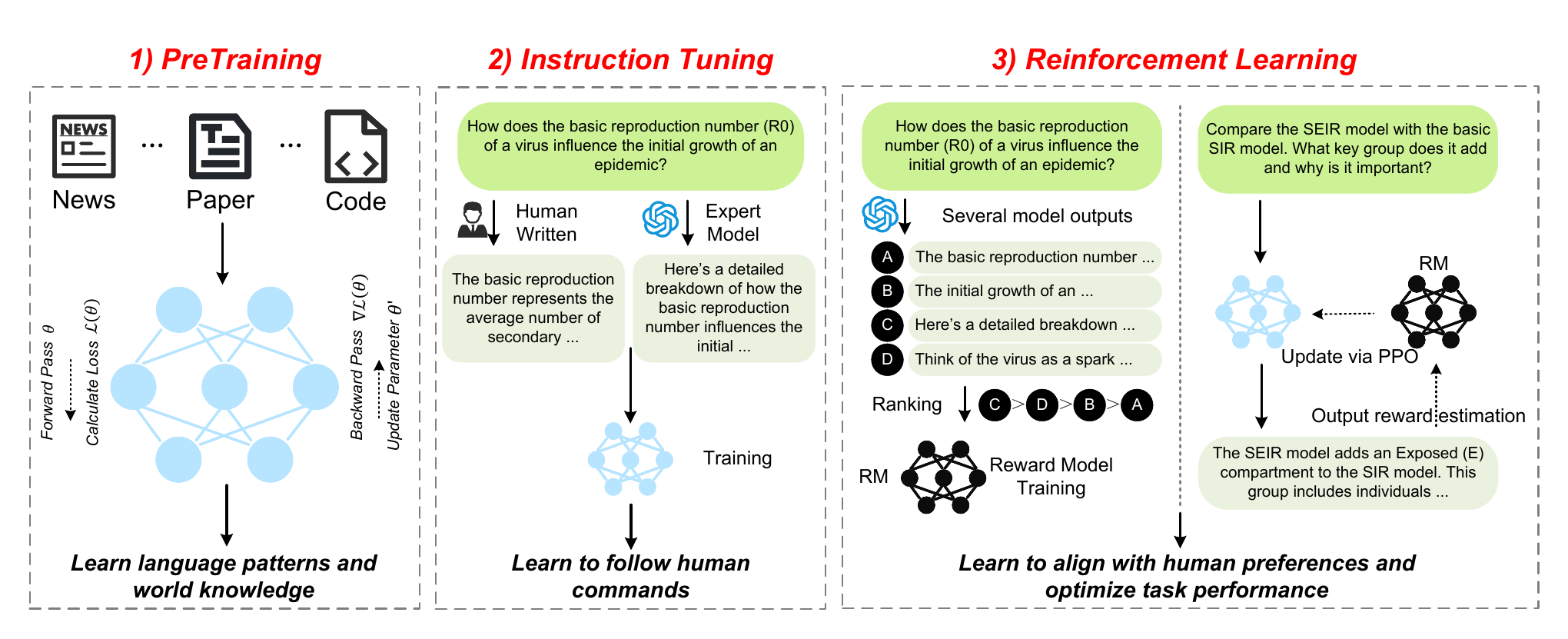}
    \caption{An illustration of how to enable LLMs to handle various complex tasks.  (1) Pre-training: Use massive datasets for self-supervised learning. This is to teach the model to master language rules and knowledge, and form a basic model. (2) Instruction Tuning: Use instruction data for supervised training. This is to teach the model to understand and follow human instructions during conversations. (3) Reinforcement Learning: Use preference data for further optimization to ensure that the output is useful, safe, and as expected.}
    \label{fig2}
\end{figure*}

To enable the model to gradually learn the optimal parameters required to complete various complex tasks, LLMs typically go through three stages: pre-training, instruction fine-tuning, and reinforcement learning based on human or artificial intelligence feedback, as shown in Figure \ref{fig2}.

In the pre-training stage, the model learns the fundamental structure of language and world knowledge through self-supervised learning on large-scale text data, such as news articles, academic papers, and code \cite{Bert, GPT3}. This data is input into the neural networks, where the model computes prediction errors via forward propagation, represented by the loss function $\mathcal{L}$($\theta$). The model's parameters $\theta$ are then updated via backpropagation, yielding a pre-trained LLM.

In the instruction fine-tuning stage, the pre-trained model further learns how to execute tasks according to given instructions. This stage utilizes training datasets composed of both human-written and expert-model-generated question-answer pairs \cite{InstructGPT, Self-Instruct, chung2024scaling}.
For example, the model will learn to answer specific questions like ``\textit{how the basic reproduction number of a virus affects the early growth of an epidemic?}'' in this stage. This fine-tuning enables the model to adjust its output according to the task at hand, improving its ability to respond accurately to diverse requirements in complex systems.

In the reinforcement learning phase, the LLM's performance is optimized through feedback signals \cite{ouyang2022training, lee2024rlaif}. First, human evaluators will rank multiple candidate responses generated by models based on quality. This ranking information is used to train a reward model that assigns rewards to different answers according to human preferences  \cite{PPO}. During this stage, the reward model scores each generated response and sends feedback to the LLM, enabling it to adjust its output accordingly. For instance, in the task of generating epidemic management strategies, the model first generates several management suggestions, each with a varying potential impact on epidemic spread~\cite{bushaj2023simulation, rakhshandehroo2025reward}. Next, human evaluators rank these strategies based on their scientific validity, accuracy, and alignment with actual data. Using this feedback, a reward model is trained to assign rewards to each new suggestion. In this process, the LLM continuously adjusts its strategy generation based on the feedback from the reward model, optimizing the management strategies it generates. Ultimately, through this reinforcement learning mechanism, the model becomes better equipped to address complex epidemic propagation problems and provide management strategies that meet real-world needs.

\subsubsection{General workflow of LLMs in epidemic-related tasks}
After completing the three training stages, LLMs acquire the capability to handle a wide range of complex tasks. By adjusting the form of inputs and the format of outputs, they can be flexibly adapted to epidemic-related applications. For example, in epidemic modeling, the model can generate agents with specific roles or functions based on different prompts; in monitoring tasks, it can produce structured early-warning signals according to predefined formats. Through such input-output customization, LLMs can transform their general capacities in language understanding and knowledge modeling into concrete functionalities for studying spreading dynamics.
\begin{figure*}[htbp]
    \centering
    \includegraphics[width=1.\linewidth]{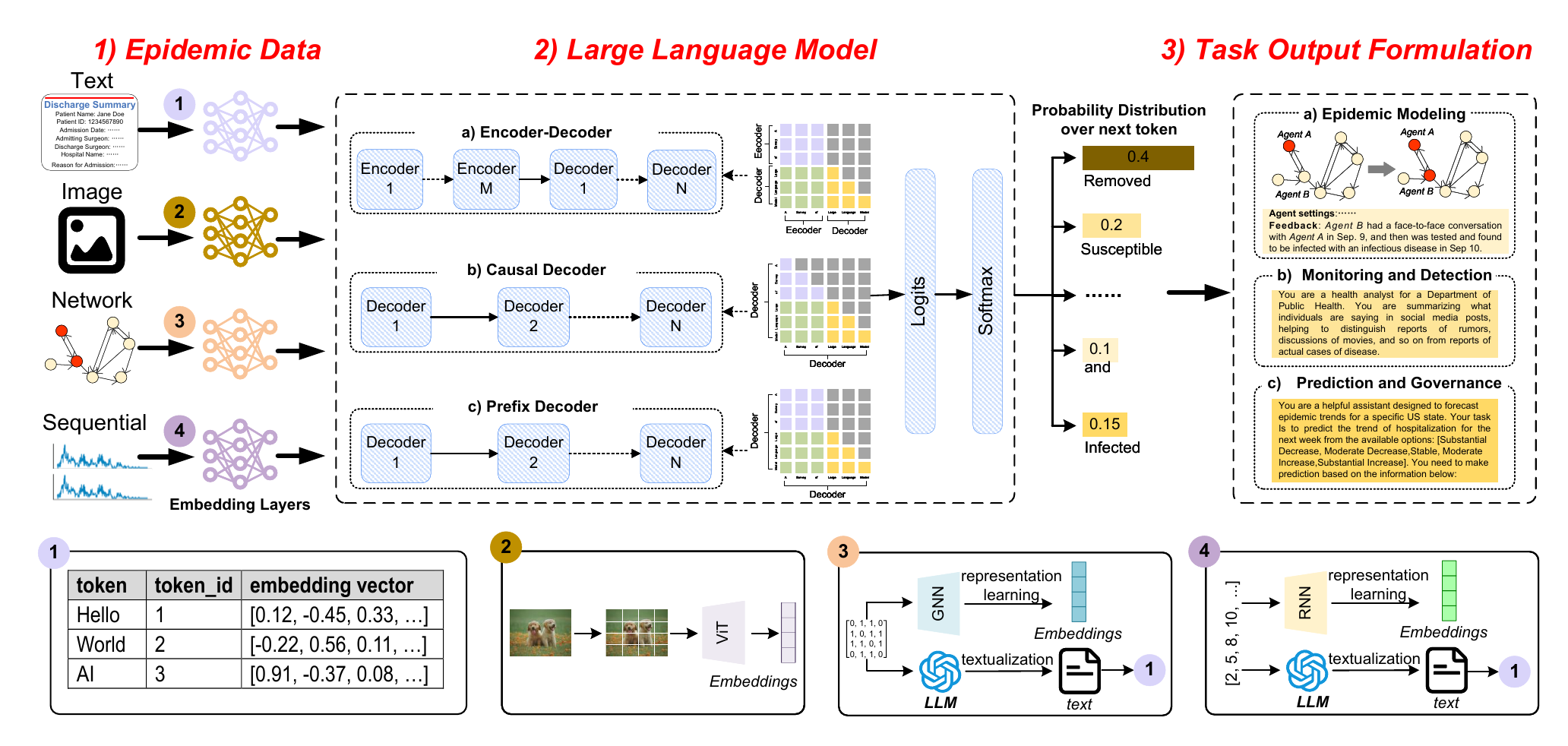}
    \caption{An illustration of LLM workflow. (1) Data embedding: Various types of input data, including text, images, networks, and time series, are transformed into high-dimensional vectors known as embedding vectors, enabling their processing by the LLM. (2) Large Language Model: The LLM processes the embedding vectors to predict the next token, using the neural networks with Transformer architectures like encoder-decoder, causal decoder, or prefix decoder. (3) Task output formulation: The final step involves customizing the model's output format for different tasks, ensuring that the LLM is adapted to a wide range of applications, such as epidemic modeling, monitoring, detection, prediction, and management.}
    \label{fig1}
\end{figure*}

During the process where LLMs handle multimodal inputs such as text, images, graph structures, and sequential data, the raw input is first transformed into high-dimensional vector representations, known as embeddings. These vectors can be considered as coordinates that capture the semantic or structural features of the input data. The embeddings are then processed through multiple Transformer layers to produce logits, which can be thought of as unnormalized scores representing the relative preference of each possible outcome. These logits are then normalized into a probability distribution over the vocabulary space using the softmax function. Finally, decoding strategies, such as top-$k$ sampling, top-$p$ sampling, or beam search, are used to select specific tokens from these probability distributions as outputs. The selected output is then added to the input, and this process is repeated iteratively to generate the next token until a stopping criterion is met.

As shown in the left and bottom of Fig. \ref{fig1}, the data embedding process differs between modalities~\cite{liu2023medical, kraemer2025artificial, du2025advancing}:
\begin{itemize}
    \item \textbf{Text}: Epidemic-related research often involves processing textual data such as disease surveillance reports or patient symptom logs. The input text is first segmented into tokens, followed by mapping these tokens to token\_ids, and finally generating their corresponding embedding vectors through the text embedding layers.
    
    \item \textbf{Images}: In epidemic data, images include medical imaging such as CT scans to detect disease symptoms, satellite images used to track disease spread patterns, etc. These images are typically divided into fixed-size pixel blocks, such as $16 \times 16$ patches, and features are extracted through a visual encoder like Vision Transformer~\cite{ViT}  to produce an image embedding vector.
    
    \item \textbf{Network data}: For epidemic modeling, network data can represent the spread of a disease through a social network, where nodes represent individuals and edges represent interactions or transmission events. There are two main approaches for processing network data. One is based on Graph Neural Networks (GNNs)~\cite{scarselli2008graph}, which learn low-dimensional node or network representations from adjacency matrices and node features, using random walks or attention mechanisms. The other approach is based on LLMs, where the network structure is converted into textual descriptions, which are then processed using text embedding layers to generate embeddings~\cite{liu2023medical}.
    
    \item \textbf{Sequential data}: In the context of epidemics, sequential data could involve time-series data on the number of reported cases over time, human mobility behavior, or daily reports of disease spread or vaccination rates. One approach for processing sequential data is to use recurrent neural networks (RNNs) to encode sequential patterns step by step through temporal layers. Another approach is LLM-based, where the sequential data is transformed into textual descriptions and processed by text embedding layers \cite{du2025advancing}.
\end{itemize}

For mapping transformations from input embeddings to outputs, LLMs commonly adopt one of the causal decoder, prefix decoder, or encoder-decoder of Transformer-based designs. As shown in the middle of Fig. \ref{fig1}, the LLMs like GPT~\cite{GPT-4} and LLaMA~\cite{Llama} series use the causal decoder framework, which adopts the unidirectional masked self-attention to allow the model to generate text in an autoregressive manner, meaning that each token attends only to preceding tokens. In contrast, LLMs consisting of prefix decoders (e.g., ChatGLM \cite{du2022glm}) use bidirectional attention to process input information while maintaining unidirectional decoding for output generation. This design allows the model to effectively handle contextual information while avoiding conflicts during generation. Other LLMs, such as T5  \cite{T5} and T0  \cite{T0}, use encoder-decoder architectures, which combine bidirectional self-attention in the encoder and cross-attention in the decoder to capture intricate dependencies between input and output sequences.

As shown on the right of Fig. \ref{fig1},  when LLMs are applied to spreading dynamics tasks in complex systems, they typically use generative outputs to transform the expected outcomes of various tasks into textual descriptions. For instance, in epidemic modeling tasks, the attributes of individual nodes, such as age, gender, and social relationships, are described in natural language, such as ``\textit{You are a 35-year-old female teacher living in a large city, leaning toward liberal politics, valuing educational equality and environmental protection...}''. This textual description is treated as the model input \(x\) and then passed to the LLM. Based on this input, the LLM generates corresponding textual responses \(y\) that reflect changes in node states, such as ``\textit{Person A was contacted and infected by Person B}'', thus capturing the evolving dynamics of transmission between individuals.
During the inference process, the LLM receives the input prompt \(x\) and generates a response sequence autoregressively \cite{InstructGPT}. Specifically, given an initial input prompt \( x_i \), the model generates a response by sequentially predicting the next token conditioned on the input and the previously generated context. The probability of producing an output sequence \( \mathbf{w}_{>|x|} = \{ w_{|x|+1}, \dots, w_{|x|+T} \} \) can be expressed as
\begin{equation}
P_\theta(\mathbf{w}_{>|x|} \mid x_i)
= \prod_{t=1}^{T} P_\theta\bigl(w_{|x|+t} \mid w_{<|x|+t}, x_i\bigr),
\end{equation}
where \( w_{|x|+t} \) denotes the token generated at step \( t \), conditioned on the input prompt \( x_i \) and all previously generated tokens \( w_{<|x|+t} \).

Compared with classical compartmental or social-contagion models that hard-code transmission mechanisms and rely on structured variables, LLMs offer a complementary paradigm along three axes: (a) representation: by ingesting heterogeneous, high-dimensional, and weakly structured signals (e.g., posts, reports, clinical notes) via prompting and retrieval without bespoke feature engineering; (b) adaptation: by few-shot or zero-shot generalization, tool use, and counterfactual role conditioning that accommodate non-stationarity and rapid regime shifts; (c) decision support: by producing human-readable rationales, policy options, and multi-agent narratives that link micro-level behaviors to macro-level dynamics. These differences create clear advantages for early warning, monitoring, and intervention design, while also raising open questions about reliability, bias, and management that traditional error metrics do not fully capture. In the following sections, we review the current landscape of LLM-based approaches for spreading dynamics, organizing the literature by domain (digital vs. biological epidemics) and task family (modeling, perception, and regulation). We highlight their methodologies, key findings, and the unique advantages they bring to the analysis of epidemic spreading.

\subsection{Empirical data for validation} 
In this section, we will introduce how empirical data is used for the validation of LLM-based spreading simulations, as well as the main evaluation metrics and datasets employed.

Generally, the role of empirical data in LLM-based spreading simulation can be divided into three aspects: initialization of the simulation, driving the simulation, and evaluating the simulation.
First, empirical data can be used in spreading simulation to determine the initial conditions of the simulation, such as social network structure, user behavior characteristics, and time-series data. Zhang et al. ~\cite{zhang2025llm} utilized a subset of the ego-Facebook dataset from the Stanford Network Analysis Project (SNAP) group to represent the network structure. In this dataset, the total number of nodes (representing individual users) is 333, and the total number of edges (representing connections or relationships between users) is 2,500. Törnberg et al. ~\cite{tornberg2023simulating} used data from the American National Election Study (ANES) to create personas that guide the behavior of each agent in their model. Specifically, they used this data to determine the news sources each agent consumes, their demographic characteristics, political beliefs and interests, attitudes toward specific groups and political figures, as well as certain non-political interests.

In addition, empirical data can not only serve as the starting point for the simulation but also dynamically guide the generation process, making the spreading behavior more realistic. Zheng et al. ~\cite{yangoasis} introduced a time engine into the simulation. This module activates agents based on their temporal characteristics, enabling them to perform various actions such as commenting, posting, and interacting with other agents and the environment. These temporal features are derived from real behavioral data, making the overall spreading process more consistent with users’ everyday activity patterns on real-world social media platforms. Yao et al. ~\cite{yao2025social} developed a simulated social media platform using Python. They took the real-world timing of truth exposure as a reference point and released offline news containing the truth to the agents at that moment.

Finally, empirical data are also essential for validating whether simulated spreading processes faithfully reproduce observed dynamics. 
Validation in this context is not limited to pointwise accuracy, but aims to assess the alignment between simulated outcomes and real-world observations at multiple levels, including individual states, temporal trajectories, and aggregate spreading trends. Accordingly, a diverse set of evaluation metrics is employed. Specifically, classification-oriented metrics such as accuracy and macro F1 score are used to evaluate the correctness of discrete state predictions; temporal alignment metrics, including dynamic time warping (DTW), assess the similarity between simulated and observed spreading trajectories over time; correlation-based and vector-based measures, such as the Pearson correlation coefficient and cosine similarity, are applied to quantify the consistency between simulated and empirical trends at a more global level. The calculations of these metrics are below.

\begin{itemize}
    \item Accuracy: In a multi-class classification setting, accuracy measures the overall proportion of correctly predicted samples across all classes and is defined as:
    \begin{equation}
        \mathrm{Accuracy} = \frac{1}{N} \sum_{i=1}^{N} \mathbb{I}\left( \hat{y}_i = y_i \right)
    \end{equation}
    where \(N\) denotes the total number of samples, $\hat{y}_i$ and $y_i$ represent the predicted label and the ground-truth label of the $i$-th sample, respectively, and $\mathbb{I}$ ($\cdot$) is an indicator function that equals 1 if the condition is true and 0 otherwise.
    
    \item Macro-F1: In a multi-class classification setting, Macro-F1 evaluates the model performance by computing the F1 score independently for each class and then averaging them with equal weight, which is defined as:
    \begin{equation}
    \mathrm{Macro\text{-}F1} = \frac{1}{C} \sum_{c=1}^{C} \mathrm{F1}_c
    \end{equation}
    where \(C\) denotes the total number of classes and \(\mathrm{F1}_c\) is the F1 score of class \(c\).
    
    \item DTW: DTW is used to measure the similarity between the simulated attitude sequence $S$ and the real attitude sequence $T$, especially when the two sequences differ in length or have temporal changes. DTW aligns the two sequences by calculating the minimum cumulative distance. The formula is as follows:
    \begin{equation}
    DTW\left(S, T\right) = \min \sqrt{\sum_{k=1}^{K} \left( s_{i_k} - t_{j_k} \right)^2}
    \end{equation}
    where $\left(i_k, j_k\right)$ signifies the aligned points in sequences $S$ and $T$, and $K$ is the length of the alignment path. A smaller DTW distance indicates a higher similarity between the time series of the simulated and real attitudes, signifying that the model performs better in simulating complex time series patterns. 
    
    \item Pearson correlation coefficient: Pearson correlation coefficient is used to measure the linear correlation between the simulated attitude sequence $S = \left(s_1, s_2, \dots, s_n\right)$ and the real attitude sequence $T = \left(t_1, t_2, \dots, t_n\right)$. The formula is as follows:
    \begin{equation}
    r = \frac{
        \sum_{i=1}^{n} \left(s_i - \bar{s}\right) \left(t_i - \bar{t}\right)
    }{
        \sqrt{\sum_{i=1}^{n} \left(s_i - \bar{s}\right)^2} \cdot \sqrt{\sum_{i=1}^{n} \left(t_i - \bar{t}\right)^2}
    }
    \end{equation}
    where $\bar{s}$ and $\bar{t}$ are the mean values of the simulated and real attitudes, respectively. The Pearson correlation coefficient ranges from $[-1, 1]$, and the closer the value to $1$, the stronger the linear correlation between the simulated and real attitudes, indicating that the model better captures changes in real attitudes.

    \item Cosine similarity: The cosine similarity metric reflects the semantic similarity between the simulated content and the real content, and is computed as follows:
    \begin{equation}
    \mathrm{Cos\text{-}Sim}\left(\mathbf{s}, \mathbf{r}\right) =
    \frac{\mathbf{s} \cdot \mathbf{r}}
    {\|\mathbf{s}\|_2 \, \|\mathbf{r}\|_2}
    \end{equation}
    where $\mathbf{s} \in \mathbb{R}^d$ and $\mathbf{r} \in \mathbb{R}^d$ denote the $d$-dimensional embedding vectors of the simulated content and the real content, respectively. The $\|\mathbf{s}\|_2$ and $\|\mathbf{r}\|_2$ denote the Euclidean norms of the corresponding vectors. This metric reflects how closely the simulated content matches the real content in terms of semantic meaning.
\end{itemize}

Representatively, Mou et al. ~\cite{mou2024unveiling} consider both micro-level and macro-level evaluations, focusing respectively on individual user alignment and overall system-level outcomes. At the micro level, they provide each core user agent with authentic contextual information in single-round simulations and evaluate their decision-making behaviors, focusing on stance alignment, content alignment, and behavior alignment. Specifically, stance alignment is measured by the accuracy and macro F1 score of stance classification, as well as the mean absolute error (MAE) of the attitude scores; content alignment is assessed through the classification accuracy of content types, the macro F1 score, and the cosine similarity between simulated and real content; behavior alignment is evaluated based on the classification accuracy and macro F1 score of posting and retweeting actions. At the macro level, they quantify the attitude distribution in a complete multi-round simulation, taking into account both the static attitude distribution and the time series of the average attitude. Specifically, the static attitude distribution is measured by bias (the deviation of the average attitude from the neutral attitude) and diversity (the standard deviation of attitudes), while the time series of the average attitude is evaluated for similarity to real data using DTW and the Pearson correlation coefficient. Yao et al. ~\cite{yao2025social} also used DTW and Pearson correlation coefficient to evaluate the effectiveness of the model in simulation. In addition, Zhang et al. \cite{zhang2025llm} used BERT to calculate the evolution of semantic similarity between topics extracted from real-world comments and those generated by simulations across multiple iterations.

In current studies, the use of empirical data for validation remains limited, but the candidate datasets shown in Table \ref{tab:datasets} can be considered as potential sources for validation.

\begin{table}[htbp]
\centering
     
\caption{Candidate datasets for validation of LLM-based spreading simulations.} 
\label{tab:datasets}
\begin{tabular}{l l l}
\toprule
\textbf{Dataset} & \textbf{Domain} & \textbf{Scale} \\
\midrule
Twitter15 \cite{liu2015real} & Rumors & ~842 events\\
Twitter16 \cite{ma2016detecting} & Rumors & ~5,656 events; 4,907,641 posts; 3,238,047 users \\
PHEME \cite{zubiaga2016learning} & Rumors & ~5 events; 5,802 posts \\
RumourEval 2017 \cite{derczynski2017semeval}  & Rumors & ~297 events; 7,100 posts  \\
RumourEval 2019 \cite{gorrell-etal-2019-semeval}  & Rumors & ~446 events; 7,995 posts\\
MR$^2$ \cite{hu2023mr2}& Rumors & ~14,700 events; 1,073,487 posts; 107,892 users \\
FakeNewsNet \cite{shu2020fakenewsnet} & Fake News & ~23,196 news articles; 690,732 posts \\
Weibo21 \cite{nan2021mdfend}  & Fake News & ~9,128 news articles\\ 
MC-Fake \cite{min2022divide} & Fake News & ~27,155 news articles; $\sim$ 5M posts; $\sim$ 2M users \\
FakeSV \cite{qi2023fakesv} & Fake News & ~738 events; 5,538 news videos \\
FTT \cite{hu2023learn}  & Fake News & ~ -\\
TruthSeeker \cite{dadkhah2023largest}  & Fake News & ~1,400 news articles; $\sim$ 186,000 posts\\
MCFEND \cite{li2024mcfend}  & Fake News & ~23,789 news articles; 170,713 posts; 803,779 users\\
FauxBuster \cite{zhang2018fauxbuster}  & Misinformation & ~917 posts; 1,988,493 comments; 621,983 users \\
MuMiN \cite{nielsen2022mumin}  & Misinformation & ~$\sim$ 26k threads; $\sim$21M posts \\
NETINF \cite{gomez2012inferring}  & Information Diffusion & ~$\sim$172M news articles and blog posts \\
DeepHawkes \cite{10.1145/3132847.3132973} & Information Diffusion & ~55,108 posts \\
COVID-19 \cite{cinelli2020covid} & Information Diffusion & ~$\sim$8M comments and posts \\
Weibo-COV \cite{hu2020weibo}  & Information Diffusion & ~$\sim$40M posts; $\sim$20M users \\
\bottomrule
\end{tabular}
\end{table}

\section{LLMs for Epidemic Modeling}
\label{sec3}
Spreading dynamics has long been a central theme in statistical physics, encompassing processes such as percolation, diffusion, and transport on networks.  These frameworks provide fundamental insights into how local interactions give rise to large-scale collective behavior,  offering powerful abstractions for diverse phenomena ranging from information diffusion to epidemic outbreaks. 
Within this framework, epidemic spreading is modeled as a dynamic process that unfolds on an interacting network of individuals, analogous to the propagation of states or particles through physical media.
Traditional epidemic modeling approaches, such as compartmental models and agent-based simulations, largely inherit the deterministic and rule-based paradigms of physics, relying on fixed parameters and simplifying assumptions of homogeneity.
These models describe epidemic spreading using determined transmission and recovery rates, with individual behaviors often simplified to a uniform response under mean-field theory. While analytically tractable, such methods often struggle to capture the heterogeneity, stochasticity, and multimodal influences that characterize real-world epidemic scenarios~\cite{gao2024large,williams2023epidemic}. 

\begin{figure*}[htbp]
    \centering
    \includegraphics[width=1.\linewidth]{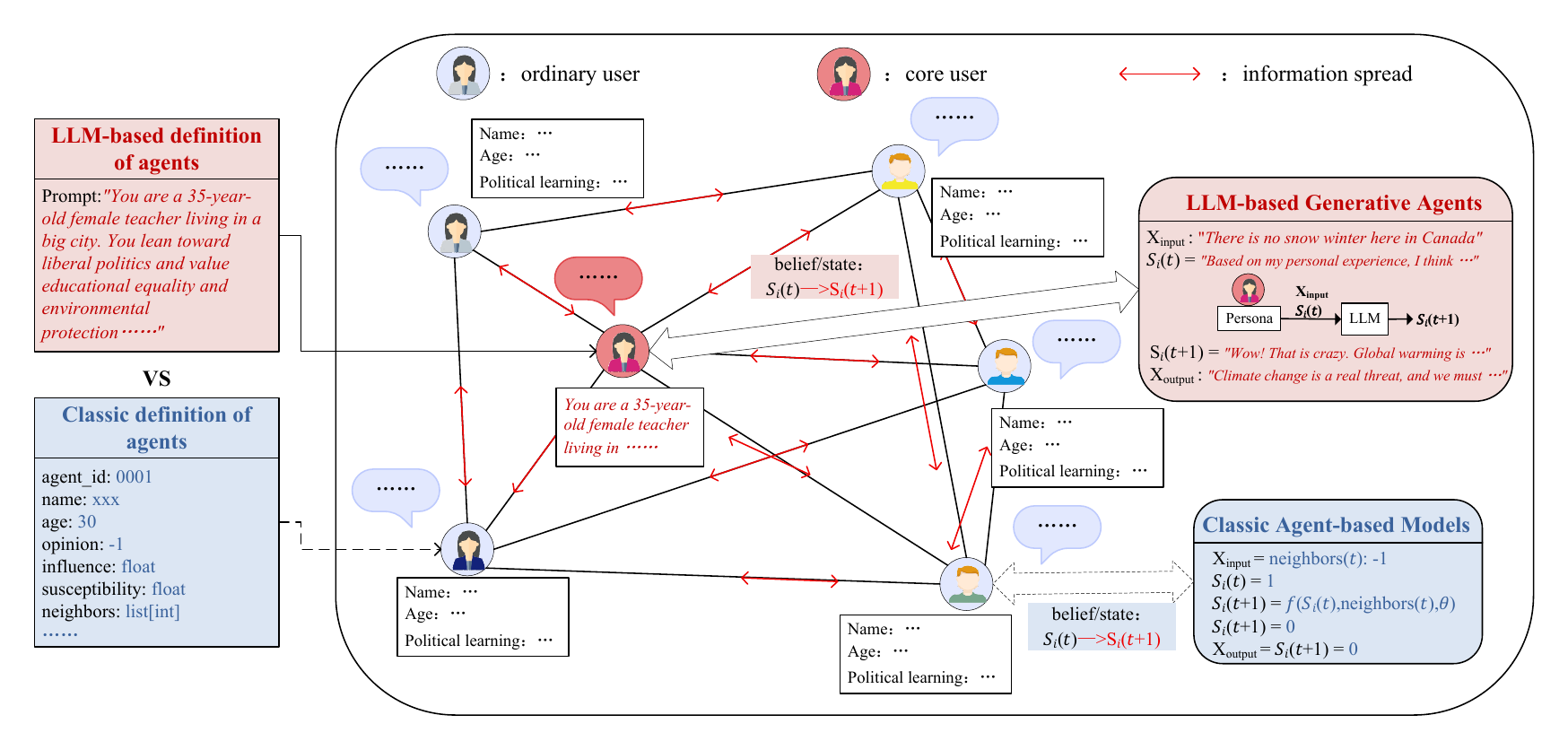}
    \caption{Comparison of classical epidemic modeling and LLM-based epidemic modeling. { In classical agent-based models, each agent is characterized by a predefined set of attributes, such as influence, susceptibility, and network neighbors, and state transitions are computed through explicitly specified rules or equations based on these parameters and local interactions. In contrast, LLM-based generative agents are modeled as cognitively driven individuals whose state transitions emerge from internal reasoning processes and contextual perception. Rather than following fixed transition equations, these agents interpret environmental cues and generate context-dependent behaviors, allowing transmission dynamics and individual responses to adapt flexibly to cognitive, social, and environmental factors.}}
    \label{fig3-1}
\end{figure*}

In contrast, LLMs possess human-like internal representation capabilities, enabling them to spontaneously form concept systems that closely resemble human cognition, thereby supporting the capture of complex, human-like internal features of individuals~\cite{du2025human, gao2024large, dillion2023can, argyle2023out}. As illustrated in Figure~\ref{fig3-1}, in LLM-based epidemic models, the state transitions of each agent do not rely on explicit equations or parameterized rules. Instead, they depend on the agent's human-like internal cognitive processes and its perception of changes in the external environment.  
Meanwhile, depending on the powerful comprehension and reasoning abilities, LLM-based agents can dynamically perceive multimodal external environments, including text, images, and network data, and generate context-dependent behavioral responses through natural language reasoning. Within this framework, disease transmission rates and individual behavioral patterns are dynamically adjusted according to cognitive processes, past behaviors, cultural backgrounds, and so on~\cite{williams2023epidemic}.
This approach shifts the prediction of individual behavior beyond fixed rules to real-time generation shaped by the interplay of multiple factors, thereby better reflecting the complexity and uncertainty of real-world epidemics. By allowing agents to make adaptive decisions under the influence of social, cultural, and environmental cues, LLM-based modeling offers a more realistic reproduction of epidemic spreading dynamics in complex human populations. 
Table~\ref{tab3} summarizes LLM-based agent modeling methods for epidemics from multiple aspects, including agent-related settings (LLM, scale, and memory), network structure, state representation, and interaction mechanisms.
In the following subsections, we will survey how LLMs are being applied to model digital and biological epidemics, respectively. 

\begin{table}[htbp]
\centering
\caption{Summarization of LLM agent-based modeling methods for epidemics.\textcolor{black}{The third column records the specific LLM used in the modeling process. Scale refers to the number of agents involved in the modeling process. Memory in agents is categorized into three types: sensory memory, which enables perception; short-term memory, which supports real-time decision-making; and long-term memory, which facilitates sustained knowledge retention~\cite{liu2025advances}. Network structure refers to the topological form that defines how agents are connected and interact with one another, including random networks, scale-free networks, and others. Interaction describes the exchange of information among agents. Homogeneous interaction occurs only between agents of the same type, while heterogeneous interaction involves communication between agents of different types.}}
\label{tab3}
\label{tab:agent_benchmarks}\resizebox{\textwidth}{!}{
\begin{tabular}{cllllp{3cm}lll}
\toprule
\rowcolor{blue!10}
\textbf{Scenario} & \textbf{Research} & \textbf{LLM} & \textbf{Scale} & \textbf{Memory} & \textbf{Network structure} & \textbf{State Representation} & \textbf{Interaction} \\ 
\midrule
\multirow{14}{0.08\textwidth}{Digital} & Törnberg et al.~\cite{tornberg2023simulating} & gpt-3.5 & 500 & Sensory & - & Textual & Homogeneous \\
& Li et al.~\cite{li2024large} & gpt-3.5-turbo-1106 & 288/300 & Sensory & Random, scale-free, high brokerage network & Rule & Homogeneous \\
& Mou et al.~\cite{mou2024unveiling} & gpt-3.5-turbo & 1K & Long-term & Real world network & Textual, rule & Heterogeneous \\ 
& FUSE~\cite{liu2025stepwise} & gpt-4o-mini & 40 & Short-term, long-term & Random, scale-free, high clustering network & Textual & Heterogeneous \\ 
& Hu et al.~\cite{hu2025simulating} & gpt-4o-mini & 100/168 & Sensory & Random, scale-free, small world, real world networks& Textual & Homogeneous \\ 
& FDE-LLM~\cite{yao2025social} & ChatGLM-4 & $[206,9890]$ & Sensory & - & Textual, rule & Heterogeneous \\ 
& Li et al.~\cite{li2023quantifying} & - & - & - & - & Rule & Homogeneous \\ 
& Ju et al.~\cite{ju2024flooding} & \begin{tabular}[l]{@{}c@{}}Vicuna-7B,\\ Llama-3-8B,\\ Gemma-7B\end{tabular} & $[2,10]$ & Long-term & - & Textual & Heterogeneous \\
& Zheng et al.~\cite{zheng2024simulating} & gpt-4o & 15 & Long-term & Small world, scale-free networks & Textual & Homogeneous \\ 
& TrendSim~\cite{zhang2024trendsim} & GLM-3-turbo & 1K & Sensory, short-term, long-term & - & Textual, rule & Homogeneous \\ 
& Y Social~\cite{rossetti2024social} & Llama-3-7B & $[1\text{K},2\text{K}]$ & Sensory & - & Textual & Homogeneous \\ 
& OASIS~\cite{yangoasis} & Llama-3-8B-Instruct & 1M & Short-term, long-term & Scale-free network & Textual, rule & Homogeneous \\ 
& LLM-AIDSim~\cite{zhang2025llm} & Llama 3 & 333 & Sensory & Real world network & Textual & Homogeneous \\ 
& MOSAIC~\cite{liu2025mosaic} & gpt-4o & $[50,130]$ & Long-term & Scale-free network & Textual & Heterogeneous \\ 
& Nasim et al.~\cite{nasim2025simulating} & gpt & - & - & - & Textual, rule & Heterogeneous \\ 
& S$^3$~\cite{gao2023s3} & gpt-3.5, ChatGLM-6B & 8,563/17,945 & Long-term & Real world network & Textual & Heterogeneous \\ 
& Nudo et al.~\cite{nudo2026generative} & Gemini, Mistral, DeepSeek & 1,186 & Sensory & - & Textual & Homogeneous \\
\midrule
\multirow{2}{0.08\textwidth}{Biological} & Williams et al.~\cite{williams2023epidemic} & gpt-3.5-turbo-0301 & 100/1K & Sensory & - & Textual, rule &  Homogeneous \\ 
& Villaplana et al.~\cite{villaplana2024application} & gpt-3.5 & 576 & Sensory & - & Textual & Homogeneous \\ 
\bottomrule
\end{tabular}
}
\end{table}

\subsection{LLM-based digital epidemic modeling}
\label{Sec: Single epidemic on simple networks}
In the epidemiological modeling of digital information spreading, such as rumors, misinformation, conspiracy theories, or opinions, an infected individual represents an agent who has received and possesses a specific piece of information and can transmit it to others. A susceptible individual refers to one who has not yet encountered the information but may become infected upon exposure. A recovered individual is considered immune to the information, meaning that upon encountering it, they will not be influenced by it. For instance, in rumor spreading, infected individuals are those who have heard and believe the rumor, susceptible individuals have not yet encountered it, and recovered individuals have recognized its falsehood.

Recently, the rapid development of LLMs has gradually permeated research on digital epidemiological modeling and spreading dynamics, reshaping traditional research paradigms to extent. On the one hand, LLM-based agents are replacing rule-driven or statistically driven conventional agents by employing natural language to simulate network nodes' attributes, state transitions, interactions, etc., thereby enhancing the complexity and realism of spreading simulations. On the other hand, LLMs themselves can be regarded as novel information producers and amplifiers within spreading processes. Their large-scale content-generation capabilities alter key aspects of information spread, such as its speed and scale, thereby constituting critical external variables that influence information spreading. In the following sections, we elaborate on these two aspects and summarize relevant studies as well as key findings.

\subsubsection{LLMs as augmented tools for digital epidemic modeling}
Traditional digital epidemic modeling typically relies on predefined mathematical equations or rules to predict the spreading trends of different types of information. Although straightforward,  these approaches often depend excessively on numerical abstractions of opinions and messages, which oversimplify the complexity of real-world scenarios. As a result, they fail to capture the semantic content of information and cannot fully reflect the flexibility of human reasoning, communication, and decision-making. Nowadays, an increasing number of researchers have begun leveraging LLM capabilities for environmental perception, logical reasoning, and autonomous planning to address these shortcomings~\cite{gao2024large}. Within ABM frameworks, LLM-driven agents have been proposed as substitutes for traditional rule-based agents, enabling simulations of individual behavioral responses to information diffusion in a manner that more closely reflects human cognition~\cite{li2024large,liu2024skepticism,ji2025llm,taillandier2025integrating}.

In early attempts, Li et al.~\cite{li2024large} employed simple prompts such as ``\textit{You are a [persona]. What is your opinion on the [news]? Then would you like to share the [news] with your friends based on your personality? Reply `Yes' or `No' along with reasons.}'' to instruct LLMs to reason about whether the pre-defined agent would disseminate the given misinformation. They treated personality as a control variable and focused on how three network structures (random, scale-free, and high-brokerage) affect the spread of fake news.
Their experimental results revealed that participants with high scores in extraversion and openness were more likely to spread information~\cite{sampat2022fake, ahmed2022social}. At the structural level, scale-free networks exhibited the fastest spreading due to the accelerating effect of highly connected nodes, whereas high-brokerage networks showed the slowest spreading as sparse inter-cluster connections created bottlenecks.

\begin{figure*}
        \centering
    \includegraphics[width=0.9\linewidth]{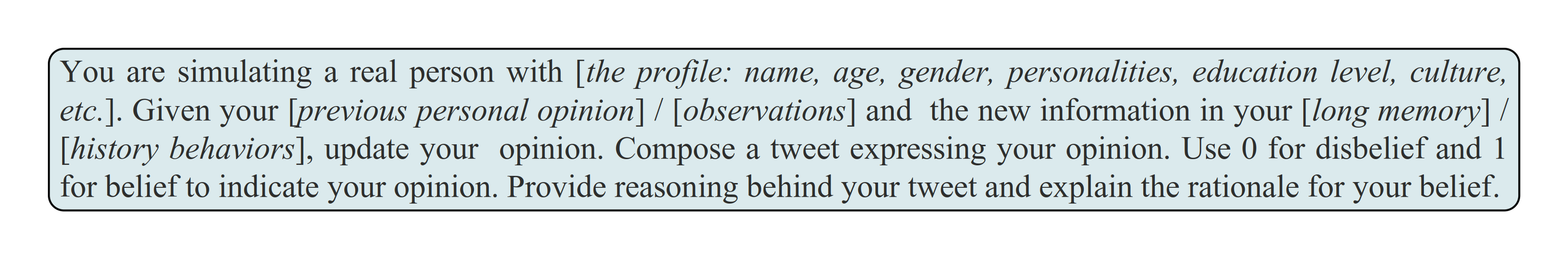}
    \caption{An example of opinion reasoning prompt templates for LLM-based agents.}
    \label{table:opinion-reasoning}
\end{figure*}

\begin{figure*}[htbp]
    \centering
    \includegraphics[width=0.9\linewidth]{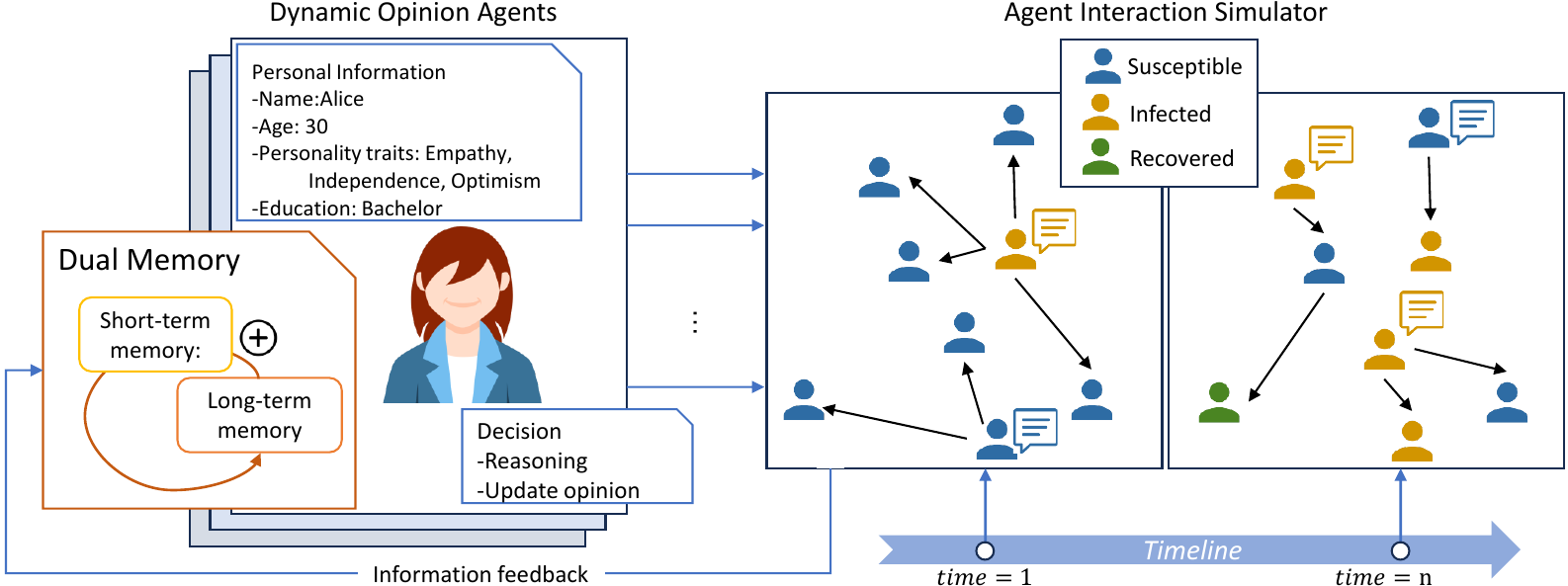}
    \caption{An illustration of the FPS framework. { The framework simulates the process of digital epidemic spreading and consists of two main components: dynamic opinion agents and an agent interaction simulator. The dynamic opinion agents leverage short-term memory to record recent interactions with other agents, long-term memory to store background information, and a reflection mechanism to emulate human-like cognitive processes and update their attitudes toward specific information. The agent interaction simulator constructs an interactive environment in which agents communicate with one another through natural language exchanges, rather than relying on predefined parameters or rules.}
   \\Source: Reproduced from Ref.~\cite{liu2024skepticism}.}
    \label{fig:model-FPS}
\end{figure*}

Nevertheless, in real social networks, an individual's decision to adopt and disseminate information is influenced not only by age, gender, and personality, but also by historical behaviors, social identity, and environmental factors. 
To tackle this limitation, Liu et al. proposed a propagation simulation framework (FPS) based on LLMs, which leverages LLMs to emulate human-like cognition and behavior by systematically integrating historical memory, environmental interaction, and identity traits~\cite{liu2024skepticism}. As illustrated in Fig.~\ref{fig:model-FPS}, FPS consists of two core modules: the dynamic opinion agent (DOA) and the agent interaction simulator (AIS). 
The DOA equips each agent with short-term memory to record daily interactions, long-term memory to capture broader context, and a reflection mechanism to simulate human thought processes and update attitudes toward news. All these functions are implemented by the prompt technology of LLMs. AIS, on the other hand, constructs the environment in which agents interact. 
As shown in Fig.~\ref{table:opinion-reasoning}, opinion exchange in FPS is conducted through natural language communication, similar to the real world, rather than abstract numerical representation. Their results found that political misinformation spreads significantly faster than misinformation related to terrorism, natural disasters, science, folklore, or finance. Furthermore, individuals with high agreeableness and high neuroticism were found to be more susceptible to believing and propagating false information. Wang et al.~\cite{wang2025decoding} proposed a social simulation framework which adopted nearly the same LLM-based agent attributes and propagation action settings as FPS. It incorporated commonly used recommendation algorithms to push similar opinions, aiming to investigate the echo chamber effect and the phenomenon of opinion polarization.

Similar to the aforementioned studies, Chuang et al.~\cite{chuang2024simulating} also employed LLMs to mitigate the distortion and oversimplification inherent in traditional ABMs that rely on numerical abstraction of information and rule-based interactions. Unlike prior approaches, their work incorporated confirmation bias into agents' opinion evolution and introduced a distinct memory updating mechanism to examine how these factors shape opinion dynamics. Their simulation framework adopted a speaker-listener setting, where LLM-based agents update their opinions by receiving information from the current LLM-based speaker in a one-to-one manner, rather than by aggregating all the opinions they have previously encountered (one-to-many). Under this framework, the study employed a cumulative memory mechanism, in which each new interaction was sequentially retained in memory, thereby fully recording the trajectory of agents' opinion evolution. By assigning different levels of confirmation bias to LLM-based agents, the study found that, without confirmation bias, LLM-driven agents tended to converge rapidly toward scientific consensus. However, once confirmation bias was introduced, opinion diversity increased significantly, reproducing the polarization phenomena commonly observed in traditional ABMs, with stronger bias leading to greater diversity. 
At the same time, the study revealed a limitation of constructing agents with LLMs: regardless of the roles assigned, LLM-based agents generally tended to reject inaccurate information and endorse factual content, which limits their authenticity when simulating individuals who persistently resist factual evidence. { It is noteworthy that consensus refers to an ordered state that emerges from interactions among social agents~\cite{castellano2009statistical}. Ashery et al.~\cite{ashery2025emergent} showed that agents controlled by LLMs can spontaneously form consensus through their interactions. Similarly, De Marzo et al.~\cite{de2024ai} found that interacting agents can influence one another such that groups may spontaneously coordinate on arbitrary decisions. Related studies provide valuable insights into understanding LLM-based digital epidemic spreading, and further discussions can be found in Refs.~\cite{flint2025group,jeon2025simulating,cirulli2025large,verginer2025irrational,bouleimen2025collective,orlando2025emergent,yu2024large}}.

Excluding the two types of speaker-listener LLM-based agents, several other studies have constructed more diverse agent roles using LLMs to better capture individual heterogeneity and behavioral complexity in the spread of digital epidemics. For example, the FUSE framework~\cite{liu2025stepwise} employed prompt engineering to construct four types of LLM-based agents: spreaders, commentators, verifiers, and bystanders. These agents respectively perform information diffusion, opinion interpretation, fact-checking, and passive observation in the evolution of misinformation in social networks. Meanwhile, Ma et al.~\cite{ma2024simulated} designed a set of agents based on LLMs that incorporate demographic differences (e.g., age, gender, education level, political orientation, and trust in science, government, and news) together with diverse life experiences, serving as virtual participants in the sensitivity to misinformation simulation test. These studies demonstrate that LLMs can be used to generate multiple types of agent roles, thereby reproducing complex social dissemination phenomena at both semantic and behavioral levels.

However,  compared to traditional rule-based or probability-based ABMs, LLM-based agent modeling exhibit disadvantages in terms of computational efficiency and resource consumption. 
Classic ABM methods can achieve large-scale population spreading simulations on a short time. For example, EpiSimdemics can complete spreading simulations of hundreds of millions of people on NCSA's Blue Waters supercomputing system (with 352,000 computing cores) over a period of 120 days in just 12 seconds~\cite{bissett2021agent}; and Epiabm can simulate the spread of infection in the Gibraltar region (population about 34,000) for 80 days, with the simulation taking only 8 seconds on an AMD 3600X processor (6 cores, 3.8 GHz)~\cite{gallagher2022epidemiological}.
In contrast, LLM-based agent systems often require invoking large-scale language models for inference in each round of interaction or state update, with computational costs increasing rapidly with agent size and the number of interaction rounds. As shown in Table ~\ref{tab:computational_cost}, existing LLM-based agent systems typically require several hours or even tens of hours of runtime at agent scales of thousands to tens of thousands, relying on high-performance computing resources or expensive inference costs. This high computational overhead limits the scalability of LLM-based methods in large-scale, long-duration spreading simulations.
\begin{table}[htbp]
\centering
 
\caption{Computational cost of LLM-based agent systems.}
\label{tab:computational_cost}
\begin{tabular}{l l l p{4cm} l l}
\toprule
\textbf{Method} & \textbf{Model} & \textbf{Scale} & \textbf{Hardware} & \textbf{Latency} & \textbf{Cost} \\
\midrule
AgentSociety \cite{piao2025agentsociety} & DeepSeek-V3 & 10K & Huawei Cloud c7.16xlarge.4 & 458.82 s/round & - \\
GABM \cite{williams2023epidemic} & gpt-3.5-turbo & 1K & 32 GB of RAM CPU & 80 h/round & \$20/round \\
TrendSim\cite{zhang2024trendsim}  & GLM-3-turbo & 1K &  Intel Xeon Gold-5118 (48 Core) CPU & 16h & - \\
OASIS \cite{yangoasis} & Llama3-8b-instruct & 1M & 24 A100 GPUs & 2.8h/round & - \\
\bottomrule
\end{tabular}
\end{table}

To address the challenges of cost and efficiency in simulating large-scale social network users, Mou et al.~\cite{mou2024unveiling} proposed a hybrid simulation framework. This framework leverages the inherent Pareto distribution characteristics of social network users to categorize them into two groups: core users, like opinion leaders and ordinary users. Differentiated modeling approaches are employed for these roles: core users are characterized and driven by LLMs to capture human complex behavioral patterns, whereas the large population of ordinary users is controlled by traditional ABMs, enabling more efficient large-scale user simulation.

Similarly, the fusing dynamics equation-LLM (FDE-LLM) framework~\cite{yao2025social} also classifies social network users into opinion leaders and followers, where the former are modeled through LLM-based role-playing, and the latter are simulated using traditional ABM models. To maintain the natural decay of group attitudes under the guidance of LLM and improve the efficiency of opinion propagation within the simulated group, the FDE-LLM framework integrates constraints from both cellular automata (CA) and the SIR model. In this design, opinion leaders are implemented via LLM-driven role-playing, with opinion shifts constrained by CA, whereas followers are embedded in a dynamic system that combines CA with the SIR model. Experimental results demonstrate that this design significantly improves simulation accuracy and predictive efficiency, particularly in scenarios involving reversal news.

In the aforementioned two studies, the division between core and ordinary users was typically fixed, lacking dynamic adaptability and thus constraining the flexibility of the spreading simulations. To address this limitation, Liu et al.~\cite{liu2025rumorsphere} proposed a dynamic and hierarchical social network simulation framework capable of modeling millions of individuals. This framework introduces a multi-agent dynamic interaction strategy based on the theory of information cocoons, which adaptively identifies core and ordinary agents at each time step according to the degree of informational disorder. Similarly, core agents are driven by LLMs to capture complex reasoning processes, whereas ordinary agents are controlled by ABMs to enable large-scale scalability. Both the number of core agents and the interaction patterns among agents are dynamically adjusted over time, thereby enhancing the flexibility and realism of the simulation. Furthermore, the framework incorporates a hierarchical collaborative network, where priority connections foster opinion leaders and triangular connections facilitate the formation of tightly knit local communities, accelerating information diffusion. Experimental results demonstrate that this approach achieves higher consistency and accuracy in simulating real-world rumor propagation, reducing the average opinion deviation by 64\% compared to existing models, and further reveal that tightly connected local community structures in social networks constitute one of the key factors driving the rapid spread of rumors. 

In addition to modeling efficiency, generating realistic, scalable social networks is also essential for digital epidemic modeling, as the network structure largely determines spreading pathways and scale. Unfortunately, traditional rule-based models struggle to capture the complexity of real-world interactions, while deep learning approaches often rely on large-scale annotated datasets. 
To overcome these limitations, Ji et al.~\cite{ji2025llm} introduced the zero-shot social network generation framework, \textit{GraphAgentGenerator (GAG)}. By introducing LLM-driven multi-agent mechanisms into the graph generation process, GAG enables agents to dynamically generate nodes and edges through simulated interactions, producing social graphs that more closely resemble reality.  
The GAG framework consists of two types of nodes: (i) actor nodes, instantiated as LLM-based agents, and (ii) item nodes representing content. Edges emerge progressively through agent-content interactions. For each actor, its profile is generated by LLMs,
while the agent maintains a textual memory component that records its activity history. In each simulation round, agents are guided by the designed action prompt to act based on historical memory and contextual retrieval, updating states through reflection. Agent activities in GAG follow a Pareto distribution, with about $20\%$ of highly active agents generating most interactions.  
Experimental results demonstrate that GAG generates social networks exhibiting seven macroscopic properties observed in real-world systems, including power-law degree distributions, small-world characteristics, and shrinking diameters. Moreover, the framework shows strong scalability, supporting the generation of diverse network types with up to $10^5$ nodes and $10^7$ edges. 
This work indicates that incorporating LLMs into social network generation not only enables semantic-level simulation of individual interactions but also reproduces statistical regularities of complex networks well established in physics. Consequently, GAG provides a realistic and scalable network environment for modeling the spread of the digital epidemic.

Cheng et al.~\cite{cheng2025interactive} proposed a social network simulation system that integrates LLM-driven agents with interactive visual analytics. By combining modeling with visualization techniques, the system introduces an interactive approach to modeling and studying digital epidemics, enabling researchers to conduct iterative exploration and analysis in complex, dynamic scenarios. The system demonstrates how LLM-driven agents with different characteristics vary in processing digital information, regulating emotions and stances, providing feedback, and influencing the evolution of events, thus visually uncovering the dynamic mechanisms of public opinion events on social networks.

Beyond the ABM-inspired modeling approaches discussed above, other studies have integrated LLMs into specific modules of different modeling frameworks to enhance their functionality. For instance, Jiang et al.~\cite{jiang2025epidemiology} employed LLMs to infer nodes' stances toward posts within propagation trees, thereby assigning state labels to nodes and facilitating the optimization objectives of epidemiological information representation learning. Meanwhile, the BotSim~\cite{qiao2025botsim} system leveraged LLMs to construct user nodes that generate and disseminate harmful information in digital epidemics, and further utilized tailored prompts to command LLMs to accomplish tasks such as task decomposition and perception of real social network environments.

\subsubsection{LLMs as influential factors in digital epidemic spreading}

In the modeling of digital epidemic spreading dynamics, the emergence of LLMs introduces a set of novel external factors that influence propagation landscape. { These factors include LLMs acting as sources of digital epidemics, LLM-controlled agents participating in human-centered spreading processes, and propagation occurring among networks of LLM-driven agents. We subsequently analyze each scenario in detail and discuss how these external factors collectively affect the modeling of digital epidemic spreading.} 

First, LLMs expand the set of ``information sources'' by becoming new content producers~\cite{lakshmi2025rebranding, jiang2024catching,farquhar2024detecting,lucas2023fighting}. They can generate content that is virtually indistinguishable from human writings at a very low cost and on a large scale \cite{park2022social, bai2025llm,HUANG2025114529}, producing misinformation unintentionally through hallucinations or intentionally when exploited for persuasive and misleading narratives \cite{barman2024dark, radivojevic2024llms}. From the perspective of inoculation theory \cite{roozenbeek2022psychological}, such LLM-generated outputs can be conceptualized as new ``pathogens'', constantly evolving, adaptive, and difficult to detect \cite{jiang2024catching}. Consequently, LLMs not only increase the volume of such ``pathogens'' in circulation but also accelerate misinformation diffusion by enhancing their diversity and credibility \cite{barman2024dark, radivojevic2024llms}. From a dynamical viewpoint, the continuous generation of content by LLMs functions as an exogenous injection term, elevating the reproduction rate of misinformation and its effective infection rate, thereby potentially strengthening system-level transmissibility and altering outbreak thresholds and steady states.  

Second, LLM-controlled agents may participate in human-centered digital epidemic spreading, interacting with human users and influencing the diffusion of information, opinions, or misinformation. This participation introduces two effects.
On the one hand, LLMs internalize and amplify social biases during training, which makes their generated outputs inherently structured with ideological or polarizing tendencies ~\cite{omar2025sociodemographic, tan2025unmasking, barfar2026propaganda}. 
As many  open-ended question-answering systems and interactive platforms rely on LLMs to produce responses, these biased outputs can directly influence users and are further amplified through reshare mechanisms on social networks, propagating specific narratives. 
This implies that individual susceptibility and transmission probabilities are no longer homogeneous but heterogeneously modulated by bias. On the other hand, 
LLM-driven social bots can be deployed in online networks, mimicking human-like interactions and altering diffusion pathways through targeted engagement and personalized content injection  \cite{uyheng2022bots,zhang2024toward,radivojevic2024llms}. Recent work demonstrates that such bots can even infer user attributes with high accuracy and perform targeted interventions,  acting as ``super-spreader nodes'' with intelligent intervention strategies~\cite{staabbeyond2024}. Hitz et al.~\cite{hitz2025amplifier} find that LLM-driven agents tend to have lower adoption thresholds compared to human subjects and may therefore act as amplifiers, leading to faster and broader digital epidemic spreading.

{ Third, spreading dynamics can also arise within networks formed by large numbers of LLM-driven agents, potentially controlled by different LLMs, where interactions among artificial agents themselves give rise to complex propagation patterns. The dynamics and characteristics of such propagation processes constitute an emerging research topic in their own right. This line of inquiry falls within the emerging interdisciplinary field of machine behavior, as articulated by Rahwan et al.~\cite{rahwan2019machine}, which calls for the scientific study of intelligent machines not as engineered artifacts, but as a class of actors with characteristic behavioral patterns and interaction ecologies. Recent studies have begun to explore this direction empirically. For example, Marzo et al.~\cite{de2023emergence} investigate the interactions among multiple generative agents implemented using GPT-3.5-turbo, analyzing how information and behaviors propagate within populations composed entirely of artificial agents. They find that such agents are able not only to mimic individual human linguistic behavior, but also to exhibit collective phenomena intrinsic to human societies. Similar studies can be found in Refs.~\cite{chang2025llms, papachristou2025network}. These studies can contribute to an improved understanding of the spreading behaviors and propagation patterns exhibited by LLM-enhanced agents as machines, within the broader context of machine behavior. At the same time, such agent–agent spreading processes are not isolated from human participation. Information exchanged and amplified among artificial agents in cyberspace may ultimately be observed, consumed, or further disseminated by human users, who remain embedded in the broader socio-technical system. Understanding the spreading dynamics among LLM-driven agents therefore provides important insights into how machine-generated information may interact with, and feed back into, human-centered digital epidemic processes.}

The three factors discussed above all influence digital epidemic spreading. In practice, they are not isolated influences, but may coexist in cyberspace complex system settings. For example, Lu et al.~\cite{lu2025understanding} investigated the role of LLM-driven adversarial social influences in online information diffusion. Their study highlights that, as generative AI advances, LLMs can produce highly realistic, human-like text, significantly enhancing the effectiveness of social bots. As a result, LLM-based bots can not only mass-produce comments that contradict the veracity of news but also generate adversarial replies to human users' posts, thereby constructing a false sense of a dominant opposing viewpoint. In a randomized controlled experiment with 176 participants, the authors found that such LLM-driven adversarial influences reduced individuals' ability to distinguish between true and false news and decreased their willingness to share authentic information. Notably, adversarial comments posted directly against the truthfulness of news were more effective at suppressing the dissemination of real news compared to adversarial replies. Furthermore, the effects were moderated by participants' political orientation: when the news content conflicted with their political stance, LLM-driven adversarial influence significantly amplified misjudgments. These findings underscore the dual role of LLMs in the information diffusion process: they not only generate misleading content but also disrupt human cognition and behavior through adversarial social interactions, thereby exacerbating the large-scale spread of misinformation.  
From the perspective of spreading dynamics, this study provides important insights. First, LLM-driven social influence can be regarded as an external dynamical factor that is not neutral random noise but a systematic perturbation that propagates adversarial viewpoints at local nodes through ``comments'' and ``replies,'' effectively introducing a directed disturbance field into diffusion models. Second, since adversarial influence reduces the willingness to share true information while increasing the effective transmission rate of false information, it may lower the critical threshold of information diffusion, making misinformation outbreaks more likely to surpass the tipping point. Traditional approaches in physics-inspired ``digital epidemiology'' and complex systems often assume perturbations to be uniform or random, whereas LLM-driven disturbances are intelligent, directional, and dynamically adaptive. This suggests that future spreading models need to explicitly incorporate the dynamical coupling of LLM-based bots as active propagators and consider how such directed manipulations alter system stability and criticality.  

In addition, Bandara~\cite{bandara2024hallucination} investigated how hallucinations generated by LLMs can serve as sources of disinformation, and how they systematically amplify the spread of conspiracy theories and fake news. In short, LLM hallucinations operate through a three-level progression: beginning with technical deficiencies in model architecture, proceeding to the generation of false content, and culminating in social dissemination. This makes LLM hallucinations a central driving force in a new form of disinformation production and distribution. To explain how LLMs contribute to the amplification of disinformation ecosystems, the study categorizes their influence into three major aspects. (a) Conspiracy theory propagation: LLMs construct false associations by linking unrelated events as supposed ``hidden conspiracies'', creating the illusion of evidence. Within closed communities, collective interpretation of such content, combined with algorithmic recommendation mechanisms, reinforces these hallucinations as shared group beliefs. Additionally, historical grafting occurs when hallucinated content merges factual historical events with contemporary rumors, enhancing the perceived credibility of disinformation. (b) Erosion of the public sphere: the intensification of ideological polarization results from algorithmic recommendations and echo chamber effects, trapping users in feedback loops that reinforce existing biases. Meanwhile, the collapse of institutional trust emerges as hallucinated content diminishes the perceived authority of mainstream media and academic institutions. This deterioration of trust further leads to democratic risks, where the public becomes unable to form a consensus based on factual information, thereby obstructing rational policy-making. (c) Large-scale fake news production: through automated content pipelines, LLMs integrated with orchestration tools enable the coordinated, cross-platform dissemination of fabricated news. Traffic-driven optimization adjusts generated content based on real-time engagement metrics, prioritizing emotionally charged and extreme narratives. At the same time, search engine manipulation through keyword optimization can allow disinformation to dominate prominent positions in search results. In summary, the integration of LLMs not only spawns new pathogens but also deeply intervenes in network dynamics as new transmission agents, posing new challenges to the modeling of digital epidemic spreading.

\subsection{LLM-based biological epidemic modeling}
Biological epidemic modeling is often closely integrated with downstream tasks such as prediction and intervention, aiming to capture the underlying transmission dynamics and support epidemiological forecasting and {  management} strategies. With the increasing availability of large-scale data and cross-disciplinary techniques, researchers have begun exploring how LLMs can enhance the intelligence and automation of this process, since their strong capabilities in context-aware inference, cross-source semantic integration, and latent pattern abstraction provide a promising foundation for understanding the complex dynamics of disease spread~\cite{bzdok2024data}. Broadly, existing studies fall into two categories. The first involves directly engaging LLMs in the modeling process, where they play an active role in transmission modeling, dynamic inference, or prediction. The second focuses on indirectly leveraging LLMs as auxiliary tools to assist in tasks such as code generation, model optimization, and data fitting. While the former highlights the reasoning and mechanistic capabilities of LLMs, the latter emphasizes their potential to enhance modeling efficiency and automation. The following subsections elaborate on these two lines of research.

\subsubsection{LLMs directly involved in the modeling process}

Similar to LLM-based ABM approaches in digital epidemics, several studies in biological epidemic modeling have employed LLM-based agents to directly simulate epidemic dynamics~\cite{villaplana2024application}. For example, Villaplana et al.~\cite{villaplana2024application} utilized ChatGPT-3.5 to construct a group of agents endowed with heterogeneous attributes and information backgrounds. Through prompt engineering, each agent was assigned distinct behavioral traits to emulate human decision-making processes, thereby addressing the absence of behavioral feedback mechanisms in traditional SIR models. In their experiments, they explored two key questions about pathogen characteristics and their impact on biological epidemic spreading: the number of people the agent interacts with daily and whether social distance is maintained during these interactions. The results showed that as the agent receives more information about the epidemic, the case growth curve may gradually flatten, with peaks becoming less pronounced or even absent, indicating that information dissemination can significantly influence the dynamics of disease transmission.

Apart from the LLM-based agent modeling paradigm, several studies have employed LLMs to directly model the temporal evolution of epidemic transmission. The EpiLLM framework~\cite{gong2025epillm} utilizes an LLM as the foundational model for spatiotemporal epidemic modeling, adopting an autoregressive approach to characterize the transmission process. The transmission process is divided into temporal segments, during which the GNN encodes epidemiological features (e.g., infection counts) into tokenized inputs for the LLM. The model then predicts the transmission outcomes of the $N$th segment based on the preceding $N-1$ segments, enabling generative modeling of spatiotemporal epidemic evolution.

Furthermore, Moon et al.~\cite{moon2025miflu} and Du et al.~\cite{du2025advancing} proposed multimodal epidemic modeling frameworks based on LLMs that integrate multiple sources of contextual information, such as epidemic statistics, public health policy texts, and temporal sequences, to support more accurate predictions. Similarly, Kang et al.~\cite{kang2025llm} leveraged LLMs to learn from multimodal information involved in epidemic modeling, enriching the semantic representations of patients, diseases, and discharge summaries to construct more detailed and higher-order health state representation matrices. Distinctively, they introduced an LLM-enhanced intra-patient multimodal modeling strategy by constructing a disease co-occurrence network to better understand the progression and interactions of diseases within individual patients.

\subsubsection{LLMs indirectly involved in epidemic modeling as an auxiliary tool}

Beyond directly engaging LLMs in epidemic modeling, some studies employ them as auxiliary tools to indirectly support the modeling process. For example, Kwok et al.~\cite{kwok2024utilizing} proposed using ChatGPT to assist in developing infectious disease transmission models. Through natural language interaction, ChatGPT performs code generation, optimization, and debugging tasks, enabling researchers to build an SEIR model and fit it to epidemiological data. This approach substantially lowers the technical barrier, allowing non-programming experts to rapidly construct transmission models.

Moreover, LLMs have also been applied after  modeling models, providing functionality for making judgments and decisions based on modeling information. Zaslavsky et al.\cite{zaslavsky2024enhancing} integrate LLMs with spatially-aware ABM and system dynamics (SD) models, leveraging the models' spatial reasoning and knowledge representation capabilities to enhance sensitivity to population distribution patterns. For example, in COVID-19 transmission simulations, the LLM generates behaviors with spatial preferences for participants and dynamically adjusts the contact probability based on participants' health status and the viral load in the environment, thereby making the transmission process more in line with real social spatial interaction characteristics.

\begin{figure*}[htbp!]
  \centering
  \begin{minipage}[t]{0.45\textwidth}
    \centering
    \includegraphics[width=\linewidth]{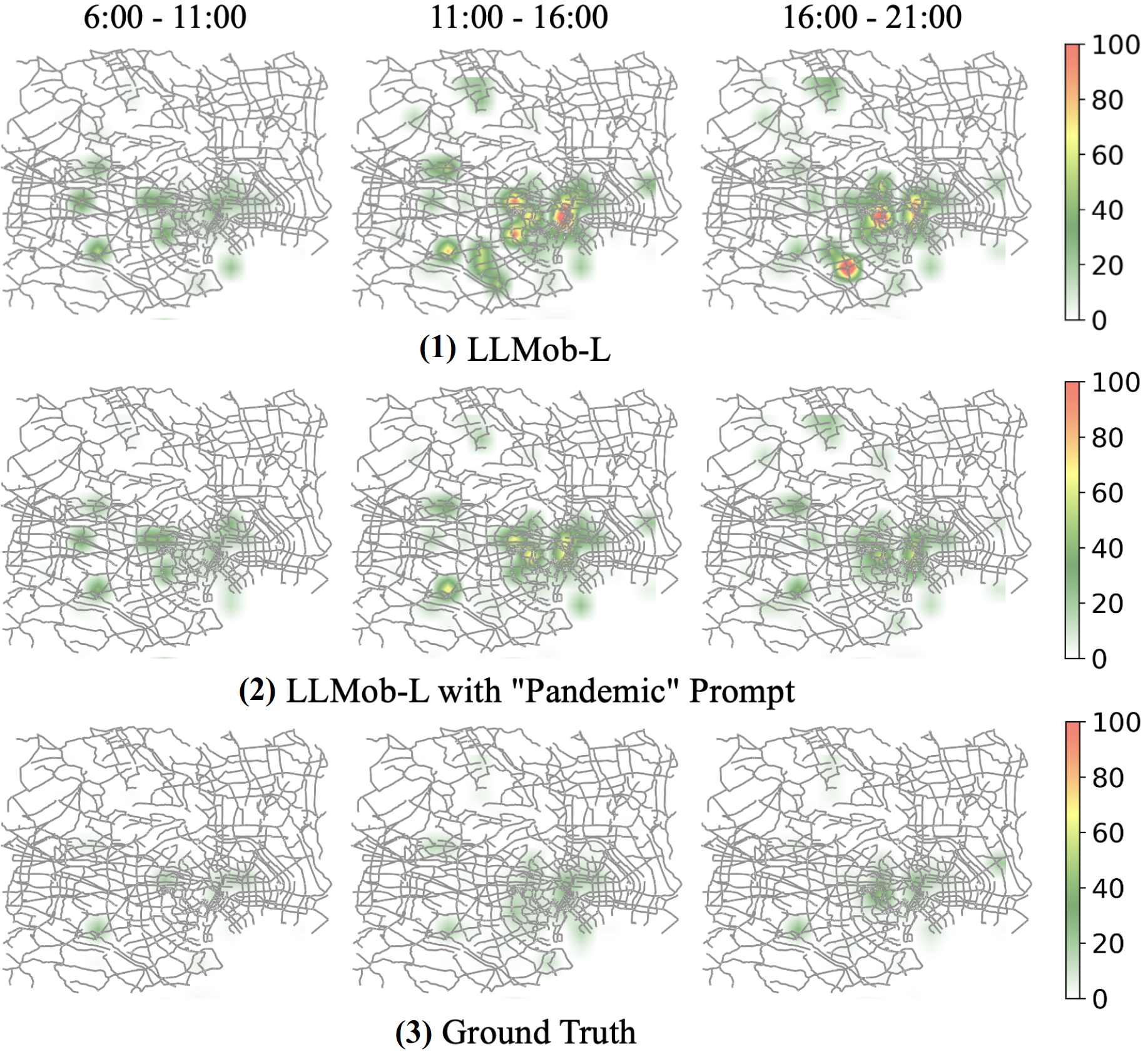}
    \vspace{0.5ex}
    \textit{(a) Arts \& entertainment.}
  \end{minipage}
  \hfill
  \begin{minipage}[t]{0.45\textwidth}
    \centering
    \includegraphics[width=\linewidth]{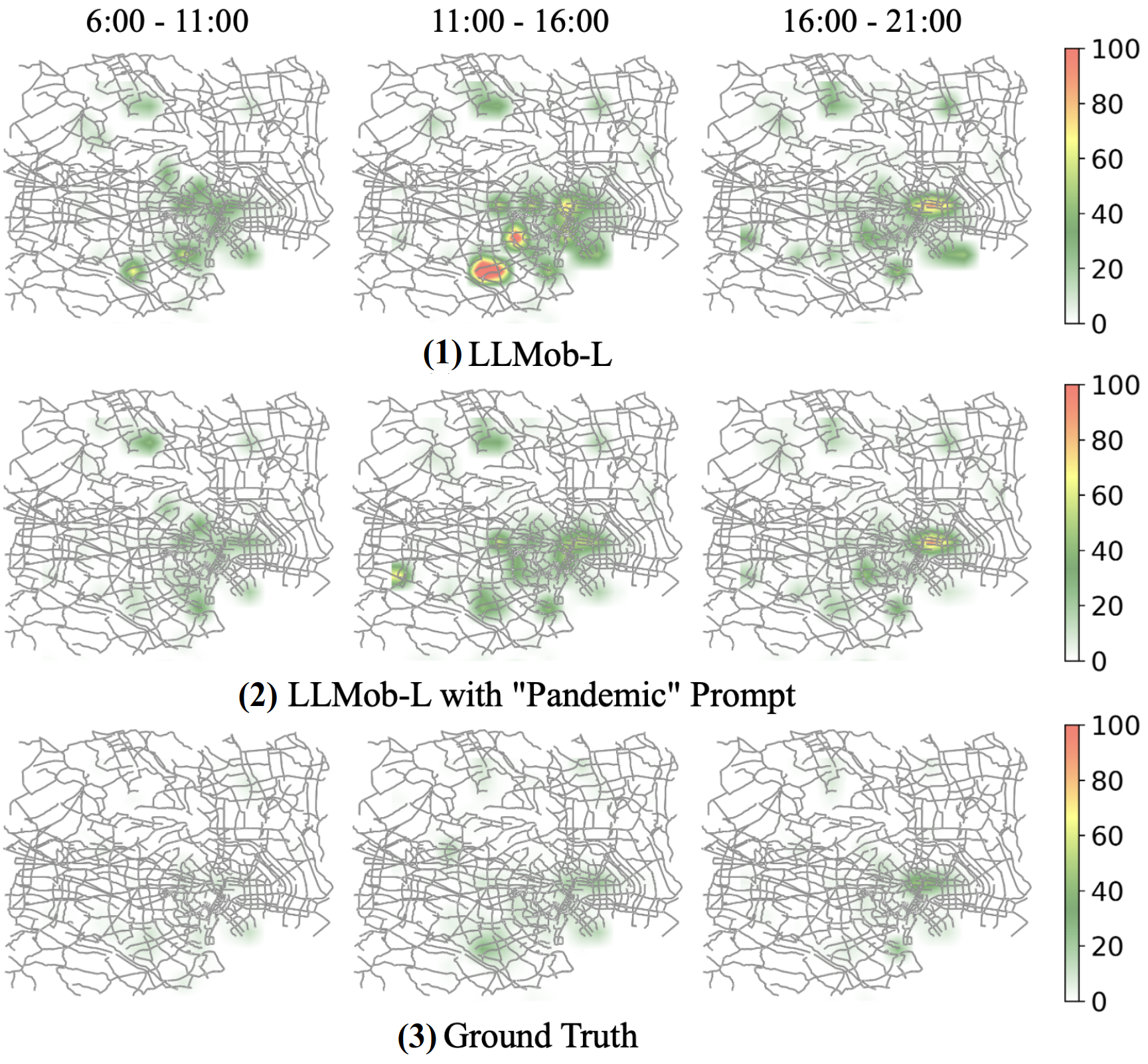}
    \vspace{0.5ex}
    \textit{(b) Professional \& other places.}
  \end{minipage}
  \caption{Activity heatmaps for the pandemic scenario in ``arts \& entertainment''and ``professional \& other places'' activities. 
  ``LLMob-L'' denotes the variant of the LLMob framework that integrates a learning-driven motivation retrieval mechanism.
  The ``pandemic'' prompt is used to enable the LLM agent to plan activities in pandemic situations, as follows: ``\textit{Now it is the pandemic period. The government has asked residents to postpone travel and events and to telecommute as much as possible}''.
  \\Source: Reproduced from Ref.~\cite{wang2024large}}
  \label{fig:exp:real-world-application}
\end{figure*}

Building on this integration of spatial reasoning and behavioral adaptation, researchers have further extended the application of LLMs to human mobility prediction, a crucial component of epidemic modeling, as the spread of infectious diseases largely depends on the spatial distribution and interactions of individuals and groups.
Accurate simulation and prediction of human movement patterns have thus become key to improving the fidelity of biological epidemic models. Recently, several studies have proposed LLM-based frameworks dedicated to high-fidelity human mobility modeling and prediction, providing new impetus for epidemic transmission modeling~\cite{wang2024large}.  
For example, Wang et al. proposed an
LLM Agent Framework for Personal Mobility Generation (LLMob), which treats individuals in urban environments as LLM-based agents capable of generating socially constrained and spatiotemporally consistent mobility trajectories by jointly modeling habitual activity patterns and context-driven motivations~\cite{wang2024large,zhang2025web}. 
They found that their model can reproduce more realistic spatio-temporal activity distributions under the pandemic prompt, as illustrated in Fig.~\ref{fig:exp:real-world-application}. This enhanced realism stems from the model's incorporation of prior knowledge about epidemic impacts and government responses, enabling LLM agents to act in ways that align with real-world behavioral adaptations.

\section{Epidemic Detection and Surveillance: LLMs-Enhanced Perception of Spreading Dynamics}
\label{sec4}
The detection and surveillance of { epidemic} spreading processes are primarily used to \textcolor{black}{perceive} the objects, behaviors, and evolution characteristics of contagions occurring in the real world, including identifying the source or origins of propagation~\cite{pinto2012locating,zhu2017catch}, recognizing early-warning signals~\cite{scheffer2009early} from network structures, and monitoring ongoing trends in disease or information diffusion~\cite{berkelman1994infectious,mello2020ethics,brownstein2023advances}. From the perspective of physicists, spreading dynamics mainly refers to the theoretical modeling, simulation, and analysis of how contagions propagate through complex networks~\cite{pastor2015epidemic}. \textcolor{black}{Thus, through the perception of the spreading process, detection and monitoring can provide a realistic basis for the theoretical characterization and parameter calibration of spreading dynamics models,} thus playing a crucial bridging role between theoretical research and practical applications of spreading dynamics.

This section investigates the role of LLMs in perceiving the emergent feature and ongoing trends of the spreading process. By analyzing massive volumes of textual and multimodal data, LLMs can detect weak signals, trace outbreak sources, and characterize propagation states. Specifically, this section reviews how LLMs enhance the perception of digital epidemics by identifying early signals of rumors or misinformation, and how they assist in biological epidemic surveillance for early warning, symptom signal detection, and ongoing trends monitoring.

\subsection{LLM-enhanced digital epidemic perception}
\begin{figure*}[htbp]
    \centering
    \includegraphics[width=1.\linewidth]{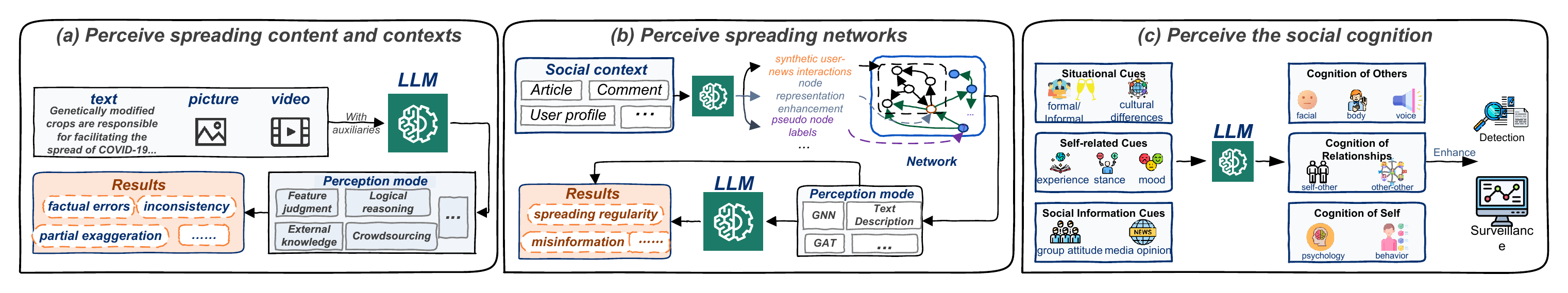}
    \caption{\textcolor{black}{Different perception targets in digital epidemics.}}
    \label{fig:digital_percept}
\end{figure*}
Benefiting from the comprehensive capabilities in language understanding, cross-modal information processing, and contextual reasoning, LLMs have been increasingly applied to the detection and surveillance of digital epidemics spreading. LLMs can not only recognize and interpret spreading content within complex and rapidly evolving online environments, but also integrate information from heterogeneous platforms and multimodality to uncover propagation paths, semantic associations, and latent dynamic characteristics, providing new technical means for perceiving the process of digital epidemic spreading.  
\textcolor{black}{Overall, the enhancement of digital epidemic perception enabled by LLMs manifests across multiple dimensions.} On the one hand, LLMs can perform fine-grained analysis of information spreading at the content and contextual levels by examining narrative logic, stance orientation, and background knowledge, thereby assisting in the identification of potentially false, misleading, or manipulative content. On the other hand, the generative and inferential capabilities of LLMs can be leveraged to supplement, reconstruct, and infer the structures of spreading networks, enabling the characterization of spreading paths, structural patterns, and their evolutionary processes among different information units.
\textcolor{black}{Moreover, LLMs offer a possible means for perceiving social cognition. They could capture subtle social-cognitive cues related to spreading from large-scale textual and multimodal data, such as changes in emotional expression, stance evolution, and interactional context, thereby introducing characterization of social cognitive states into spreading perception beyond traditional structures and behavioral perceptions.}

Based on different perceptual objectives, existing studies can be broadly categorized into \textcolor{black}{three} directions: (a) perceiving spreading content and contexts, which focuses on leveraging the understanding and reasoning abilities of LLMs to analyze textual semantics, comment-based contexts, and external knowledge in order to reveal content-level falsification and distortion; (b) perceiving spreading networks, which focuses on utilizing the generative and structural inference capabilities of LLMs to reconstruct, recover, and analyze the pathways, patterns, and evolutionary processes of information dissemination; \textcolor{black}{(c)perceiving the social cognition, which concentrates on depicting individuals' cognition of others, relationships, and the self by leveraging LLMs to capture cognition cues such as emotions, stances, and their evolving changes during dissemination, in order to supplement the deficiencies of traditional perception in terms of subjective cognitive factors. 
Fig.~\ref{fig:digital_percept} illustrates the common paradigms for perceiving above perception targets.} In the following subsections, we elaborate on these in detail.

\subsubsection{Perceive spreading content and contexts}
Studies about perceiving spreading semantics and contexts focus on perceiving both the semantic and contextual aspects of digital epidemic content, aiming to leverage the understanding and reasoning capabilities of LLMs to detect latent semantic deviations and authenticity issues in misinformation, rumors, manipulative narratives, and so on~\cite{hu2024bad,nan2024let, ma2025local, zong2025empowering, xie2024multiknowledge,zhang2025llms}. Generally, these studies can be divided into three main directions:
using LLMs to interpret thematic, stance, and discourse semantics; improving contextual awareness through external knowledge and comment-based information; and constructing agent-based frameworks with LLMs to simulate human debate for authenticity reasoning.

\begin{figure*}[htbp!]
  \centering
  \begin{minipage}[t]{0.48\textwidth}
    \centering
    \includegraphics[width=\linewidth]{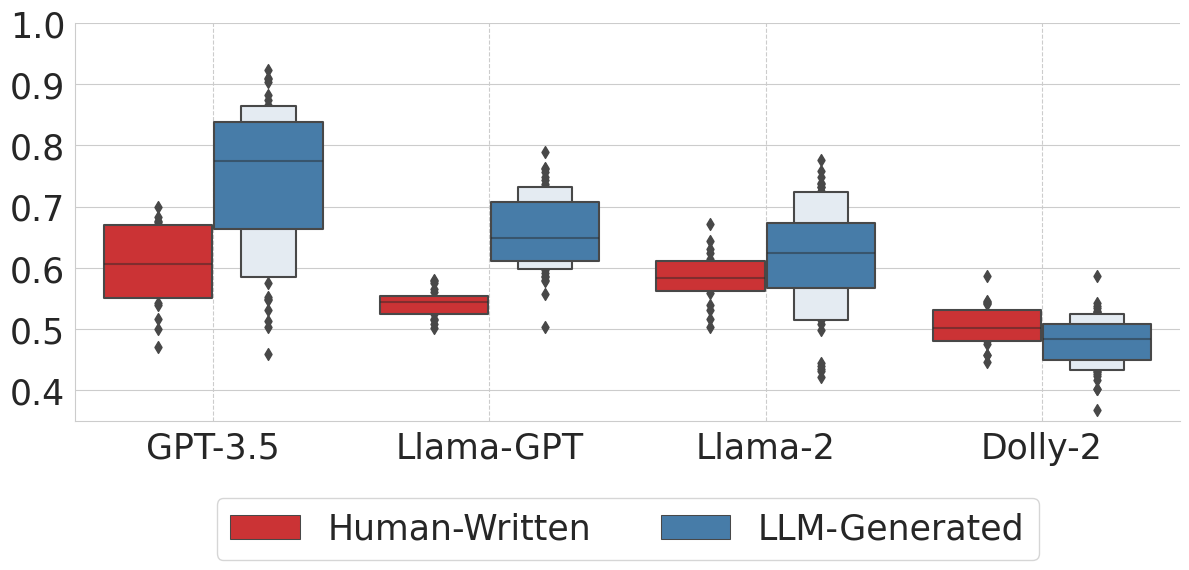}
    \vspace{0.5ex}
    \textit{(a) LLM's zero-shot detection performance on Human vs LLM-generated disinformation.}
  \end{minipage}
  \hfill
  \begin{minipage}[t]{0.48\textwidth}
    \centering
    \includegraphics[width=\linewidth]{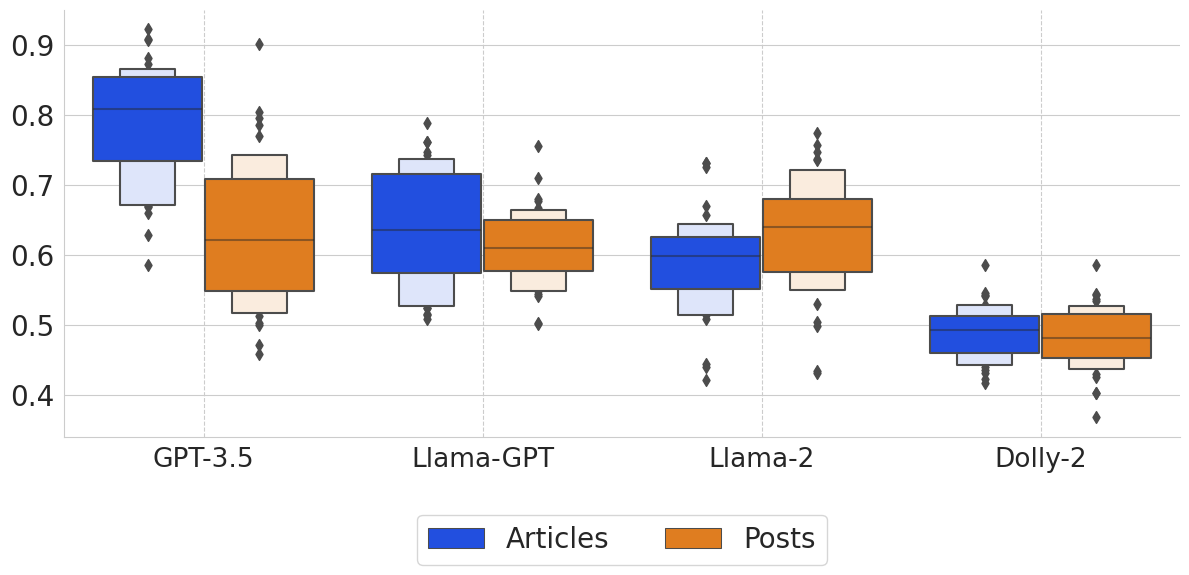}
    \vspace{0.5ex}
    \textit{(b) LLM's zero-shot disinformation detection performance on (long) news articles vs (short) social network posts.}
  \end{minipage}
  \caption{LLMs' zero-shot disinformation detection performance analysis. { In subfigure (a), the y-axis represents accuracy, while in subfigure (b), the y-axis represents the F1-score. (a) shows that LLMs achieve higher accuracy in disinformation detection than human-authored disinformation detection. (b) indicates that most LLMs exhibit stronger disinformation detection performance on articles.}
  \\Source: Reproduced from Ref.~\cite{lucas2023fighting}}
  \label{fig:llm_selfdetect}
\end{figure*}

Firstly, researchers have employed LLMs to interpret the thematic, stance, and discourse semantics of spreading content, thereby capturing implicit semantic cues underlying digital epidemics. Such phenomena as rumors, fake news, and viral posts often rely on textual carriers that contain factual distortions, exaggerated narratives, or toxic content~\cite{ bodaghi2023literature,jahan2023systematic,jiang2025rezg}. Lucas et al.~\cite{lucas2023fighting} investigated the zero-shot misinformation detection capability of LLMs, examining their potential to perform self-detection through in-context reasoning without additional training. As shown in Fig.~\ref{fig:llm_selfdetect}, they observed that LLMs find it more challenging to identify human-written misinformation than machine-generated misinformation. In terms of input length, LLMs also achieved better zero-shot detection performance on news articles (long input) than on social network posts (short input). These findings suggest that LLMs' contextual reasoning mechanisms are more robust when processing information-rich and well-structured texts, yet still face limitations in perceiving fragmented or noisy social network contexts.
Ma et al.~\cite{ma2024fake} proposed the fake news detection with LLM-enhanced semantics mining (LESS4FD) method, which guides LLMs with specially designed prompts to extract and embed news topics and real entities, constructs heterogeneous news-entity-topic graphs, and propagates adaptive features through a generalized PageRank model with a consistency learning criterion. This allows the model to jointly capture both local semantics surrounding each news piece and global semantics across all related news, thereby identifying inconsistencies between individual news texts and broader knowledge contexts.  \textcolor{black}{Compared with traditional classifiers such as TextCNN, TextGCN, HAN, BERT, and Sentence-BERT, LESS4FD achieves over 17.5\% improvement in accuracy and over 16.1\% improvement in F1 score on the MM COVID dataset~\cite{ma2024fake}}
On the other hand, some studies employ LLMs as content analyzers to obtain information such as stance and topics, rather than as direct misinformation detectors. For example,
Yang et al.~\cite{yang2025llm} developed the LLM-enhanced Multiple Instance Learning (LLM-MIL) framework, which incorporates LLM-generated stance explanations into a hierarchical stance-tree attention mechanism to improve interpretability and accuracy in rumor detection. Moreover, Krykoniuk et al.~\cite{krykoniuk2024detecting} detected topic  ``derailments'' by comparing LLM-predicted linguistic trajectories of normal utterances with real user responses at the discourse level. 
The above research primarily focused on misinformation based on text or images, with relatively less attention paid to video-based fake news detection. To address this gap, Hu et al.~\cite{hu2025mage} proposed the multimodal adaptive fusion guided by LLM expertise (MAGE-fend) framework. This framework utilizes LLMs to generate image descriptions for videos, providing high-level semantic information for images. Additionally, MAGE-fend integrates information from video titles, subtitles, and cover image descriptions using LLMs to generate explanatory reasoning text for assessing news authenticity and providing key judgments. \textcolor{black}{ Compared with methods that do not use LLMs, such as Hou-SVM, FANVM, TikTec, and SV-FEND, MAGE‑fend achieves over 6.8\% improvements in both accuracy and F1 score on the TikCron dataset~\cite{hu2025mage}}

Building on semantic understanding, subsequent studies have enhanced contextual awareness by integrating background knowledge, stance explanations, and comment-based contexts, and have found that this external related information can effectively support the model's detection of digital epidemics.
Hu et al.~\cite{hu2024bad} highlighted that small language model (SLM) detectors still struggle to perceive multi-source clues and background knowledge due to their limited reasoning capacity. Thus, they introduced a paradigm where an LLM serves as an ``advisor'' to provide multi-perspective guidance and enhance contextual reasoning during fake news detection. In addition, user comments serve as an important contextual signal for evaluating misinformation, yet real-world datasets frequently suffer from comment sparsity due to exposure bias or user silence. To fill this gap, Nan et al.~\cite{nan2024let} proposed employing LLMs as ``user simulators'' and ``comment generators''. By prompting LLMs with diverse user profiles and aggregating generated comments from multiple subgroups, their approach simulates rich and varied comment contexts, improving models' semantic comprehension of fake news. 
Similarly, Chen et al.~\cite{chen2025you} also focus on enhancing fake news detection through the diversity of user comments or public opinions. To fully leverage public opinions, Chen et al.~\cite{chen2025you} developed a dual-perspective reasoner module, which uses LLMs to simultaneously generate supporting and refuting evidence, analyzing multimodal content including images, text, and user responses from both positive and negative perspectives, ensuring the comprehensiveness of the evidence.
For external fact verification, Bai et al.~\cite{bai2024large} introduced a full-context retrieval and verification (FCRV) framework, which combines LLM-based claim and entity extraction with a retrieval-augmented generation (RAG) mechanism. The model retrieves relevant evidence online and synthesizes both internal contextual information and external facts to assess the veracity of spreading content, thereby improving detection reliability. Furthermore, Zeng et al.~\cite{zeng2024combining} proposed a hybrid human-model reasoning framework that integrates ``model first'', ``worker first'', and ``meta vote'' strategies, combining model predictions with crowdsourced evaluations to achieve more robust authenticity assessment under complex social contexts.

Finally, researchers have explored LLM-based agent systems to simulate human debate and reasoning for authenticity assessment. Liu et al.~\cite{liu2025truth} presented the Truth Enhanced through Debate (TED) framework, where LLMs act as debating agents engaging in cross-examination, rebuttal, and conclusion to emulate human cognitive reasoning and refine veracity judgments. Li et al.~\cite{li2024largefake} proposed FactAgent, which enables LLM to mimic the behavior of human experts in verifying news claims. It decomposes news verification into multi-step reasoning workflows, allowing LLMs to sequentially verify subclaims through internal knowledge or external tools and synthesize the final verdict. 
To address the challenges of limited open-domain knowledge acquisition and understanding in multimodal fake news detection, Xie et al.~\cite{xie2024integrating} implemented two LLM-based agents to integrate open-domain knowledge: the query generation agent, which generates knowledge retrieval queries by extracting key information from text and images; and the knowledge filtering agent, which filters the retrieval results to retain key information that helps improve accuracy. These two agents collaborate through feedback to enhance overall reliability, achieving over 83\% detection accuracy on data from the Weibo and Twitter social networks.

\begin{figure*}[htbp]
    \centering
    \includegraphics[width=0.8\linewidth]{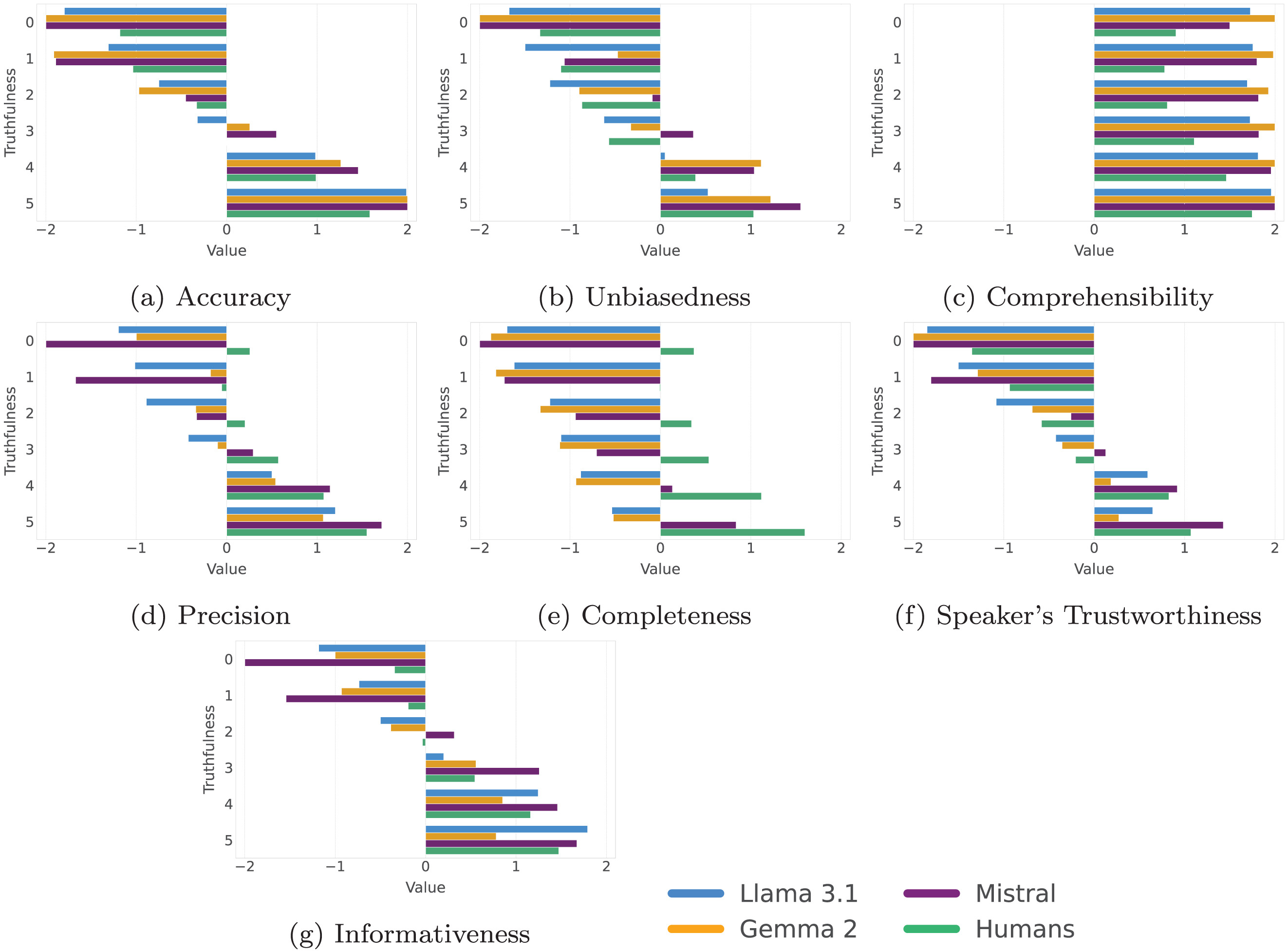}
    \caption{Average scores  \textcolor{black}{in fact-checking tasks} assigned by LLM-based agents and human evaluators for each quality dimension under varying levels of truthfulness. \textcolor{black}{The vertical axis represents the truthfulness score, and the horizontal axis represents the scores of the different metrics shown in the subheadings.}
    \\Source: Reproduced from Ref.\cite{costabile2025assessing}}
    \label{fig:agent_human_eval}
\end{figure*}

Furthermore, Costabile et al.~\cite{costabile2025assessing} explored the feasibility of using LLM-based agents as substitutes for human crowds in fact-checking tasks. They constructed a population of LLM-driven agents modeled on demographic and ideological characteristics (e.g., gender, age, political orientation) to simulate human annotators, and instructed these agents to evaluate the veracity of claims through two sequential stages: ``evidence selection'' and ``evaluation questionnaire filling''. During evaluation, each agent selected external resources as evidence and assessed claims across multiple dimensions, including accuracy, fairness, credibility, etc. Figure~\ref{fig:agent_human_eval} illustrates the average scores of LLM agents and human evaluators for each quality dimension, revealing distinct evaluation tendencies between the two groups. The results show that LLM agents tend to produce more polarized judgments, whereas human evaluators prefer moderate or neutral ratings and exhibit greater variability, reflecting broader subjective interpretations. Finally, the study concluded that generative agents demonstrate substantially higher internal consistency than human annotators and can effectively leverage external evidence to assess claim veracity, often achieving performance comparable to or even exceeding that of human groups. These findings suggest that LLM-based agents help ease the consistency and scalability challenges in misinformation mitigation.

\subsubsection{Perceive spreading networks}
In digital epidemics, information spreading is determined not only by message content but also by the spreading paths, structural patterns, and dynamic evolution within social networks. Spreading networks capture how information is received, responded to, and retransmitted, revealing the formation of group consensus, the amplification of emotions, and mechanisms that escalate misinformation~\cite{yuan2020early, qureshi2022deception, sun2023fighting, chen2020proactive}. However, real-world user–interaction data are often incomplete, fragmented, or difficult to obtain, rendering traditional models that assume static graphs or full network visibility less applicable. Consequently, recent studies leverage LLMs' generative, inferential, and structural-abstraction capabilities to reconstruct, recover, and infer spreading structures, thereby supporting the detection and monitoring of digital epidemics.

First, to address missing or partially observable networks, Wan et al.~\cite{wan2024dell} employ LLMs to generate multi-perspective comments and user reactions associated with news, constructing a synthetic user-news interaction network. They further enrich node representations with LLM-derived implicit background signals (e.g., stance, sentiment, external knowledge), and then perform information propagation over this network using GNN.

Second, to mitigate narrative fragmentation and divergent cognitive perspectives in digital epidemics, Li et al.~\cite{li2025semantic} propose a narrative reordering and perspective fusion (NRPF) method. By conducting multi-round perspective refinement and narrative reordering with LLMs, NRPF aggregates semantic cues dispersed across comments, citations, and spreading paths, thereby mapping cognitive evolution during the spreading effectively and  capturing rumor-spreading patterns and structures.

Further, recognizing that LLMs reason more effectively on moderately sized structured contexts, Zeng et al.~\cite{zeng2025exploring} prune rumor spreading networks and convert the retained core nodes into LLM-friendly textual representations. They then design a chain-of-clues prompt to guide LLMs to gradually analyze local comments, global comments, and propagation patterns, culminating in rumor labels and interpretable reasoning outputs. This alleviates the problem of LLMs' insufficient processing capacity for large-scale structured networks.

Under the assumption that linked nodes are more likely to share the same label, Hu et al.~\cite{hu2025synergizing} proposed a LLM-based global label propagation network to enhance the accuracy of multimodal misinformation detection. They constructed a cross-modal news network in which each news item is modeled as a node, and edges are created when intra- or inter-modal similarities exceed a threshold.  To alleviate the issue of limited labeled samples, LLMs are employed to generate high-confidence pseudo-labels for unlabeled nodes, thereby providing more supervisory signals and stabilizing the propagation process.

{ 
\subsubsection{Perceive the social cognition}
In real-world social spreading processes, individuals' behaviors and state transitions are shaped not only by external contact relations or spreading structures, but also by their social cognition regarding the self, others, and interpersonal relationships~\cite{simon1990invariants}. An individual's social cognition is influenced by various unstructured factors such as the current content of their attention, past experiences, and social background~\cite{smith1992exemplar}. As a result, even when embedded in the same social network structure, different individuals may develop distinct interpretations of spreading information and respond differently to the social environment~\cite{galesic2021human, cialdini2004social}.

Traditional spreading perception methods, as well as most social dynamics models of belief evolution and collective behavior, tend to emphasize observable external behaviors and structural features—such as contact relations, spreading paths, or node state changes—and simplify individuals as rational or rule-driven units that respond to external stimuli within interaction contexts. Such assumptions often overlook individuals' internal psychological processes and social cognitive mechanisms~\cite{galesic2021human}. Recent studies in computational social science on social sensing have shown that subjective representations of social networks can enhance the descriptive validity and predictive power of social dynamics models~\cite{galesic2021human}. Unlike ``objective'' measurements based on structure or behavior, these subjective representations capture how individuals actually experience their social relationships and are jointly influenced by affective attitudes, social identity, and degrees of interaction dependence. Incorporating such social cognitive factors into spreading perception and modeling frameworks can therefore improve the ability of social dynamics models to describe and predict real-world spreading processes.

Building on this perspective, LLMs provide a scalable technological pathway for perceiving social cognition in spreading processes. Acting as perception engines with capabilities in language understanding, contextual reasoning, and cross-modal processing, LLMs can extract cognition-related cues from large-scale textual and multimodal data associated with spreading, such as changes in emotional expression, stance orientation, and their dynamic evolution over the course of spreading~\cite{yang2025llm,lan2024stance,liu2024emollms,zhang2025dialoguellm}. By integrating heterogeneous observations across platforms and time scales, LLMs enable the modeling of social cognitive states during spreading processes beyond traditional structure- and behavior-based perception, thereby offering a richer and more realistic perceptual foundation for detection and surveillance tasks, as well as valuable complementary information for the characterization and calibration of spreading dynamics models.}

\subsection{LLM-enhanced biological epidemic perception}
Considering the significant threat posed by infectious diseases to public health, the primary objectives of infectious disease detection~\cite{browne2024evaluating} and surveillance~\cite{herrera2016disease,si2020epidemiological,lemon2007global,yuan2019systematic} are to characterize the current burden and epidemiological patterns of the disease, monitor temporal trends, and identify outbreaks and emerging pathogens~\cite{murray2016infectious}. These efforts provide critical support for assessing the urgency of public health risks and informing the development of effective response strategies and intervention policies, thus increasing the likelihood of containing outbreaks at an early stage~\cite{world2023future}. { As defined on the Web pages of the WHO, public health surveillance is the continuous, systematic collection, analysis, and interpretation of health-related data}\footnotemark\footnotetext{https://www.who.int/emergencies/surveillance}. Such data has already been directly used to inform policy decisions and to communicate timely information to the public~\cite{clark2024changes}. To achieve these goals, practitioners have developed a variety of surveillance systems, including case-based, event-based, laboratory-based, syndromic-based, and web search-based approaches~\cite{yang2017early}. These systems rely on various data sources, such as mortality and morbidity statistics, clinical records, laboratory reports, field investigations, demographic information, and environmental data, to build analytical models that support real-time monitoring and situational assessment~\cite{declich1994public}. To derive accurate surveillance results and actionable insights from these data, researchers from multiple disciplines have made significant contributions~\cite{birello2024estimates,lo2020genomics,ecker2005rapid,moura2016whole}. For instance, some physicists use available data to estimate system parameters and compute key indicators, thereby constructing dynamical models that capture the mechanisms of epidemic spread~\cite{hu2024global}. Such dynamical models provide a theoretical foundation for quantifying transmission processes, predicting epidemic trajectories, and identifying high-risk populations or regions, thus enhancing the precision and timeliness of infectious disease detection and surveillance. 

With the rapid development of the Internet and smart devices, massive multi-source surveillance data, { such as epidemiological reports, mobility aggregates, anonymized digital traces or social network and web search data related to epidemics}, have become available~\cite{yang2013mining}. These heterogeneous data sources enable more accurate epidemic detection and monitoring; however, traditional approaches, such as dynamical modeling, struggle to integrate and utilize such complex data effectively, leading to limitations in timeliness and sensitivity~\cite{hu2024global}. In contrast, AI-driven approaches, particularly those enhanced by LLMs, have strong potential to fuse, interpret, and reason over these multi-source epidemic data. In the remaining part of this subsection, we review the progress of biological epidemic detection and surveillance across different phases of the epidemic. We first discuss early stage detection~\cite{hashimoto2000detection}, followed by ongoing epidemic monitoring.

\subsubsection{Early stage detection}
Early-stage detection for epidemic spreading aims to identify the onset of an outbreak as early as possible, ideally before widespread transmission occurs, thereby serving as the basis for epidemic early warning~\cite{meckawy2022effectiveness,southall2021early,macintyre2023artificial,villanueva2025artificial}. Such detection is critical to an effective public health response and containment, facilitating timely alerts and management actions, reducing transmission and facilitating early intervention~\cite{alavi2022real}. Next, we categorize existing studies by methodological paradigm. We first provide a brief overview of traditional model-based approaches, as well as machine learning or deep learning-based methods for early-stage detection. We then present a detailed review of early-stage biological disease detection methods enhanced with LLM. 

\textbf{(a) Traditional model-based detection methods}. 

These types of methods can be broadly categorized into temporal warning models, spatial warning models, spatiotemporal warning models, and dynamic models. Temporal, spatial, and spatiotemporal models aim to detect potential outbreak hotspots by analyzing temporal trends, spatial distributions, or their joint evolution over time and space. In contrast, dynamic models focus on capturing the transmission mechanisms within a population to infer the underlying epidemic dynamics. These approaches typically rely on data related to the current burden and epidemiology of the disease, the temporal variation of the cases, and the identification of outbreaks and emerging pathogens~\cite{hu2024global}. Further details on such traditional model-based detection methods can be found in Refs.~\cite{unkel2012statistical,siettos2013mathematical,kulldorff2005space,hohl2020daily,whiteman2019detecting}.

\textbf{(b) Machine learning or deep learning-based methods.} 

With the widespread adoption of the Internet and smart devices, open-source data, including news articles, social network posts, online forum discussions, and public health bulletins or reports, have become an indispensable supplement to traditional early warning systems. These data may contain weak but timely outbreak signals, enabling near-real-time situational awareness. Using machine learning, deep learning and natural language processing, a series of AI-driven early warning systems have been developed to detect early epidemic signals from such heterogeneous sources~\cite{bhatia2021using, li2024reviewing}. Representative examples include Epidemic Intelligence from Open Sources (EIOS)~\cite{williams2025evaluation, EIOS}, EPIWATCH~\cite{quigley2025epiwatch}, and the Boston University Center on Emerging Infectious Diseases (BEACON) system~\cite{bhadelia2024eager, hariri_institute_beacon_2025,consoli2024epidemic}.

Among them, EIOS, developed by the World Health Organization (WHO), applies machine learning and multilingual text mining techniques to identify potential outbreak signals from a wide range of nontraditional sources, including news articles, social network posts, official announcements, and online health forums. By automatically classifying and clustering open-source reports, EIOS supports early detection of infectious disease events and forwards likely signals to experts for verification. Deployed in more than 160 countries, it enhances global epidemic intelligence through its real-time monitoring and multilingual adaptability. 

EPIWATCH, developed at the University of New South Wales, leverages machine learning and NLP to continuously scan global online data such as news, blogs, and social networks for unusual disease patterns. It assesses the severity and credibility of detected signals and issues alerts that frequently precede official notifications. Compared with EIOS, EPIWATCH is research-oriented and is particularly suited for academic and public health applications, offering a scalable and cost-effective framework for early epidemic intelligence. 

BEACON uses AI-driven event-based surveillance to detect and contextualize biological threats in human, animal, and environmental health domains. The system aggregates open source information, such as news feeds, advisories, and online health content, and routes filtered signals to subject matter experts for verification. BEACON emphasizes transparency, real-time accessibility, and expert validation to improve the precision and timeliness of epidemic alerts. Other representative AI-based early warning systems include ProMED-mail~\cite{carrion2017promed}, EPIWATCH~\cite{macintyre2022preventing}, and HealthMap~\cite{freifeld2008healthmap}. In addition, beyond these general early warning systems, some studies directly apply machine learning or deep learning techniques to perform early detection for specific diseases, which can be found in Refs.~\cite{ong2018mapping,chen2014non,lampos2017enhancing,ding2018mapping}.

\textbf{(c) LLM-enhanced methods}

While these AI-driven systems have significantly enhanced automation and responsiveness in early detection, LLM-enhanced approaches can improve contextual reasoning, cross-lingual comprehension, and integration of heterogeneous data. In a recent infoveillance study on ocular infection, Deiner et al.~\cite{deiner2025use} evaluated seven LLMs using thousands of synthetic and real-world conjunctivitis-related posts collected from X, online forums, and YouTube. They designed prompts that required models to infer outbreak-relevant attributes rather than making only binary ``outbreak vs. non-outbreak'' decisions, and then compared model outputs with expert assessments using correlation analysis, sensitivity/specificity, and inter-rater reliability. The results show that GPT-4 and Mixtral-$8×22$B provided the most reliable performance: GPT-4's estimated outbreak probability correlated at $0.73$ with expert labels, and Mixtral's estimated outbreak size correlated at $0.82$ with ground-truth case counts. When infectious etiology was validated on real posts, GPT-4 achieved high specificity ($0.83–1.00$) and moderate sensitivity ($0.32–0.71$). These findings indicate that LLMs are capable of recovering multiple epidemiological dimensions, such as outbreak probability, outbreak magnitude, and likely etiology, rather than merely generating coarse binary warnings. Consequently, LLMs can serve as effective tools for early-stage disease detection, supporting situational awareness before traditional reporting systems detect the outbreak.

Deiner et al.~\cite{deiner2024use} also explored the use of LLMs to monitor the posts of tweets and assess the likelihood of epidemic outbreaks. They collected 12,194 tweets related to a specific infectious disease and provided them to two LLMs (GPT-3.5 and GPT-4), prompting the models to evaluate whether an outbreak was likely to occur. Their results show that LLMs can effectively interpret social network posts and achieve prediction accuracy comparable to human assessments. Although the performance is similar to manual evaluation, this approach requires substantially less human effort, as it only involves supplying relevant social network posts to the model. This demonstrates the potential of LLMs as a scalable and low-cost component in early epidemic warning workflows.

Moreover, Kaur and Butt~\cite{kaur2025ai} proposed an LLM-driven integrated framework for epidemic intelligence that combines multi-source data streams, including news reports, social network content, and web search logs, with modules for resource optimization and cross-lingual processing. The framework emphasizes real-time analysis, cross-source data fusion, and automated generation of situational assessments, allowing it to detect weak epidemic signals and emerging threats at an early stage. Their study systematically articulates the potential of LLMs in enhancing early detection, situational awareness, and response preparedness, while also identifying several key challenges, such as system architecture design, model validation, cross-lingual adaptation, and false alarm control.

\subsubsection{Ongoing epidemic monitoring}
For researchers studying the spreading of biological diseases, it is essential to analyze and process relevant data efficiently to maintain continuous surveillance and generate insights that meaningfully inform epidemic trend assessment and the development of effective intervention strategies. 
Multiple heterogeneous data can also be leveraged for { ongoing epidemic monitoring}, ranging from traditional public health indicators, such as hospital emergency department admissions and over-the-counter medication sales, to modern digital traces, including web search queries and social network activity~\cite{jiao2023application}. As with early-stage detection, we first provide a brief overview of traditional model-based approaches, as well as machine learning or deep learning-based methods for ongoing disease surveillance. We then present a detailed review of { ongoing epidemic monitoring} methods enhanced with LLM.

\textbf{(a) Traditional model-based methods.}

Ongoing epidemic monitoring can be supported through biological epidemic modeling. For example, Han et al.~\cite{han2023sars} simulated a typical COVID-19 epidemic scenario in low- and middle-income countries to examine how testing rates, sampling strategies, and sequencing coverage jointly influence surveillance outcomes. Their results show that low testing rates and spatiotemporal biases can delay the identification of new variants by several weeks to months and may lead to unreliable estimates of variant prevalence. Allard~\cite{allard1998use} introduces the practical application of autoregressive, integrated, moving average modeling of time series in infectious disease surveillance, discussing the full workflow from model specification and transformation to parameter estimation, forecasting, and iterative updating. More similar studies can be found in Refs.~\cite{tomov2023applications, katris2021time,lemey2020accommodating}.

\textbf{(b) Machine learning or deep learning-based methods.}

Subsequently, many researchers began to employ artificial intelligence to support continuous surveillance. These methods integrate noisy and complex data, transforming them into effective information, hence generating meaningful insights from difficult-to-interpret multidimensional data. Such AI-driven approaches facilitate the tracking of incidence, mortality, transmission sources, and other key epidemiological indicators, thereby advancing the development of infectious disease surveillance~\cite{brownstein2023advances}. For example, during the COVID-19 pandemic, researchers conducted surveillance of disease transmission, vaccination uptake, substance use, and misinformation propagation by analyzing mobile phone mobility data, social network content, and search engine query logs, thus allowing the identification of COVID-19 transmission signals\footnotemark\footnotetext{{ Analyses of disease transmission, vaccination uptake, and related processes typically rely on aggregated, anonymized, or otherwise privacy-preserving data, and are conducted within established public health and ethical frameworks.}}~\cite{clark2024changes}. Li et al.~\cite{li2020substantial} perform ongoing epidemic monitoring of SARS-CoV-2 by combining reported case observations with human mobility data. They construct a networked dynamic metapopulation model and employ Bayesian inference to infer key epidemiological characteristics, including the fraction of undocumented infections and their relative transmissibility. Wu et al.~\cite{wu2025federated} considered the potential health information leakage risks that may arise when exploring relationships between infectious diseases and patients using knowledge graphs~\cite{bang2023biomedical,consoli2025epidemiological} for ongoing epidemic monitoring. They proposed a framework based on federated learning~\cite{stadtmann2024physics,song2024quantum} that constructs locally differentially private~\cite{qi2023differentially,ma2023differential} individual symptoms relationship graphs for epidemic risk surveillance, effectively representing the potential transmission relationships between individuals while preserving privacy.
\begin{figure}[th]
    \centering    \includegraphics[width=0.99\linewidth]{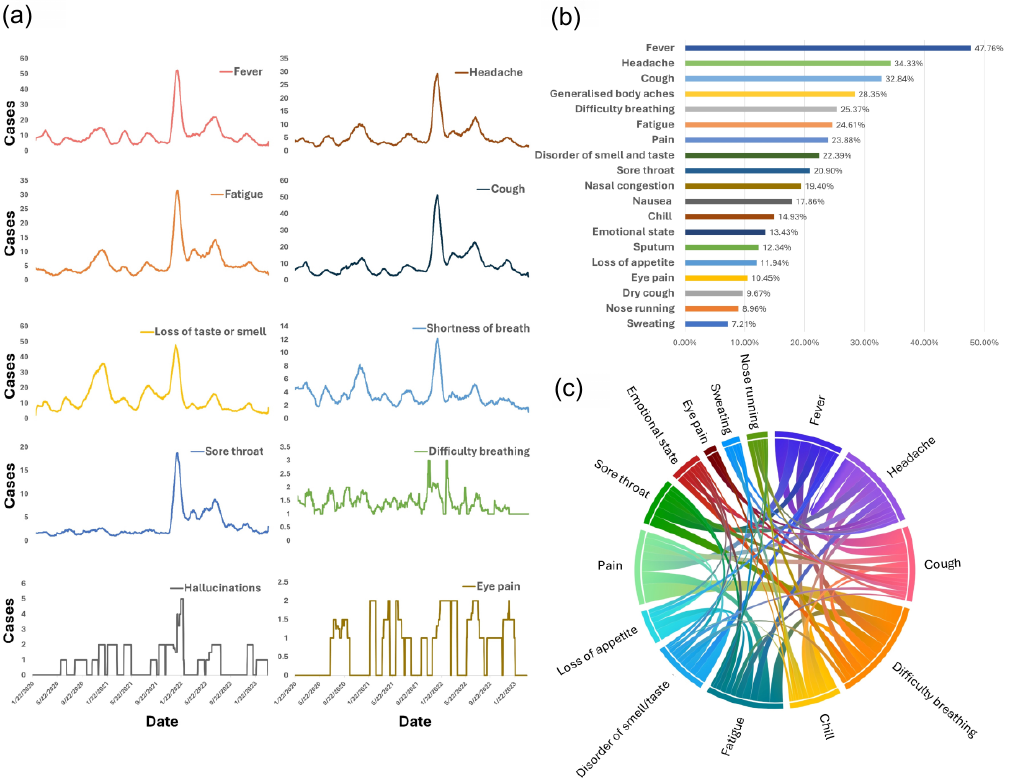}
    \caption{Example of leveraging LLMs for ongoing disease surveillance. LLMs are used to identify whether social media posts correspond to users with positive infections and to extract their reported symptoms, which are then aggregated for statistical analysis. (a) Temporal trends in symptom prevalence over time. (b) Proportion of different symptoms among reported cases. (c) Correlations among symptoms. 
    \\Source: Reproduced from Ref.~\cite{xie2025leveraging}}
    \label{fig:SurveillanceResults}
\end{figure}

\textbf{(c) LLM-enhanced methods.}

LLMs have substantial capability for infectious disease surveillance~\cite{cheng2023potential}, as they can accurately process and analyze large volumes of unstructured text and can help professionals streamline labor-intensive workflows and uncover latent trends~\cite{brownstein2023advances}. Xie et al.~\cite{xie2025leveraging} observed that many individuals with mild COVID-19 symptoms conduct self-testing at home and report their conditions on social media. To leverage these data effectively, they employ LLMs to identify and analyze posts related to COVID-19 infection, enabling continuous monitoring of epidemic development and symptom evolution. Specifically, they first collect potentially relevant posts from platform X, then manually annotate a subset of posts indicating verified infection, and fine-tune an LLM on these labeled examples. The fine-tuned model is subsequently used to detect likely positive cases and extract reported symptoms, which are aggregated for statistical analysis and real-time surveillance. As illustrated in Fig.~\ref{fig:SurveillanceResults}, this LLM-enhanced surveillance framework enables: (i) Tracking temporal trends of symptoms, which can reveal the emergence of new variants (e.g., loss of taste or smell became prevalent around September 2020, aligning with the rise of the Beta variant). (ii) Estimating symptom prevalence which can support resource allocation and preparedness planning.
(iii) Analyzing correlations among symptoms, offering insights into clinical progression patterns. Moreover, their longitudinal findings indicate that: a) Recovery time has gradually shortened from 2020 to 2023; b) Most COVID-19 cases in 2023 were mild, suggesting decreasing viral severity; c) Reinfection within 180 days is relatively uncommon.
d) Fewer than $10\%$ of individuals experienced reinfection twice, around $1\%$ experienced three infections, and an even smaller proportion reported four or more reinfections. Consoli et al.~\cite{consoli2024epidemic} systematically evaluate the capability of nine LLMs in ongoing epidemic monitoring tasks, examining their performance in extracting outbreak-relevant information from unstructured text. They further apply in-context learning to enhance the models' abilities, which leads to marked improvements in both accuracy and processing efficiency. Sufi et al.~\cite{sufi2024innovative} explored the use of LLMs for COVID-19 surveillance. They also leveraged LLMs to extract and analyze epidemic-related signals from social networks.

\section{Epidemic Prediction and Management: LLM-Enhanced Regulation of Spreading Dynamics}
\label{sec5}
An ultimate objective of studying spreading dynamics is to enable accurate prediction and effective intervention to mitigate or prevent undesired propagation.~\cite{pastor2015epidemic,vespignani2012modelling}. 
This section categorizes and synthesizes existing research on the applications of LLMs in the prediction and management of spreading processes.
Specifically, it reviews recent LLM-enhanced studies on both digital epidemics and biological epidemics, summarizing how LLMs assist in forecasting and managing propagation.
Each subsection first examines how LLMs contribute to epidemic prediction and then discusses how they facilitate epidemic management.

\subsection{LLM-enhanced digital epidemic regulation}
In physics, particularly in statistical physics, complex systems, and nonlinear dynamics, information spreading is not studied from the perspective of media and audiences, as in communication studies, but rather treated as a dynamical process on complex networks. The focus is on how information, rumors, or opinions diffuse across the ``node-edge'' structures of networks. It is widely regarded as a percolation-like process, drawing parallels with phase transitions in theoretical physics~\cite{xie2021detecting}. Common modeling approaches include SIR/SEIR epidemic models, threshold models~\cite{granovetter1978threshold}, rumor models (e.g., Daley-Kendall, Maki-Thompson), the voter model, Ising model and spin dynamics, percolation theory, independent cascades model~\cite{goldenberg2001talk}, and agent-based simulations. Research themes cover the laws of propagation across different network topologies (small-world, scale-free, community structures); identifying influential nodes to maximize spread; strategies to suppress the diffusion of misinformation or viruses; competition and coexistence of multiple pieces of information or opinions within the same network; temporal evolution of network structures; and factors affecting the efficiency and stability of spreading. Physicists treat information diffusion as a dynamical process and, through modeling, analysis, and simulation, aim to understand, predict, and ultimately guide - or, in broader contexts, manage - the spreading.
In this subsection, we first provide a brief review of traditional methods for information spread prediction, followed by a detailed discussion of LLM-based prediction approaches to highlight the evolution from classical models to LLM-driven paradigms.

\begin{figure}[th]
    \centering    \includegraphics[width=0.98\linewidth]{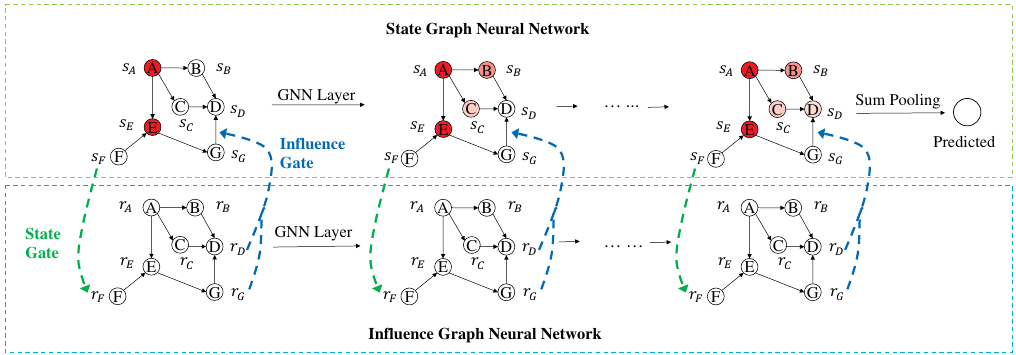}
    \caption{Example of leveraging the states of early adopters together with the underlying network structure for the infection prediction of other nodes. { Given a directed network consisting of seven nodes, each node is associated with an initial infection state, and directed edges encode influence relationships among nodes. The method employs a graph neural network to learn the spreading dynamics from the underlying network structure, initial node states, and directed interactions. Based on the learned propagation patterns, the model predicts the infection states of other nodes at subsequent time steps.}
    \\Source: Reproduced from Ref.~\cite{cao2020popularity}}
    \label{fig:popularity_prop}
\end{figure}

Before the advent of LLMs, traditional approaches to information spread prediction relied primarily on mathematical models such as SIR and SEIR. These models offer a strong generalization ability as they are independent of specific datasets and can be applied to arbitrarily complex networks. However, they rely on strong assumptions and struggle to capture the heterogeneity of real-world social network environments. Their input information is limited, often focusing only on network structures and initial infected nodes, with limited capability to handle textual content and semantic information. For example, Cao et al.~\cite{cao2020popularity} predicted infection of other nodes by leveraging the states of early adopters together with the underlying network structure. Specifically, they modeled and predicted the activation state of a node by considering both the activation status of the target user and that of its neighbors, along with their respective influence. As illustrated in Fig.~\ref{fig:popularity_prop}, given that nodes A and E are already infected and their positions in the network are known, they employed a graph GNN to iteratively predict which nodes are likely to be infected next. 

\begin{figure}[th]
    \centering    \includegraphics[width=0.5\linewidth]{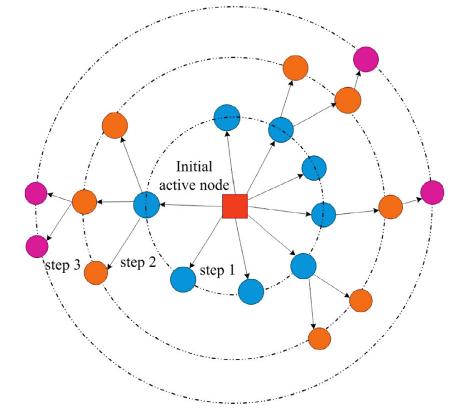}
    \caption{Example of directional label propagation. { The network consists of multiple nodes, where only the centrally located rectangular node is initially activated, while the remaining circular nodes are inactive. The inactive nodes are grouped into multiple layers according to different colors. Nodes within the same layer are treated as mutually independent and update their states synchronously and independently. Each layer corresponds to one infection round over a discrete time interval. In each round, all susceptible nodes, defined as inactive neighbors of activated nodes, update their states based on a microscopic behavior estimation model. By performing such layer-by-layer estimation, the spreading process is simulated, enabling the analysis of potential influencing factors and key driving mechanisms of diffusion, as well as the prediction of nodes that may become infected at subsequent time steps.}
    \\Source: Reproduced from Ref.~\cite{wu2016evolution}}
    \label{fig:Label_prop}
\end{figure}

Wu et al.~\cite{wu2016evolution} investigated the problem of predicting the macro-level evolution of information diffusion. They attributed the diffusion of information to social influence, which drives the transfer of information from one node to another. Based on this insight, they proposed a directional label propagation algorithm that aggregates micro-level propagation modules of individual nodes to predict macro-level diffusion patterns, as shown in Fig.~\ref{fig:Label_prop}. In their framework, whether a node becomes infected is treated as a label, with the label selection mechanism corresponding to the micro-level decision-making process of propagation. By modeling micro-level behaviors, they identified potential influencing factors and the key driving mechanisms of diffusion. Interestingly, they found that increasing the number of features does not necessarily improve the accuracy of diffusion behavior estimation. Moreover, when a sufficient number of features are considered, the accuracy of the estimation is not strongly correlated with the choice of prediction model. In addition, Taylor et al.~\cite{taylor2023does} and Jin et al.~\cite{jin2023predicting}, noting that the dissemination of information within communities differs from that at the individual level, extended the perspective from the individual to the community level by addressing the prediction between groups of nodes. Although their approaches incorporated textual information, the capability to process such content remained limited.
 
Compared with traditional prediction methods, the emergence of GPT, LLaMA, and other LLMs has transformed approaches to forecasting the spread of information. Driven by data, LLMs can process massive amounts of social network content and extract topics and sentiments from text, images, and videos. They can not only predict whether a digital epidemic will break out, but also anticipate which topics will spread, among which groups, and with what emotional stance. By integrating textual semantics with network structures, LLMs can handle complex contexts, such as sarcasm and metaphor, thereby enabling more accurate predictions. Most methods for predicting and intervening in information spread can be broadly classified into two categories: direct prediction using LLMs (without modeling processes) and simulation-based forecasting grounded in the modeling process introduced in Section \ref{sec3}.

\subsubsection{LLMs for prediction}
(a) \textbf{Utilizing LLMs to perform prediction}

\begin{figure}[th]
    \centering    \includegraphics[width=0.98\linewidth]{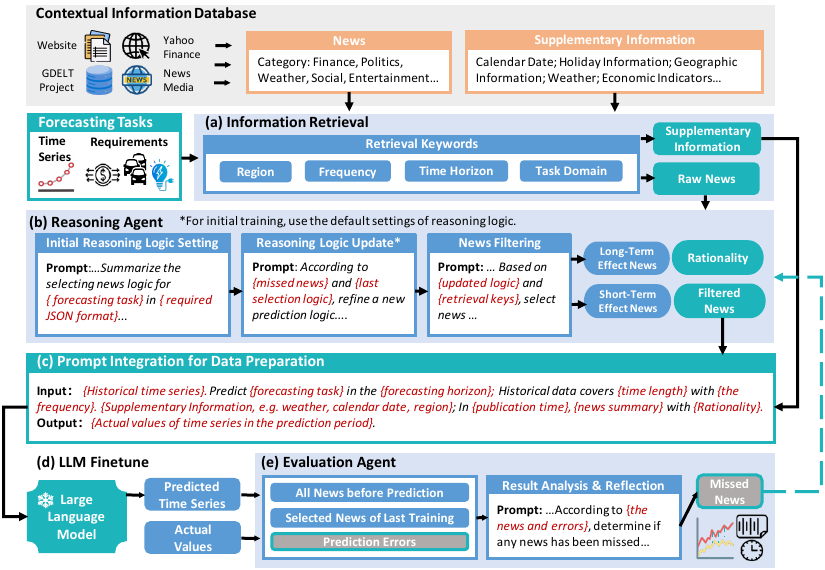}
    \caption{Example of integrating textual information in time series forecasting by LLMs. { A base LLM is adapted to a specific time-series forecasting task through fine-tuning. The overall workflow consists of the following steps. (a) Based on the geographical location, temporal scope, and contextual requirements of the forecasting task, relevant raw news articles and supplementary information are retrieved from the internet or from an integrated proprietary database. (b) An LLM-based agent analyzes and selects relevant news, filtering out news that is unrelated to the forecasting task. (c) The selected news is combined with predefined prompt templates to construct task-specific prompts. (d) The LLM is then fine-tuned using these prompts and employed to perform the forecasting task. (e) Discrepancies between the model’s predictions and observed ground truth trigger a retrospective review of historical news and data, allowing the reasoning process to be refined.}    
    \\Source: Reproduced from Ref.~\cite{wang2024news}}
    \label{fig:LLM_TimeseriesPred}
\end{figure}

As previously discussed, compared with traditional methods, LLM-based approaches possess a stronger capability to process textual content and semantic information. An LLM can be viewed as a black box: once textual information is provided as prompts, it exhibits semantic understanding that is reflected in its outputs. In the context of information spread prediction, a key question is how to leverage LLMs to capture such semantic information. To illustrate this, we take the study of Wang et al.~\cite{wang2024news} as an example, which explores using LLMs to extract content from news articles for time series forecasting. As shown in Fig.~\ref{fig:LLM_TimeseriesPred}, the researchers embedded news and supplementary information (such as weather, calendar dates, and geographical data) into time series data via textual prompts and then fine-tuned a pre-trained LLM to predict numerical sequences within the series. The fine-tuning process begins with retrieving relevant original news and supplementary information. LLM-based reasoning agents then iteratively filter out irrelevant news from the prediction task and extract valuable insights from relevant events. Next, the filtered news is integrated with prompt templates to form prompts, which serve as the inputs to the fine-tuned LLM. The model then predicts the numerical sequences in the time series. Finally, the framework evaluates the accuracy of prediction results and continuously refines both the logic of news selection and the robustness of agent outputs through iteration. Once the LLM has been fine-tuned, subsequent prediction only requires feeding the corresponding news and time series task into the model, which can then automatically generate prediction results. Let the set of historical data be denoted as $\mathcal{H}$, the set of historical prompts as $\mathcal{Q}$, and the set of historical outputs as $\mathcal{O}$. Let the training dataset be defined as 
$\mathcal{D} = \{(h_i, q_i, o_i)\}_{i=1}^{|\mathcal{H}|}$, 
where $h_i \in \mathcal{H}$ denotes historical data, $q_i \in \mathcal{Q}$ denotes historical prompts, and $o_i \in \mathcal{O}$ denotes the corresponding outputs. Given a pretrained LLM $\mathcal{M}_\theta$ with parameters $\theta$, the fine-tuning process can be formulated as minimizing the training objective  
\begin{equation}
\begin{aligned}
\hat{\theta} &= \arg\min_{\theta'} \mathcal{L}(\mathcal{M}_{\theta'}; \mathcal{D}), \\
\hat{\mathcal{M}} &= \mathcal{M}_{\hat{\theta}},
\end{aligned}
\end{equation}
where $\mathcal{L}$ is a suitable loss function that measures the discrepancy between predicted and target outputs. After fine-tuning, for a new prediction task, a prompt $q$ is provided as input to the fine-tuned model $\hat{\mathcal{M}}$, which then generates the prediction by
$o^{q} = \hat{\mathcal{M}}(q)$.

Next, we provide a detailed overview of LLM-based approaches to information spread prediction. When applied to this task, LLMs typically reformulate it as future node activation prediction, i.e., given the already infected nodes and the underlying network structure, predict which nodes will be infected next. Zhong et al.~\cite{zhong2024information} proposed a framework called cascade-retrieved in-context learning (CARE), which uses LLMs to infer the next activation of users in social networks. Here, a cascade refers to the path along which information is transmitted from one user to another. CARE consists of three key components: prompt pool construction, prompt retrieval, and in-context learning for prediction. First, CARE extracts and constructs a prompt pool from historical cascades, which contains diverse patterns of information spread. These prompts serve as past examples of spreading behavior. Next, using ranking-based search engine techniques, the system retrieves prompts from the pool that exhibit spread patterns similar to the target cascade, thereby identifying historical cases with diffusion mechanisms analogous to the ongoing spread. Finally, the retrieved prompts are integrated as contextual input to the LLM, which then performs spread prediction for the target information, such as inferring the infection behavior of individual users. Compared with time series forecasting, the training dataset $\mathcal{D}$ in CARE additionally includes the network structure $G$, i.e., $\mathcal{D} = \{\mathcal{H}, \mathcal{Q}, \mathcal{O}, G\}$.

Shang et al.~\cite{shang2025leveraging} addressed the challenge of enabling LLMs to effectively capture information from network structures and user profile texts, and proposed a fused attention-aware prediction method. They extract cascade paths from the network structure using a constrained BFS algorithm, which provides the structural basis for modeling cascades. User posts are integrated with these paths to construct training prompts for fine-tuning the Diffusion-LLM, enabling it to capture both structural and semantic information and to identify user pairs with diffusion influence. User profiles are encoded with a transformer to generate a similarity matrix. Finally, the diffusion influence matrix from the Diffusion-LLM and the similarity matrix from user profiles are fused, and the resulting representation is fed into multi-head attention layers to produce cascade representations for prediction.

Chen et al.~\cite{chen2024predicting} addressed the problem that semantic relationships between field names and field values are often ignored, leading to loss of contextual information and insufficient feature representation. They proposed a method that combines metadata transformation with LoRA-based fine-tuning of LLMs. The metadata transformation converts structured social network metadata (e.g., UID, Pid, Geoaccuracy, Title) into semantically enriched, contextualized natural language descriptions. For instance, a record such as ``Uid: 25893@N22, Pid: 565381, Title: Beautiful tree.'' is transformed into ``A social network user (5893@N22) has uploaded a post (565381) with a geo-accuracy of 16, titled `Beautiful tree''' This transformation explicitly conveys implicit meanings and links different metadata elements. Subsequently, LoRA technology is used to fine-tune a pretrained LLM for the prediction task.

Zheng et al.~\cite{zheng2025autocas} proposed an LLM-enhanced model called AutoCas, which formulates the prediction of information diffusion as an autoregressive modeling task and exploits the architectural strengths of LLMs. AutoCas comprises three modules: cascade tokenization, autoregressive cascade modeling, and cascade prompt learning. In the first module, three types of information are considered: the local topological structure of cascade data, its global contextual position, and temporal dynamics. The cascade diffusion process is divided into multiple time patches, where user snapshots within each patch are pooled into cascade tokens, thereby capturing temporal dependencies. Local and global embeddings are fused to construct user embeddings, which are further aggregated into cascade tokens forming sequential inputs. The second module maps the tokenized cascade data into the language embedding space, aligning it with the LLM architecture, and redefines cascade diffusion as an autoregressive process that predicts the next token in the sequence based on preceding tokens. The third module incorporates prompt learning, integrating textual prompts (e.g., timestamp annotations) to provide additional semantic cues for cascade tokens. Finally, a two-layer multilayer perceptron serves as the task head, predicting the popularity of information items at the target time point from the generated cascade token sequence. 

(b) \textbf{Using LLMs as components to enhance predictive capabilities}

\begin{figure}[th]
    \centering    \includegraphics[width=0.98\linewidth]{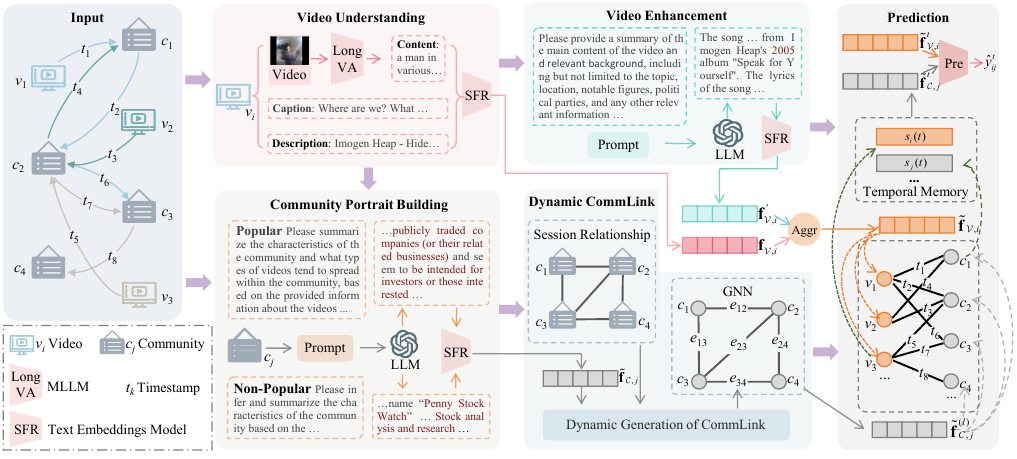}
    \caption{Illustration of using LLMs as components to enhance predictive capabilities. The framework consists of five modules. Different LLMs are employed as distinct tools: they encode video content and community information into a textual video representation and a community feature representation, respectively. These representation vectors are then used as inputs to train a dedicated prediction model.
    \\Source: Reproduced from Ref.~\cite{gao2025social}}
    \label{fig:LLM_CommPred}
\end{figure}

Unlike approaches that directly rely on LLMs to generate prediction outputs, some methods use LLMs to assist in training a dedicated prediction model. Gao et al.~\cite{gao2025social} followed this strategy to forecast the spread of videos within communities. As shown in Fig.~\ref{fig:LLM_CommPred}, their framework consists of five modules: video understanding, video enhancement, community portrait building, dynamic common link, and prediction. In the video understanding module, a multimodal LLM processes raw video files to produce textual descriptions, which are then combined with the video's title, subtitles, and user-provided descriptions to form a unified textual video representation. This representation is encoded as a low-dimensional vector enriched with semantic information. In the video enhancement module, the LLM injects relevant historical, cultural, and social knowledge into the video representation, providing richer contextual grounding for subsequent prediction. In the community portrait building module, the LLM analyzes each community to construct a feature profile, represented as a community feature vector. In the dynamic common link module, special handling is applied to less active communities to improve robustness. Finally, in the prediction module, the textual video representation and the community feature representation are integrated as inputs to train a lightweight prediction model for the prediction of community-level video spread. In summary, LLMs are employed in the video understanding, video enhancement, and community portrait building modules to transform videos and community characteristics into semantically enriched representations, which then serve as inputs to the final predictive model. Importantly, LLMs are not used as the predictors themselves, but rather as feature extractors and knowledge integrators to support downstream prediction.

Similarly, Sun et al.~\cite{sun2025learning} proposed adapting a pre-trained LLM for repost prediction to improve the accuracy of modeling repost behaviors on social networks. Their framework integrates the power of LLMs through two key steps: multimodal feature extraction and semantic temporal adaptation. In the first step, LLMs extract multi-dimensional features from text, visual content, and graph structures. To transfer general-domain knowledge to repost prediction tasks, lightweight adapters are inserted after the self-attention layers of each LLM block, guiding the model to capture semantic patterns indicative of repost tendencies. In the second step, temporal adapters are placed after each feed-forward layer to capture dynamic changes in user interests and preferences. User histories are segmented across temporal adapters, and a multi-head mechanism separately adapts each segment before merging the outputs. By combining LLM-based multimodal representations with semantic and temporal adaptation, this approach significantly enhances prediction accuracy and overcomes the limitations of existing methods in handling sparse data and high-dimensional temporal features.

(c) \textbf{Employing LLMs to generate agents to simulate diffusion}

\begin{figure}[th]
    \centering    \includegraphics[width=0.7\linewidth]{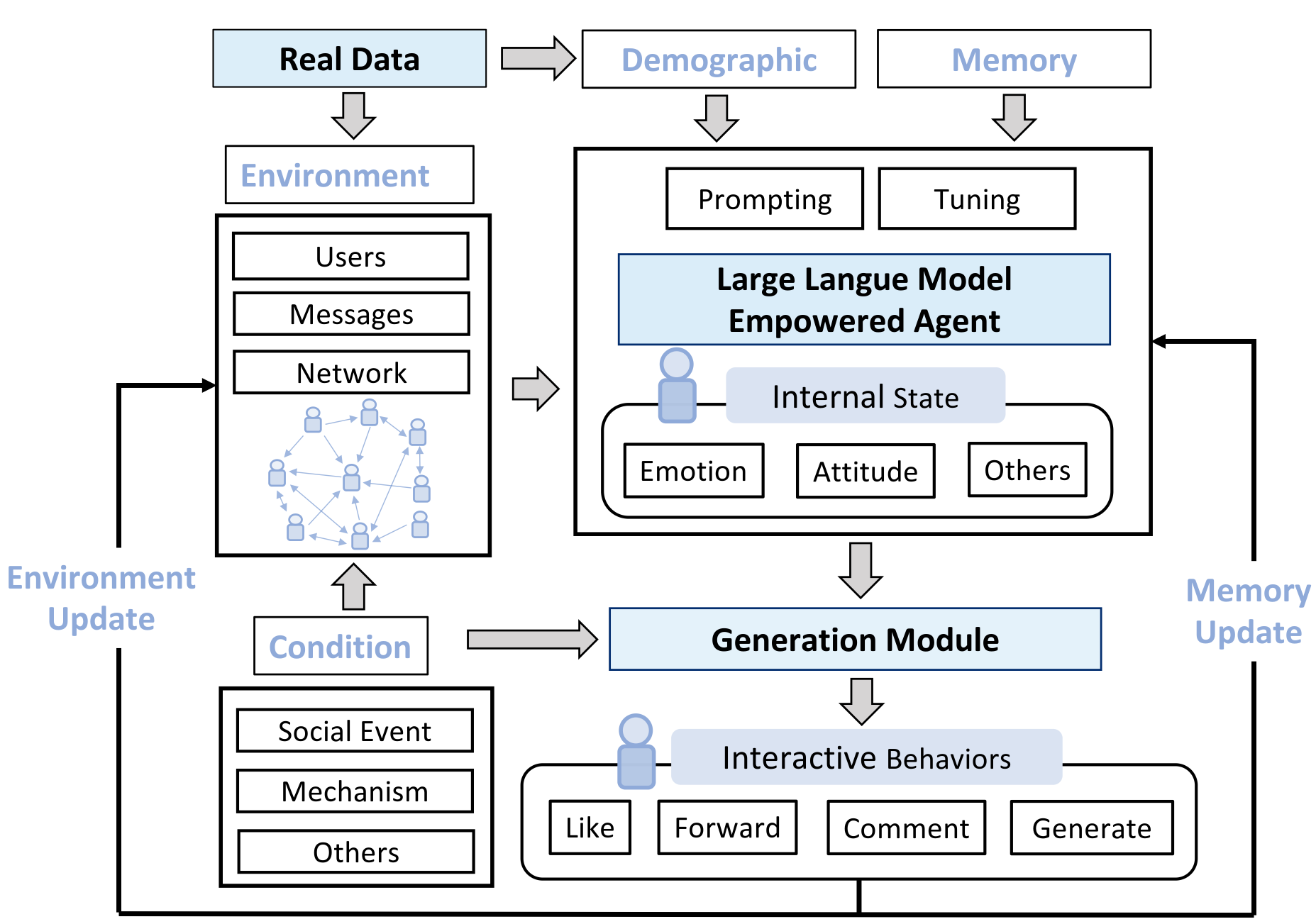}
    \caption{Illustration of employing LLMs to generate agents for simulating propagation dynamics and enabling prediction. Real-world data are fed into a simulated environment where LLM-empowered agents replicate user behaviors. These behaviors are iteratively updated and aggregated, supporting the prediction of information spread.
    \\Source: Reproduced from Ref.~\cite{gao2023s3}}
    \label{fig:LLMforAgent}
\end{figure}

This type of method employs LLMs to generate agents that replicate users in a simulated environment. By modeling how these agents interact and spread information, the system reproduces propagation dynamics and enables the prediction of future diffusion. Gao et al.~\cite{gao2023s3} proposed a social network simulation system enhanced by LLM agents, designed to model the propagation process observed on real social platforms. As shown in Figure~\ref{fig:LLMforAgent}, real-world data, including users, relationships, and messages, are fed into a simulated environment where LLM-empowered agents generate user-level behaviors such as liking, forwarding, commenting, and posting. These behaviors are iteratively updated and aggregated into population-level dynamics of information, attitude, and emotion diffusion, enabling prediction and evaluation of social processes. The system comprises three main components: (i) Environment construction. User data, follower relationships, and textual posts are collected from real social platforms to form a directed graph. Users are categorized into high-influence, regular, and low-influence groups to reflect differences in reposting and interaction behaviors. An LLM with prompt and tuning is used to infer and complete user attributes (e.g., age, gender, occupation), and each agent maintains a memory pool that filters influential posts by factors such as temporal decay and content relevance. (ii) Individual-level simulation. At each timestep, agents update their emotional states (calm, moderate, intense) and attitudes (positive, negative) via Markov processes informed by their profiles and memory pool. Based on these states, agents decide whether to repost, generate original content (with text generated by LLMs), or remain inactive. (iii) Population-level dynamics. Aggregating individual behaviors produces macro-level diffusion curves of information, emotion, and attitude, which can be compared with real-world data to evaluate the fidelity of the simulation. Through this design, the system integrates environment construction, individual simulation, and population-level aggregation into a unified framework. The system can be used for predicting social movements, public opinion shifts, or the adoption of new cultural practices.

Similarly, Chuang et al.~\cite{chuang2024simulating} employed LLM-based multi-agent systems as a replacement for traditional ABMs to simulate human opinion dynamics. Through this simulation, the model predicts the evolutionary trajectories of group opinions, examining how factors such as the initial opinion distribution, network size, and bias strength shape outcomes of consensus or polarization. For instance, simulation indicators can be used for prediction: Bias, defined as the average value of group opinions, reflects the likely direction of consensus; and Diversity, defined as the standard deviation of opinions, indicates the degree of opinion polarization. 

Li et al.~\cite{li2024large} adopted an LLM-driven multi-agent simulation framework to model the dynamics of fake news diffusion across different social network structures, with the resulting simulations enabling prediction of fake news spread trends. Their study examines the effects of three types of networks (random, scale-free, and highly clustered), as well as different agent personalities based on the Big Five traits~\cite{goldberg2013alternative}. The results show how these factors shape propagation dynamics: fake news spreads fastest in scale-free networks and slowest in highly clustered networks, while certain personality traits, such as extraversion and openness, increase the likelihood of news sharing. 

Ma et al.~\cite{ma2024simulated} proposed a simulated misinformation susceptibility testing (SMIST) method, which replaces human participants with LLM-generated counterparts. SMIST creates simulated participants by constructing life experiences and demographic profiles, and then administers questionnaires on misinformation to these agents. The core objective is to assess individuals' susceptibility to misinformation. By generating participants with diverse demographic attributes, such as age, gender, education level, political orientation, and trust in scientists, government, and journalism, SMIST enables the prediction of how these factors shape individuals' propensity to believe and disseminate misinformation.

Yao et al.~\cite{yao2025social} argued that traditional algorithms struggle to capture the complexity of real social interactions, particularly large opinion fluctuations and self-attenuation phenomena. To address this, they proposed FDE-LLM, a hybrid approach designed to more accurately simulate and predict opinion dynamics on social networks. In this framework, users are divided into two categories: opinion leaders and opinion followers. Opinion leaders are modeled using LLMs in combination with Cellular Automata (CA), while opinion followers are simulated with CA and SIR models. LLMs effectively capture drastic opinion shifts, CA accounts for interpersonal influence and continuous dynamics, and SIR models user ``fatigue'' toward repeated events, whereby opinions attenuate to neutrality. Experimental results show that FDE-LLM significantly outperforms both traditional ABMs and pure LLM approaches across evaluation metrics, demonstrating higher accuracy and robustness in simulating opinion dynamics. These findings highlight the potential of integrating LLMs with traditional dynamic models as a more precise solution for opinion prediction on social platforms. 

Rossetti et al.~\cite{rossetti2024social} presented Y Social, an LLM-driven digital twin platform for social networks that simulates complex agent behaviors, including user interactions, content dissemination, and network dynamics. By enabling fine-grained simulation of content spread and structural evolution, Y Social provides a foundation for predicting user responses and, consequently, the diffusion of information. Zhang et al.~\cite{zhang2025llm} integrated LLMs into ABMs to more accurately simulate influence diffusion in social networks. Their approach consists of two main components: an LLM-enhanced agent model and simulation execution. The former extends the traditional independent cascade (IC) model, where agents are limited to binary states (active or inactive) and spread occurs with fixed probabilities, by incorporating LLM capabilities to capture the heterogeneity of individual agents. The latter manages the overall diffusion process and conducts data analysis. This simulation framework is then applied to predict user behaviors related to news, ideas, and topics. Liu et al.~\cite{liu2025mosaic} proposed MOSAIC, an LLM-based multi-agent social network simulation framework for predicting user behaviors such as liking, sharing, and reporting content. The framework first employs LLMs to generate agents endowed with memory, reflection, and reasoning capabilities. Using predefined question-answer prompts and scenario guidance, each agent is assigned personality traits and behavioral motivations. A directed graph is then constructed to represent real social platforms (e.g., X), where nodes denote users and edges denote follow relationships, enabling the simulation of content diffusion paths. Through interactions, agents form behavior patterns based on dynamic memory, reproducing diverse human behaviors. Finally, MOSAIC incorporates three misinformation management mechanisms to adjust agent behavior under different moderation strategies, allowing analysis of content diffusion dynamics, spread speed, and scope, and the effectiveness of each strategy in suppressing misinformation while preserving user engagement.

\subsubsection{LLMs for management}
In addition to serving the purposes of understanding the principles of information diffusion, detecting and monitoring emerging events, and predicting their propagation paths, duration, and trajectories, research on information spread also plays a critical role in management, often framed in terms of management or intervention. This is particularly relevant to the regulation of harmful content, including terrorist propaganda, extremist material, misinformation, and { harmful} rumors, among others. For example, to prevent the uncontrolled dissemination of terrorist propaganda, extremist content, and hate speech online, the European Commission requires hosting service providers to remove terrorist content immediately upon receiving a removal order, and in any case within one hour~\cite{EU2021TERREG}. Likewise, the WHO has formally introduced the concept of infodemic management, which refers to the systematic application of evidence- and risk-based interventions to prevent, detect, and respond to the harmful effects of misinformation and disinformation during health emergencies~\cite{WHO2024Infodemic}. 
management can be examined from complementary perspectives, notably a regulatory perspective and an intervention strategy perspective. From the perspective of regulatory, management entails detecting the spread of harmful content and reporting it to relevant authorities, who then instruct hosting service providers to remove the content. From the perspective of the intervention strategy, platforms and regulators are concerned with how to design more effective intervention measures. A central research question is under what conditions fact-checking efforts by reputable scholars, experts, or mainstream media outlets are most effective in mitigating or eliminating rumors. In this regard, a growing body of research has explored how LLMs can be leveraged to better understand and optimize intervention strategies. 
\begin{figure}[th]
    \centering    \includegraphics[width=0.98\linewidth]{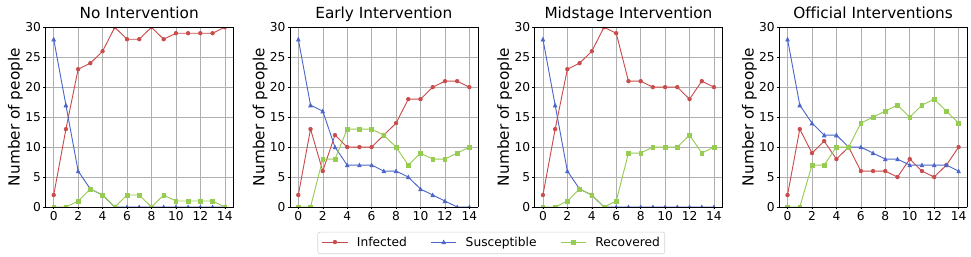}
    \caption{Comparison of different intervention strategies. In the simulated fake news propagation environment, 30 users are modeled, and the diffusion of a political news item is tracked over a 14-day period. From left to right: the first subfigure shows the baseline scenario without intervention; the second subfigure illustrates early intervention starting on day 1; the third subfigure depicts midstage intervention starting on day 7; and the fourth subfigure compares official interventions releasing fact-checking news either daily or every three days.
    \\Source: Reproduced from Ref.~\cite{liu2024skepticism}}
    \label{fig:Intervention_analysis}
\end{figure}

Liu et al.~\cite{liu2024skepticism} proposed a fake news FPS driven by LLM-based agents to model individual attitude dynamics and the spread of fake news. Each agent is equipped with unique personality traits, memory, and reflection mechanisms to approximate human cognition and behavior. Within this framework, they examine official intervention strategies in political topics, focusing on two dimensions: timing and frequency of intervention. As shown in Fig.~\ref{fig:Intervention_analysis}, for timing, experiments compare early intervention (introduced on day 1 of an outbreak) with midstage intervention (introduced on day 7). Early intervention markedly reduces the initial number of believers, though its effectiveness diminishes over time due to forgetting. In contrast, mid-stage intervention achieves more sustained suppression, but because many users are already influenced, a significant portion of the population remains affected. For frequency (official intervention), the study evaluates the release of official fact-checking news daily versus every three days, finding little difference between the two. These findings suggest that fact-checking at moderate intervals provides an effective and cost-effective means of curbing fake news propagation.

Lu et al.\cite{lu2025understanding} explore how LLM-driven adversarial social influence affects people's online information processing abilities. Based on experiments powered by LLM, they propose future directions for intervention strategies to address LLM-driven misinformation. For example, they suggest using bot indicators, descriptions of LLM capabilities (such as fluent language and broad world knowledge), or labels indicating whether content is LLM-generated, to increase user awareness and vigilance toward machine-generated content.

Li et al.~\cite{li2024large} designed and evaluated three specific intervention strategies aimed at curbing the spread of fake news. The first is encouraging commentary, which prompts agents to provide comments when sharing news, thereby stimulating critical thinking and reducing impulsive dissemination. The second is disclosing news accuracy, which simulates fact-checking mechanisms that, once the news has reached a certain level of diffusion, notify agents of its inaccuracy to prevent further spread. The third is blocking highly influential agents, which involves identifying and removing ``influencers'' in the network, those characterized by high connectivity and personality traits such as openness or extraversion, to directly disrupt key nodes of information diffusion.

Almaliki et al.\cite{almaliki2025combining} combined LLMs and persuasion techniques to detect and intervene in online hate speech. They used a template-based prompting strategy integrated with LLMs, tailoring models for hate detection and intervention suggestion recommendations. When users post comments on interactive news sites, before submission, the system intercepts the comment, applies a template-based prompt engineering process by combining the input with two predefined templates, and then generates contextual suggestions to users for intervention.

In fact, interventions can be broadly categorized into three distinct types: nudging, boosting, and refutation strategies~\cite{barman2024dark}. Existing studies employing LLMs to analyze intervention strategies remain rather limited in scope and lack fine-grained examinations of each category, especially compared to more detailed explorations found in traditional research. Previous studies have shown that even at the individual level, interventions span a wide spectrum, from correcting misinformation~\cite{lewandowsky2020debunking} and enhancing media literacy~\cite{guess2020digital} to shaping communication environments that emphasize accuracy~\cite{pennycook2022accuracy}. Therefore, there remains considerable potential for research on how LLMs can be { used more effectively} to support and advance information management strategies.

\subsection{LLM-enhanced biological epidemic regulation}
In this subsection, we first provide a brief overview of traditional research on biological epidemic prediction, also called biological epidemic forecasting~\cite{tsang2024adaptive}, and management~\cite{bak2021stewardship} in non-LLM settings, followed by a detailed discussion of recent advances enabled by LLMs.

Biological epidemic prediction aims to anticipate the potential spread, scale, and dynamics of infectious diseases, with the ultimate goal of supporting effective management. For example, during the COVID-19 outbreak, a research team at Carnegie Mellon University leveraged data from Google, Facebook, and other sources to provide real-time forecasts of the national epidemic in the United States~\cite{Cho2020AI}. Their system predicted the demand for beds and ventilators in intensive care units up to four weeks in advance, enabling proactive preparation and allocation of critical medical resources.
Kissler et al.~\cite{kissler2020projecting} used time series data to estimate the seasonality, immunity, and cross-immunity of two human coronaviruses, and based on these estimates construct a SARS-CoV-2 transmission model to conduct epidemic prediction. Their results suggest that after the initial large outbreak, SARS-CoV-2 is likely to experience recurrent winter resurgences. Then, they propose two epidemic management measures: (i) maintaining social distancing of the public over a prolonged or intermittent period, and (ii) expanding intensive care capacity and improving clinical treatment effectiveness.

Biological epidemic management entails the design and implementation of { mitigating} and { guiding} measures, which have long been a central focus for the WHO, national and regional health authorities, among others. In this context, the educational, academic, and medical communities have established the discipline of infectious disease epidemiology, in which a main focus lies on developing and evaluating strategies to prevent, { manage}, and mitigate the spread of infectious diseases~\cite{anderson1979population}. Kissler et al.~\cite{kraemer2025artificial} leveraged the' powerful capabilities of LLMs to analyze texts, case reports, and related data. The modeling and analysis of diseases can therefore be grounded in more comprehensive datasets and supported by advanced analytical capacity. Such AI-driven approaches hold great promise for reshaping infectious disease response measures and enabling interventions that are more targeted, equitable, and effective. 

management measures, such as policy or intervention strategies, can generally be categorized into two types. The first involves human-designed interventions, where authoritative government agencies or domain experts formulate strategies based on modeling and forecasting results. The second relies on model- or algorithm-driven recommendations, in which intervention strategies are generated by computational systems. The former primarily emphasizes the accuracy of modeling and prediction, which has been reviewed in the preceding sections. For the latter, how to effectively use LLMs to generate targeted and efficient intervention strategies remains an open and promising area of research. In the following, we focus on recent studies that utilize LLM to generate or support intervention recommendations.

\begin{figure}[th]
    \centering    \includegraphics[width=0.6\linewidth]{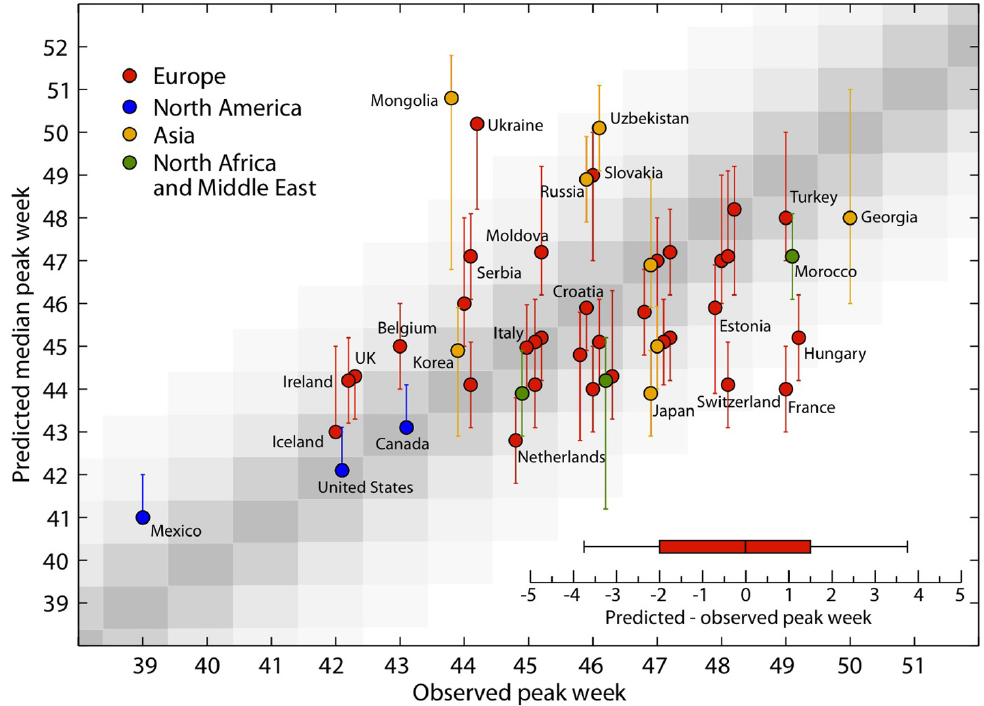}
    \caption{Relationship between the predicted and actual epidemic peak periods. { The effectiveness of the peak-time prediction model is validated using infection data collected during the 2009 H1N1 influenza pandemic. The x-axis represents the actual epidemic peak time, while the y-axis represents the predicted peak time. A strong correlation can be observed between the predicted and actual peak times across countries in Europe, North America, Asia, as well as North Africa and Middle East. Approximately half of the countries exhibit prediction errors within two weeks, and more than $95\%$ of the countries have prediction errors within four weeks.} 
    \\Source: Reproduced from Ref.~\cite{tizzoni2012real}}
    \label{fig:Diff_predandreal}
\end{figure}

Traditionally, epidemic forecasting has relied on two main approaches: mechanistic models, such as SIR or SEIR, which describe spreading dynamics through compartmental structures, and statistical models, which learn patterns from historical data~\cite{du2025advancing,li2023wastewater}. For example, Tizzoni et al.~\cite{tizzoni2012real} investigated the prediction of the maximum period of infections in different countries, which is often defined as the week with the highest number of reported cases~\cite{liu2015events}. They used a global epidemic and mobility model to predict the timing of epidemic peaks and demonstrated that an accurate prediction is feasible. As shown in Fig.~\ref{fig:Diff_predandreal}, a strong correlation was observed between the predicted and actual peak times: for approximately half of the countries, the prediction error was within 2 weeks, and for { more than} $95\%$ of the countries, the error was within 4 weeks. Although useful, these approaches often fail to capture critical real-world factors, including the heterogeneous demographic profiles of affected populations~\cite{nepomuceno2020besides}, the virological characteristics of the virus~\cite{telenti2021after}, the underlying virological dynamics that { guide} viral evolution~\cite{searls2002language}, and the dynamic interactions between public policy and human behavior~\cite{ruggeri2024synthesis}, among others~\cite{du2025advancing}. These limitations reduce the accuracy and reliability of forecasting disease spreading patterns~\cite{ioannidis2022forecasting}.  

\begin{figure}[th]
    \centering    \includegraphics[width=0.5\linewidth]{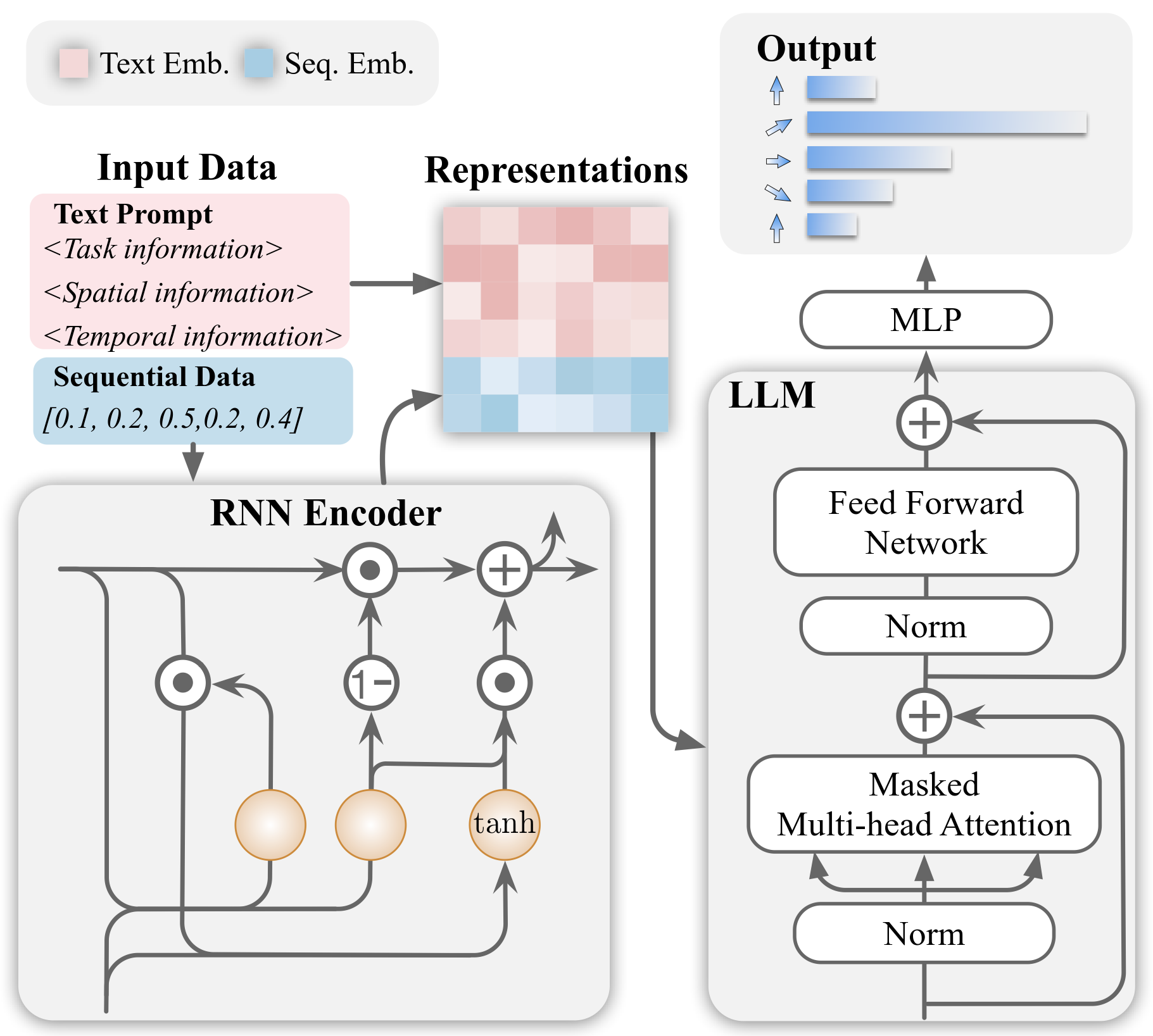}
    \caption{Processing of multimodality data in PandemicLLM. Spatial, temporal, and task-related information are transformed into textual representational vectors via prompts, while sequential data are encoded through an RNN encoder into sequential representational vectors. The two types of vectors are then concatenated and fed into the LLM to perform forecasting.  \\Source: Reproduced from supplementary information of Ref.~\cite{du2025advancing}}
    \label{fig:multimodulLLM}
\end{figure}

To overcome them, recent studies have begun to use LLMs to integrate diverse and complex data sources, ranging from epidemiological reports and biomedical literature to behavioral and policy-related information, thus improving the modeling and prediction of infectious disease spread. Du et al.~\cite{du2025advancing} introduce PandemicLLM, a fine-tuned multimodal LLM designed to predict the short-term spread of diseases by learning from heterogeneous interlinked data. These include: (i) spatial data, such as demographic information, healthcare system scores, and political affiliations; (ii) epidemiological time-series data, including reported cases, hospitalization, and vaccination rates; (iii) public health policy data, consisting of textual policy documents; and (iv) genomic surveillance data, comprising authoritative reports on variants' virological characteristics and their prevalence. Traditional epidemic models, which largely rely on structured numerical inputs, are unable to fully exploit such heterogeneous information. PandemicLLM reformulates epidemic prediction as a text-based reasoning problem, converting all data into textual form so that the LLM can process and reason over modalities inaccessible to conventional approaches. As shown in Fig.~\ref{fig:multimodulLLM}, PandemicLLM integrates multimodal inputs through two complementary pathways: task, spatial, and temporal information are transformed into textual representational vectors via prompts, while sequential data are encoded separately by a recurrent neural network into sequential representational vectors. These representations are then concatenated and fed into the LLM for forecasting. \textcolor{black}{Compared to the traditional CDC Ensemble Model, PandemicLLM improves forecasting accuracy by 119.7\% and reduces the mean square error (MSE) by 72.4\%~\cite{du2025advancing}.}

\begin{figure}[th]
    \centering    \includegraphics[width=0.99\linewidth]{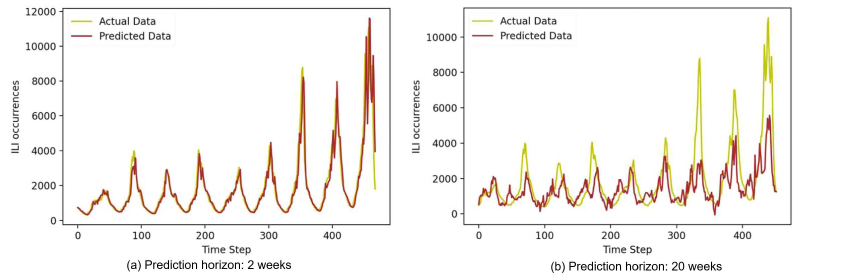}
    \caption{Prediction of ILI using LLMs to mine textual and time-series data. The horizontal axis represents different time steps within one ILI cycle, and the vertical axis denotes the number of ILI occurrences. Subfigure (a) compares the actual and predicted numbers of ILI occurrences two weeks ahead at different time steps, while subfigure (b) shows the corresponding results for twenty weeks ahead. \\Source: Reproduced from Ref.~\cite{moon2025miflu}}
    \label{fig:ILIResults}
\end{figure}

Similarly, Moon et al.~\cite{moon2025miflu} adopted LLMs to jointly learn from textual and time-series data for the prediction of influenza like illness (ILI). They employed two LLMs in total. The first was applied to textual data, where one LLM is used to transform textual data into representational vectors. The second was used in the prediction phase, where another LLM is fine-tuned with the representational vectors as input to obtain a predictive model. The input to this predictive model also includes a time series representational vector, which is concatenated with the textual representational vector and then fed into the predictive LLM as input. As shown in Fig.~\ref{fig:ILIResults}, the model is evaluated on predicting ILI occurrences two and twenty weeks ahead. The results demonstrate that short-term (two-week) forecasts achieve high accuracy, providing valuable decision support for experts making epidemic intervention decisions, while long-term (twenty-week) forecasts can still be further improved. 

For scenarios relying solely on time-series data, Saeed et al.~\cite{saeed2024llm4cast} proposed LLM4Cast, a framework that leverages pretrained LLMs to forecast viral disease trends. Multivariate time series are preprocessed by converting them into independent single-channel sequences to reduce inter-channel noise, applying z-score and instance normalization to address non-stationarity, and dividing the data into patches that are linearly embedded into high-dimensional representations. The model integrates a bi-directional Transformer encoder with a pretrained TinyLlama module, and training proceeds in two stages: (i) generalization training, where TinyLlama is adapted from textual to time-series modeling with diversified datasets across multiple domains, and (ii) domain-specific training, where TinyLlama is fine-tuned on viral disease data. This staged strategy produces a lightweight LLM framework capable of accurate { prediction of viral diseases} while supporting cross-domain generalization and efficient adaptation.

\begin{figure}[th]
    \centering    \includegraphics[width=0.9\linewidth]{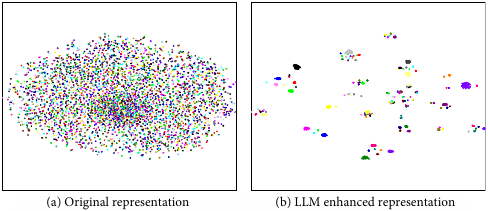}
    \caption{Visualization comparison of disease representation vectors: (a) without LLM representations, and (b) with LLM-based representations. { The representations without LLM enhancement display scattered and overlapping distributions, with indistinct boundaries between classes, making class differentiation difficult. By contrast, the LLM-based representations show substantial improvements, forming well-separated and well-defined clusters across different disease classes.} \\Source: Reproduced from Ref.~\cite{kang2025llm}}
    
\label{fig:LLM_Representation}
\end{figure}

The aforementioned studies collectively demonstrate that leveraging the powerful analytical capabilities of LLMs for textual and time series data has substantially improved the accuracy of infectious disease forecasting. This forecasting can take two main forms: one aims to directly predict future infection scales and transmission trajectories based on historical data over a given time horizon, while the other focuses on individual-level prediction, estimating the infection status of a patient at time $t$ using historical data. From an artificial intelligence perspective, the latter corresponds to a classification task, in which the model determines whether an individual is infected at time $t$ given the past data. As shown in Fig.~\ref{fig:LLM_Representation}, Kang et al.~\cite{kang2025llm} visualized the improvement achieved through LLM-based representations. It can be observed that when raw data are directly analyzed without LLM representation, the representation vectors are difficult to separate. In contrast, after being represented through an LLM, the representation vectors can be effectively classified.

In addition to the above studies, there are other works that are also helpful in the prediction and management of biological epidemics. For instance, Priya et al.\cite{priya2024smart} developed a hybrid intelligent healthcare system by integrating traditional epidemiological models with cutting-edge AI technologies to improve disease forecasting accuracy and address healthcare accessibility issues. LLMs and NLP were used to enhance analysis by analyzing social networks, news, government announcements, and other textual data in real time. This enables capturing public sentiment and risk perception (e.g., panic signals), identifying early epidemic keywords (e.g., symptom discussions), and tracking the spread of rumors to assist in formulating intervention strategies. Jo et al.\cite{jo2024understanding} explored how adding long-term memory functions to LLM-driven chatbots affects user self-disclosure behavior of health and perception of the chatbot. It supports public health monitoring by encouraging individuals to disclose health information in natural conversations. When used in conjunction with public health agencies, municipal public officials monitor call records. If the chatbot detects any negative health signals (e.g., not eating, poor sleep quality, specific health issues) or if someone repeatedly misses calls, public officials can contact the individual. McClymont et al.\cite{mcclymont2024internet} hold the opinion that advances in AI and machine learning algorithms have enhanced the ability to process complex and large-scale datasets, accelerated data mining, and improved the performance of predictive models. Mobile health applications, including AI chatbots, have the potential to capture the healthcare-seeking behavior of internet users to detect early warning signals.

\section{Conclusions and outlooks}
\label{sec6}
Spreading dynamics provides a foundational perspective for understanding collective behavior in complex systems. The emergence and growing integration of LLMs into human activities have expanded spreading dynamics research beyond traditional structure-based and state-based formulations toward richer representations that incorporate textual semantics, multimodal information, and world knowledge. Through LLMs’ powerful abilities in contextual understanding and knowledge reasoning, these previously hard-to-model factors can now be effectively captured, enabling more accurate interpretations of the underlying principles { managing} propagation. Moreover, in cyberspace, massive volumes of rumor or misinformation are now disseminated by intelligent agents guided or controlled by a small number of LLMs, resulting in propagation patterns that differ substantially from those observed prior to the LLM era~\cite{ferrara2024genai}. These changes have driven a new wave of research in spreading dynamics. By systematically analyzing recent research, this review elucidates how LLMs have been leveraged to enhance the modeling, detection, prediction, and management of spreading processes, evaluates the effectiveness of different approaches, and identifies key challenges, unexplained phenomena, and emerging trends. In summary, this review provides conceptual and methodological insights that not only support future research but also offer practical guidance for researchers, institutions, and policymakers seeking to improve the prediction and management of epidemics, rumors, and other large-scale propagation events. 
\textcolor{black}{More broadly, these lines of research are expected to contribute to informed discussions on the potential costs and benefits of policy interventions designed to protect the public and society. While it should be noted that translating insights derived from spreading models into effective and responsible policies remains a challenging task, as it inevitably involves uncertainties, context-dependent risks, and the possibility of unintended or adverse consequences.}

Looking forward, several directions appear particularly promising. 
\begin{itemize}
    \item First, the integration between LLM-driven agent modeling and established theoretical spreading frameworks remains an open area. A key challenge lies in establishing principled mappings between language-based interaction patterns and dynamical variables that are analytically tractable. 
    \item  Second, the role of LLMs as endogenous participants in spreading processes rather than external tools introduces new feedback pathways, raising questions about stability, amplification, resonance, and controllability within spreading systems. \textcolor{black}{As LLMs increasingly act as subjects in information generation and spreading, the environments they shape may in turn influence human cognition, decision-making, and collective behavior, potentially giving rise to emergent forms of ``machine culture''~\cite{brinkmann2023machine}. 
    How to further incorporate this feedback loop between the information ecosystem mediated by LLM and human behavior into a more macroscopic framework of cultural and social dynamics for modeling and analysis is an indispensable research direction for understanding the future human-machine symbiotic communication system.}
    \item Third, evaluating LLM-involved spreading processes calls for methodological advances in measurement~\cite{chang2024survey}, benchmarking~\cite{chen2024benchmarking,chen2025benchmarking,wu2025llm}, and robustness~\cite{zhong2024can}. Unlike classical spreading models with explicit transition rules, LLM-based dynamics require frameworks capable of disentangling cognitive, content-driven, and structural contributions to observed outcomes. 

\item \textcolor{black}{Fourth, how to ensure the robustness and reproducibility of scientific findings in the context of the rapid development of LLMs is an area that requires further exploration. 
As model architectures, training corpora, and inference strategies continue to change, the outcomes of LLM-based spreading analysis may become highly sensitive to specific model versions, parameter settings, and implementation details. Such instability complicates cross-study comparisons and undermines the reproducibility of scientific findings. Moreover, the LLM-driven research on spreading often involves complex perception, cognition and generation processes, making the traditional experimental reproduction paradigm based on fixed rules difficult to be directly applied. Future research may therefore establish standardized experimental protocols, version-control practices, and reproducibility benchmarks tailored to LLM-driven spreading studies, in order to improve the reliability and long-term validity of reported results.}
\item \textcolor{black}{Fifth, modeling higher-order interactions and collective effects beyond pairwise spreading constitutes another important direction. Most existing spreading models are built upon dyadic interactions between nodes, whereas real-world information diffusion is often shaped by group discussions, coordinated behaviors, and collective resonance phenomena. Incorporating such higher-order interactions into LLM-involved spreading frameworks may enable a more faithful representation of group-level coordination, nonlinear amplification, and their impact on macroscopic spreading patterns.}

    \item  Sixth, the issue of interpretability in LLM-driven spreading dynamics is particularly crucial and requires sustained attention. For instance, in biological epidemic modeling and prediction, interpretability directly affects the formulation of intervention strategies by organizations such as the WHO. This necessitates that model predictions be understandable not only to technical users but also to policymakers and the general public. However, LLMs built upon neural network architectures are often regarded as ``black boxes'', as their internal decision-making processes remain largely opaque to human understanding. If such models produce inaccurate predictions that misguide intervention decisions and result in adverse outcomes, this raises serious concerns regarding accountability and trustworthiness~\cite{kraemer2025artificial}.
    \item   Seventh, the issue of bias in LLMs also warrants serious consideration. The training and fine-tuning of large language models rely on massive datasets that may themselves contain inherent biases. For example, many biomedical datasets have historically underrepresented certain demographic groups, including women and ethnic minorities. Such biases can lead to ethical errors or misinterpretations in epidemic prediction and management, potentially compromising public trust in health information and decision-making~\cite{kraemer2025artificial}. 
    \item  Finally, LLMs are highly susceptible to adversarial attacks~\cite{qi2024visual,jiang2025decomposition,jiang2025deceiving} and prone to generating hallucinations~\cite{huang2025survey,farquhar2024detecting}. On the one hand, targeted or untargeted interferences during the training or deployment phases may induce unexpected or misleading outputs, which can severely compromise the accuracy of modeling, detection, and prediction in spreading dynamics. Thus, continuous efforts are required to develop effective defense strategies against such attacks. On the other hand, preventing erroneous judgments caused by model hallucinations in the process of epidemic modeling and analysis also remains an important challenge that warrants sustained attention.
\end{itemize}
 
Overall, the intersection of LLMs and spreading dynamics is shaping a new research landscape in which content, semantics, and behavior are treated as integral components of propagation rather than external modifiers. Advancing this direction will require close collaboration across statistical physics, network science, cognitive modeling, computational social science, 
and machine learning. Such interdisciplinary efforts may contribute not only to more expressive models of spreading but also to 
a deeper understanding of how collective behavior forms, adapts, and evolves in \textcolor{black}{increasingly interconnected human-AI systems}

\section*{Acknowledgements}
We want to acknowledge the partial support from the National Natural Science Foundation of China (No. 62202320 and No. 62402331), Program for Youth Innovation in Future Medicine, Chongqing Medical University (No, W0150), Science and Technology Research Program of Chongqing Municipal Education Commission (No.KJQN202502828), Fundamental Research Funds for the Central Universities (Grant Nos. YJ202429 and SCU2024D012), Science and Engineering Connotation Development Project of Sichuan University (No. 2020SCUNG129), Project of Humanities and Social Sciences Research of Chongqing Municipal Education Commission in 2024 (No. 24SKGH048), Natural Science  Foundation of Chongqing (No. CSTB2025YITP-QCRCX0061), and the Social Science Foundation of Chongqing (2025NDYB065).

\section*{CRediT authorship contribution statement}
Shuyu Jiang: Writing-review \& editing, Writing-original draft, Methodology, Supervision, Conceptualization. 
Hao Ren: Writing-original draft, Methodology, Investigation.
Yichang Gao: Writing-original draft, Investigation. 
Yi-Cheng Zhang: Writing-review \& editing, Supervision. 
Li Qi: Writing-review \& editing, Conceptualization. 
Dayong Xiao: Writing-original draft, Supervision. 
Jie Fan: Writing-review \& editing, Writing-original draft, Supervision, Conceptualization.
Rui Tang: Writing-review \& editing, Writing-original draft, Supervision, Conceptualization. 
Wei Wang: Writing-review \& editing, Writing-original draft, Methodology, Supervision, Conceptualization. Jie Fan(241031@cqmpc.edu.cn), Rui Tang(tangrscu@scu.edu.cn), and Wei Wang(wwzqbx@hotmail.com) are the co-corresponding authors.

\begin{spacing}{0}
\bibliography{mybibfile}
\bibliographystyle{elsarticle-num}
\end{spacing}

\end{sloppypar}
\end{document}